\documentclass[prd,aps,nofootinbib,showpacs,onecolumn, 11 pt]{revtex4}
\usepackage{graphicx}
\usepackage{epsfig}

\begin{document}
\begin{flushright}
DESY 09-081\\
June 2009
\end{flushright}
\title{Anatomy of the pQCD Approach to the Baryonic Decays $\Lambda_b \to p \pi,~p K$}
\author{Cai-Dian L\"u, Yu-Ming Wang, Hao Zou}
\affiliation{ Institute of High Energy Physics and \\
Theoretical Physics Center for Science Facilities,CAS, P.O. Box
918(4), 100049, People's Republic of China}
 \author{Ahmed Ali$^a$ and  Gustav Kramer$^b$ }
\affiliation{
$^a$ Deutsches Elektronen-Synchrotron DESY, 22607 Hamburg, Germany\\
$^b$ II. Institut f\"ur Theoretische Physik, Universit\"at Hamburg,
22761 Hamburg, Germany}
\date{\today}

\begin{abstract}
We calculate the CP-averaged branching ratios and CP-violating
asymmetries for the two-body charmless hadronic decays $\Lambda_b \to p
\pi,~p K$ in the perturbative QCD (pQCD) approach to lowest order in
$\alpha_s$. The baryon distribution amplitudes involved in the
factorization formulae are considered to the leading twist accuracy
and the distribution amplitudes of the proton are expanded to the
next-to-leading conformal spin (i.e., ``P" -waves), the moments of
which are determined from QCD sum rules. Our work shows that
the contributions from the factorizable diagrams in $\Lambda_b \to p
\pi,~pK$ decays are much smaller compared to the
non-factorizable diagrams in the conventional pQCD approach. We argue
that this reflects the estimates of the $\Lambda_b \to p$
transition form factors in the $k_T$ factorization approach,
which are found typically an order of magnitude smaller than those
estimated in the light-cone sum rules and in the non-relativistic
quark model. As an alternative, we adopt a hybrid
pQCD approach, in which we compute the factorizable contributions with the
$\Lambda_b \to p$ form factors taken from the light cone QCD sum rules.
The non-factorizable diagrams are evaluated utilizing
the conventional pQCD formalism which is free from the endpoint singularities.
The predictions worked out here are  confronted with the
recently available data from the CDF collaboration on the branching
ratios and the direct CP asymmetries for the decays $ \Lambda_b \to p
\pi$, and  $\Lambda_b \to p K$.  The asymmetry parameter
$\alpha$ relevant for the anisotropic angular distribution of the
emitted proton in the polarized $\Lambda_b$ baryon decays is also
calculated for the two decay modes.

\end{abstract}


\maketitle

\newpage
\section{Introduction}

The motivation to investigate $b$ quark decays is attributed to
their sensitivity to the quark flavor structure, which leads to
an extremely rich phenomenology, studied mostly in the context of
$B$-meson decays. However, 
heavy baryons containing a $b$-quark have been observed at the Tevatron
and they will be even more copiously
produced at the large hadron collider (LHC). Their weak decays
may provide important clues on the flavor-changing currents beyond the
standard model (SM) in a complementary fashion to the $B$ meson
decays. A particular advantage of the bottom baryons over $B$-mesons is
their spin, which provides a unique way to analyze the helicity
structure of the effective Hamiltonian for the weak transition in the SM
and beyond. Also, such baryon decays are flavor self-tagging processes
which should make their experimental reconstructions easier.

Theoretical analysis of non-leptonic decays are based on
factorization theorems, which are the fundamental tools of the QCD perturbation
theory enabling the separation of physics at different
energy scales. The theoretical basis of the factorization theorem is
a generalization of the Euclidean operator product expansion to
the time-like domain. The proof of the factorization theorem has been
worked out using the perturbative QCD  approach based on the analysis
of Feynman diagrams in the so called Collins-Soper-Sterman (CSS)
formalism \cite{Catani:1989ne,Collins:1981uk,Collins:1984kg}.
Equally importantly, the large mass of the heavy quark makes the formidable
strong interactions effects controllable and they can be studied systematically using
 methods based on  heavy quark expansion.

The basic formula for the calculation of the branching ratios for the decays of
the $\Lambda_b$ baryon into two light hadrons is based on an operator
realization of the diagrammatic analysis which can be described most easily
for the calculation of the hadronic matrix element of $B$ mesons decays into
two light hadrons $h_1$ and $h_2$. With the insertion of a set of the weak interaction
operator $O_i$ between the initial $B$ meson and the final decay products
$h_1$ and $h_2$, the decay matrix element is obtained from the following
formula~\cite{Beneke:2007zz}
\begin{equation}
\langle h_1 h_2 | {\cal O}_i | B \rangle= \Phi_{h_2}(u) \otimes
\left( T^I(u) F^{Bh_1} (0) + C^{II} (\tau, u) \otimes
\Xi^{Bh_1}(\tau,0) \right) \label{ff-SCET}
\end{equation}
involving the QCD form factor $ F^{Bh_1} (0)$ and an unknown,
non-local form factor $\Xi^{Bh_1}(\tau,0)$ at the leading power in
the $\Lambda/m_b$ expansion. Different treatments of the various
parts in the factorization formula (\ref{ff-SCET})
 have led to three popular theoretical approaches to study
the dynamics of non-leptonic two-body $B$ meson decays, which are
known as the perturbative QCD (pQCD) \cite{PQCD}, QCD factorization
(QCDF) \cite{QCDF} and SCET approaches~\cite{Bauer:2000yr,Bauer:2001yt,Bauer:2002nz}.
The function $\Xi^{Bh_1}(\tau,0)$ is supposed to be dominated by perturbative
hard-collinear interactions, and can be further factorized into
light-cone distribution amplitudes $\Phi_B(\omega)$, $\Phi_{h_1}(v)$ and a
jet function $J(\tau; \omega, v) $
\begin{equation}
\Xi^{Bh_1}(\tau,0) = J(\tau; \omega, v) \otimes \Phi_B(\omega)
\otimes \Phi_{h_1}(v) ,
\end{equation}
when the hard-collinear scale $\sqrt{m_b \Lambda_{QCD}}$ is
integrated out \cite{Beneke:2003pa}.

In contrast to these two latter approaches based on the collinear
factorization theorem, the pQCD approach, which is developed in the
framework of $k_T$ factorization, is free of the singularities from
the end-point region of the parton momentum fractions. The pQCD approach
has been widely applied for the calculation of the non-leptonic two-body
$B$ decays and it has proved itself to be successful in the description of
exclusive processes with typical momentum-transfer of
a few ${\rm GeV}$. A hallmark of this approach is that the form factor
$F^{Bh_1} (0)$ is assumed to be dominated by short-distance
contributions and it is therefore calculable in the perturbative
theory.  Soft contributions, though playing a role, are less
important because of the suppression from the Sudakov mechanism
embedded in the $k_T$ and threshold resummations~\cite{Li:2003yj}.
Current applications of the $k_T$ factorization
theorem to exclusive processes are restricted to the leading order (LO)
in the strong coupling constant $\alpha_s$. In this, the infrared
divergences involved in the radiative corrections to the weak
transition vertex are absorbed in the hadronic distribution
amplitudes in a gauge invariant manner. The factorizable,
non-factorizable and power-suppressed annihilation contributions are
calculable in this framework free of the end-point singularities.

In the case of non-leptonic two-body $B$ decays,
the decay matrix elements, in most cases, are dominated by the factorizable
 term, i.e.
the first term on the right-hand side of Eq.~(\ref{ff-SCET}),
whereas the second term, the non-factorizable one, produces a
perturbative correction.
Since the first term proportional to the form factor $F^{Bh_1}(0)$ is
in pQCD very similar to the other approaches, where the form factors are input,
the pQCD approach gives in most cases similar results for the non-leptonic
$B$ decays as the other two approaches mentioned above, though there are
differences in detail.

In the application of pQCD to two-body non-leptonic heavy baryon
decays, we do not expect a similar pattern as in the non-leptonic $B$ meson
decays on general grounds.
In particular, in the analysis
of the hadronic decays of baryons, a large number of Feynman
diagrams contribute to the hard amplitudes even in the
lowest-order. Taking the $\Lambda_b \to p \pi$ decay as an example,
some 200 Feynman diagrams need to be calculated as can be seen
in section~\ref{ff-Feynman-diagrams}. These diagrams involve the exchange of
two gluons, involving topologies where both gluons are attached to one
of the light quarks emerging from the weak interaction vertex.
As some of these diagrams build up the transition form factor, they receive contributions
in $\alpha_s^2$, yielding small values for them.
Another challenge for the baryonic transition is that the
light-cone distribution amplitudes (LCDAs) of the baryons are less known in the
literature. LCDAs  are fundamental non-perturbative input to
regularize the infrared divergence appearing in the radiative
corrections in the factorization formalism of the pQCD approach.
In view of this, applications of the pQCD approach to non-leptonic two-body
$b$-baryon decays are not worked out to a satisfactory level, and hence this area is
essentially an uncharted territory.

A first attempt to apply the pQCD approach to the baryonic
transitions was made in~\cite{Li:1992ce}, where the proton
Dirac form factor is calculated taking into account the Sudakov
suppression resulting from the resummation of the large double
logarithms involved in the radiative corrections. Subsequently, the
proton  form factor was recalculated in \cite{Kundu:1998gv} by
refining the choice of the evolution scale of the proton wave functions
and the infrared cutoffs for the Sudakov resummation, which lead to
predictions for the Dirac form factors which are consistent with the
experimental data. Following Refs.~\cite{Li:1992ce,Kundu:1998gv},
 the semileptonic charmless decays
$\Lambda_b \to p l\bar{\nu}$ \cite{Shih:1998pb}, the semileptonic
charming decay  $\Lambda_b \to \Lambda_c l\bar{\nu}$
\cite{Shih:1999yh,Guo:2005qa}, the radiative decay $\Lambda_b \to
\Lambda \gamma$ \cite{He:2006ud},
%baryonic $B$ decay of $\bar{B}_0 \to \Lambda_c \bar{p}$ \cite{He:2006vz}
and the nonleptonic charming decay $\Lambda_b \to \Lambda J/\psi$
\cite{Chou:2001bn} have been investigated in the framework of the
$k_T$ factorization scheme. However, a study of the charmless
hadronic decays $\Lambda_b \to h_1 h_2$, which has been undertaken
in the generalized factorization approach
\cite{Mohanta:2000nk,Mohanta:2000uv}, to the best of our knowledge,
is still lacking in  pQCD. Our aim is to fill in this gap and
provide further tests of the $k_T$ factorization formalism to gain
insight on the QCD dynamics of these decays. In doing this, we have
included the current information on the CKM matrix elements, updated
some input hadronic parameters, such as the distribution amplitudes
of the proton, which are systematically studied in
\cite{DAs-nucleon-1} making use of the conformal symmetry of the QCD
Lagrangian, and have used data to fix some other input
quantities. We find that the non-factorizable contributions to the
hard amplitudes overwhelm the ones from the factorizable diagrams
in the baryonic decays $\Lambda_b \to p \pi,~p K$. This
feature of the $b$-baryonic decays is at variance with what is found in the naive
factorization approximation and in the corresponding two-body $B$ meson
decays. Large
non-factorizable effects existing in the charmed baryon decays have
been pointed out in the literature \cite{Cheng:1993gf}, where it is
observed that the non-factorizable diagrams escaping from the helicity
and color suppression can be comparable to and sometimes even
dominate over the factorizable contributions.

The layout of the paper is as follows: In section II, we briefly
review the pQCD approach and give the essential input quantities
that enter this approach, including the operator basis used
subsequently and the LCDAs for the pseudoscalar
mesons, the proton as well as the $\Lambda_b$ baryon. Input values
of the various mesonic decay constants and  the baryonic wave function
at the origin in configuration space are also collected there.
Section III contains the calculation of the $\Lambda_b \to p \pi, ~p
K$ decays, making explicit the contributions from the external $W$
emission diagrams $T$, the internal $W$ emission diagrams $C$, the
$W$ exchange diagram $E$, the bow-tie contraction diagrams $B$ and
the penguin diagrams $P$, as shown in Fig.
\ref{all-Feynman-diagrams}.  Details of the calculations are
relegated to the two Appendices (Appendix A, where the Fourier
integration to derive the hard amplitudes in the impact parameter
(or $b$) space  are displayed, and Appendix B, where the
factorization formulae for the Feynman diagrams corresponding to
various toplogies are given). The decay amplitudes called $f_1$ and
$f_2$, defined in Eq.~(\ref{ff-definition}), resulting from
 the diagrams with different topologies evaluated in
the conventional pQCD approach are given numerically in Table
 \ref{results-different-topologies}. We find  that the $T$
diagrams dominate the $\Lambda_b \to p \pi, ~p K$ decays, as expected.
Numerical values of the factorizable and non-factorizable contributions from
the $T$ diagram amplitudes $f_i(\Lambda_b \to p \pi, ~p K); i=1,2$, in the
conventional pQCD approach are given  in Table \ref{fnf-effects}. From the
entries in this table we observe that the factorizable amplitudes in
these decays are essentially two orders of magnitude smaller than the
 corresponding  non-factorizable amplitudes.
The form factor $g_1$ responsible for the $\Lambda_b \to p$ transition
evaluated in various theoretical approaches are collected in Table
 \ref{ff-different-approaches}, and we find that
$g_1$ calculated in the pQCD approach is
typically an order of magnitude smaller than in other approaches
 \cite{Mohanta:2000nk,Huang:2004vf}, where the
form factors are dominated by soft dynamics. Subsequently, we employ
a hybrid prescription to deal with  the hadronic $\Lambda_b \to p
\pi,~p K$ decays. In this approach, the factorizable contributions are
 parametrized in the naive factorization approximation, and the variation of the
renormalization scale is assumed to reflect the effect of the vertex
corrections. The non-factorizable diagrams are  evaluated, as in the
conventional pQCD approach, in the
framework of the $k_T$ factorization.
Following this procedure and utilizing the from
factors calculated in the light-cone sum rules (LCSR), we reanalyze
these two channels and give the numerical results for the amplitudes
$f_i(\Lambda_b \to p \pi, ~p K); i=1,2$, for the factorizable and
non-factorizable contributions from the hybrid scheme in
 Table \ref{fnf-mixed-scheme}. We note that the factorizable contributions
are much larger in the hybrid scheme and they constitute a good fraction of the
corresponding non-factoriazable amplitudes. Numerical results for
the charge-conjugated averages of the decay branching ratios, direct
CP-asymmetries and polarization asymmetry parameter $\alpha$ are
tabulated in Table \ref{BR-CP}. A comparison of our
predictions  with the available experimental data~\cite{ Morello:2008gy}
 are also included in this
 table. Section IV contains our conclusion and an outlook.

\section{Conventions, inputs and some formulae in PQCD }

\subsection{Effective Hamiltonian}

We specify the weak effective Hamiltonian \cite{buras}:
 \begin{eqnarray}
 {\cal H}_{eff} &=& \frac{G_{F}}{\sqrt{2}}
     \Bigg\{ V_{ub} V_{uq}^{\ast} \Big[
     C_{1}({\mu}) Q^{u}_{1}({\mu})
  +  C_{2}({\mu}) Q^{u}_{2}({\mu})\Big]
  -V_{tb} V_{tq}^{\ast} \Big[{\sum\limits_{i=3}^{10}} C_{i}({\mu}) Q_{i}({\mu})
  \Big ] \Bigg\} + \mbox{h.c.} ,
 \label{eq:hamiltonian01}
 \end{eqnarray}
where $q=d,s$. The functions $Q_{i}$ ($i=1,...,10$) are the local
four-quark operators:
 \begin{itemize}
 \item  current--current (tree) operators
    \begin{eqnarray}
  Q^{u}_{1}=({\bar{u}}_{\alpha}b_{\beta} )_{V-A}
               ({\bar{q}}_{\beta} u_{\alpha})_{V-A},
    \ \ \ \ \ \ \ \ \
   Q^{u}_{2}=({\bar{u}}_{\alpha}b_{\alpha})_{V-A}
               ({\bar{q}}_{\beta} u_{\beta} )_{V-A},
    \label{eq:operator02}
    \end{eqnarray}
     \item  QCD penguin operators
    \begin{eqnarray}
      Q_{3}=({\bar{q}}_{\alpha}b_{\alpha})_{V-A}\sum\limits_{q^{\prime}}
           ({\bar{q}}^{\prime}_{\beta} q^{\prime}_{\beta} )_{V-A},
    \ \ \ \ \ \ \ \ \
    Q_{4}=({\bar{q}}_{\beta} b_{\alpha})_{V-A}\sum\limits_{q^{\prime}}
           ({\bar{q}}^{\prime}_{\alpha}q^{\prime}_{\beta} )_{V-A},
    \label{eq:operator34} \\
     \!\!\!\! \!\!\!\! \!\!\!\! \!\!\!\! \!\!\!\! \!\!\!\!
    Q_{5}=({\bar{q}}_{\alpha}b_{\alpha})_{V-A}\sum\limits_{q^{\prime}}
           ({\bar{q}}^{\prime}_{\beta} q^{\prime}_{\beta} )_{V+A},
    \ \ \ \ \ \ \ \ \
    Q_{6}=({\bar{q}}_{\beta} b_{\alpha})_{V-A}\sum\limits_{q^{\prime}}
           ({\bar{q}}^{\prime}_{\alpha}q^{\prime}_{\beta} )_{V+A},
    \label{eq:operator56}
    \end{eqnarray}
 \item electro-weak penguin operators
    \begin{eqnarray}
     Q_{7}=\frac{3}{2}({\bar{q}}_{\alpha}b_{\alpha})_{V-A}
           \sum\limits_{q^{\prime}}e_{q^{\prime}}
           ({\bar{q}}^{\prime}_{\beta} q^{\prime}_{\beta} )_{V+A},
    \ \ \ \
    Q_{8}=\frac{3}{2}({\bar{q}}_{\beta} b_{\alpha})_{V-A}
           \sum\limits_{q^{\prime}}e_{q^{\prime}}
           ({\bar{q}}^{\prime}_{\alpha}q^{\prime}_{\beta} )_{V+A},
    \label{eq:operator78} \\
     Q_{9}=\frac{3}{2}({\bar{q}}_{\alpha}b_{\alpha})_{V-A}
           \sum\limits_{q^{\prime}}e_{q^{\prime}}
           ({\bar{q}}^{\prime}_{\beta} q^{\prime}_{\beta} )_{V-A},
    \ \ \ \
    Q_{10}=\frac{3}{2}({\bar{q}}_{\beta} b_{\alpha})_{V-A}
           \sum\limits_{q^{\prime}}e_{q^{\prime}}
           ({\bar{q}}^{\prime}_{\alpha}q^{\prime}_{\beta} )_{V-A},
    \label{eq:operator9x}
    \end{eqnarray}
 \end{itemize}
where $\alpha$ and $\beta$ are the color indices and $q^\prime$ are
the active quarks at the scale $m_b$, i.e.
 $q^\prime=(u,d,s,c,b)$.
The left handed current is defined as $({\bar{q}}^{\prime}_{\alpha}
q^{\prime}_{\beta} )_{V-A}= {\bar{q}}^{\prime}_{\alpha} \gamma_\nu
(1-\gamma_5) q^{\prime}_{\beta}  $ and the right handed current as
$({\bar{q}}^{\prime}_{\alpha} q^{\prime}_{\beta} )_{V+A}=
{\bar{q}}^{\prime}_{\alpha} \gamma_\nu (1+\gamma_5)
q^{\prime}_{\beta}  $.
For later applications it will be convenient to use the following combinations
of the Wilson coefficients $Q_i$ \cite{Ali:1998eb}:
\begin{eqnarray}
a_1= C_2+C_1/3, & a_3= C_3+C_4/3,~a_5= C_5+C_6/3,~a_7=
C_7+C_8/3,~a_9= C_9+C_{10}/3,\nonumber \\
 a_2= C_1+C_2/3, & a_4=
C_4+C_3/3,~a_6= C_6+C_5/3,~a_8= C_8+C_7/3,~a_{10}= C_{10}+C_{9}/3,
\end{eqnarray}
where the scale dependence for the Wilson coefficients has been
suppressed here. For convenience, we have given the
combinations $a_i$ of the Wilson coefficients at three different values
of the energy scale in Table. \ref{Wilson-coefficients}.
\begin{table}[tb]
\caption{Numerical values of the effective Wilson coefficients
defined in the text at three different scales $\mu$, where $m_b$
is taken as $4.8~{\rm GeV}$.}
\begin{center}
  \begin{tabular}{ccccc}
  \hline\hline
  $\mu$ (GeV)                 & \hspace{2 cm} $0.5 m_b$    &  $\hspace{2 cm} m_b$       & \hspace{2 cm} $1.5 m_b$          \\  \hline
   $a_1$                      & \hspace{2 cm} $1.06$       &  $\hspace{2 cm} 1.03$      & $\hspace{2 cm} 1.02$         \\  \hline
   $a_2$($\times 10^{-2}$)    & \hspace{2 cm} $0.40$       &  $\hspace{2 cm} 10.3$      & $\hspace{2 cm} 14.8$       \\  \hline
   $a_3$($\times 10^{-3}$)    & \hspace{2 cm} $6.41$       &  $\hspace{2 cm} 3.60$      & $\hspace{2 cm} 2.63$         \\  \hline
   $a_4$($\times 10^{-3}$)    & \hspace{2 cm} $-32.6$      &  $\hspace{2 cm} -22.8$     & $\hspace{2 cm} -18.3$      \\  \hline
   $a_5$($\times 10^{-3}$)    & \hspace{2 cm} $-5.87$      &  $\hspace{2 cm} -2.29$     & $\hspace{2 cm} -1.20$      \\  \hline
   $a_6$($\times 10^{-3}$)    & \hspace{2 cm} $-48.2$      &  $\hspace{2 cm} -29.8$     & $\hspace{2 cm} -22.5$    \\  \hline
   $a_7$($\times 10^{-4}$)    & \hspace{2 cm} $12.6$       &  $\hspace{2 cm} 12.2$      & $\hspace{2 cm} 12.0$        \\  \hline
   $a_8$($\times 10^{-4}$)    & \hspace{2 cm} $9.79$       &  $\hspace{2 cm} 7.57$      & $\hspace{2 cm} 6.69$       \\  \hline
   $a_9$($\times 10^{-4}$)    & \hspace{2 cm} $-84.5$      &  $\hspace{2 cm} -82.2$     & $\hspace{2 cm} -81.4$      \\  \hline
   $a_{10}$($\times 10^{-4}$) & \hspace{2 cm} $-0.32$      &  $\hspace{2 cm} -8.20$      & $\hspace{2 cm} -11.9$        \\  \hline
    \hline
    \end{tabular}
    \label{Wilson-coefficients}
\end{center}
\end{table}

\subsection{Kinematics}

The kinematic variables of the initial and final hadrons can be defined as
follows. The $\Lambda_b$ baryon is assumed to be at rest, and the
proton recoils in the minus $z$ direction.  $p$, $p'$ and $q=p-p'$
denote the momentum of the $\Lambda_b$ baryon, the proton and the
light meson, respectively. The momenta of their valence quarks are
parametrized as
\begin{eqnarray}
& &p=(p^+,p^-,{\bf 0})=\frac{M_{\Lambda_b}}{\sqrt{2}}(1,1,{\bf
0})\;,
\nonumber\\
& &k_1=(x_1 p^+,p^-,{\bf k}_{1T})\;,\;\;\;\; k_2=(x_2 p^+,0,{\bf
k}_{2T})\;,\;\;\;\; k_3=(x_3 p^+,0,{\bf k}_{3T})\;,
\nonumber\\
& &p'=(0,{p'}^-,{\bf 0})=(0, {p}^-,{\bf 0})\;,
\nonumber\\
& &k'_1=(0,x'_1 {p'}^-,{\bf k'}_{1T})\;,\;\;\;\; k'_2=(0,x'_2
{p'}^-,{\bf k'}_{2T})\;,\;\;\;\; k'_3=(0,x'_3 {p'}^-,{\bf
k'}_{3T})\;,
\nonumber\\
& &q=(q^+, 0,{\bf 0})=(p^+,0,{\bf 0})\;,
\nonumber\\
& &q_1=(y q^+, 0,{\bf q}_{T})\;,\;\;\;\; q_2=((1-y) q^+, 0,-{\bf
q}_{T})\;,
\end{eqnarray}
where $k_1$ ($k'_1$) is the $b$ ($u$) quark momentum, $x_i$ ($x'_i$)
are their longitudinal momentum fractions, and ${\bf k}^{(\prime)}_{iT}$ are the
corresponding transverse momenta, satisfying $\sum_l{\bf
k}^{(\prime)}_{lT}=0$. $y$ is the longitudinal momentum fraction
carried by the quark in the emitted light meson and  ${\bf q}_T$ is
its transverse momentum. The kinematics of the non-leptonic two body
decays of $\Lambda_b$ is described in Fig. 1.

\begin{figure}[tb]
\begin{center}
\begin{tabular}{ccc}
\hspace{-2 cm}
\includegraphics[scale=0.6]{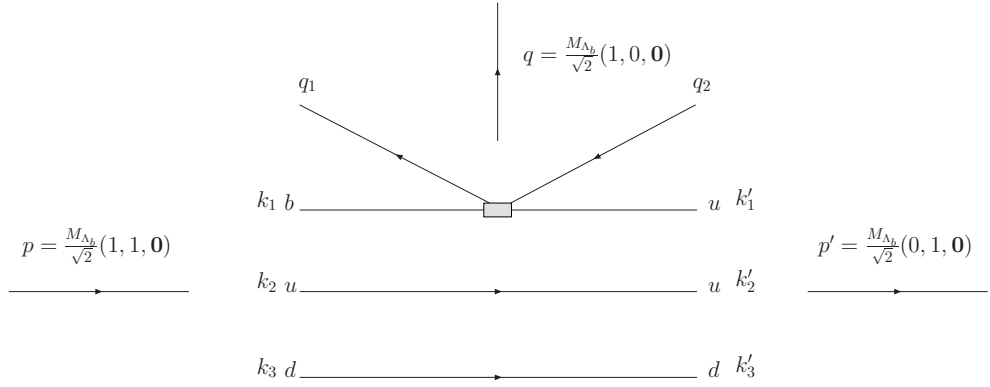}
\end{tabular}
\caption{Kinematics of the non-leptonic two-body decays of $\Lambda_b$
 in the pQCD approach.}\label{kinematics}
\end{center}
\end{figure}

\subsection{Distribution amplitudes of pseudoscalar mesons}
The light-cone  distribution amplitudes for the pseudoscalar meson
are given by \cite{PseudoscalarWV,Li:2005kt}
\begin{eqnarray}
\langle P(P)|{\bar q}_{2\beta}(z)q_{1\alpha}(0)|0\rangle &=&
-\frac{i}{\sqrt{6}}\int_0^1 dx e^{ixP\cdot z} \left[\gamma_5\not\!
P\phi^A(x) +m_0\gamma_5\phi^P(x) -m_0\sigma^{\mu\nu}\gamma_5
P_{\mu}z_{\nu} \frac{\phi^{\sigma}(x)}{6}\right]_{\alpha\beta}  \nonumber\\
&=&-\frac{i}{\sqrt{6}} \int_0^1dx e^{ixP\cdot z}\left[\gamma_5\not
\! P\phi^A(x) + \gamma_5m_0\phi^P(x) +m_0\gamma_5(\not \! n\not
\!v-1) \phi^T(x)\right]_{\alpha\beta}\;, \nonumber\\
&&
 \label{fpd}
\end{eqnarray}
where
\begin{eqnarray}
 \phi_{\pi}^A(x) &=& \frac{3f_{\pi}}{\sqrt{6}} x(1-x)[ 1 +0.44C_2^{3/2}(t)], \\
 \phi_{\pi}^P(x) &=& \frac{f_{\pi}}{2\sqrt{6}}[1 +0.43C_2^{1/2}(t) ], \\
 \phi_{\pi}^T(x) &=& -\frac{f_{\pi}}{2\sqrt{6}}[C_1^{1/2} (t)+0.55 C_3^{1/2} (t) ]
 ,\\
 \phi_{K}^A(x) &=& \frac{3f_{K}}{\sqrt{6}}x(1-x)[1+0.17C_1^{3/2}(t)+0.115C_2^{3/2}(t)], \\
 \phi_{K}^P(x) &=& \frac{f_{K}}{2\sqrt{6}} [1+0.24C_2^{1/2}(t)], \\
 \phi_{K}^T(x)
 &=&-\frac{f_{K}}{2\sqrt{6}}[C_1^{1/2} (t)+0.35 C_3^{1/2} (t)] ,
\end{eqnarray}
and the Gegenbauer polynomials are defined as:
\begin{equation}
\begin{array}{ll}
 C^{1/2}_{1}(t)=t ,&C^{3/2}_{1}(t)=3t  \\
 C_2^{1/2}(t)=\frac{1}{2} (3t^2-1),& C_2^{3/2} (t)=\frac{3}{2}
(5t^2-1),
 \\
C_3^{1/2} (t) = \frac{1}{2} t (5t^2 -3), & \\
 C_4^{1/2} (t)=\frac{1}{8} (35 t^4 -30 t^2 +3), & C_4^{3/2}
(t)=\frac{15}{8} (21 t^4 -14 t^2 +1)    ,
\end{array}
\end{equation}
and $t=2x-1$. The decay constants of these mesons are fixed as
$f_{\pi} = 130~{\rm MeV}$ and $f_{K} = 160~{\rm MeV}$ in our
numerical calculations.

\subsection{Distribution amplitudes of baryons}
\subsubsection{Distribution amplitudes of the $\Lambda_b$ baryon}
The Lorentz structure of the $\Lambda_b$ baryon wave function
$Y_{\Lambda_b}$ can be simplified using the Bargmann-Wigner equation
\cite{Hussain:1990uu} in the heavy quark limit, where the spin and
orbital degrees of freedom of the light quark system are decoupled.
In the transverse momentum space,  the wave function of the $\Lambda_b$
baryon is defined as \cite{RA,h.n.li_lambda}
\begin{eqnarray}
(Y_{\Lambda_b})_{\alpha\beta\gamma}(k_i,\mu) & = &
\frac{1}{2\sqrt{2}N_c} \int \prod_{l=2}^3 \frac{dw^-_ld{\bf w}_l}
{(2\pi)^3} e^{ik_l\cdot w_l} \epsilon^{ijk} \langle
0|T[b_\alpha^i(0)u_\beta^j(w_2)
d_\gamma^k(w_3)]|\Lambda_b(p)\rangle\;,
\nonumber \\
& = & \frac{f_{\Lambda_b}}{8\sqrt{2}N_c}
[(p\!\!\!\slash+M_{\Lambda_b})\gamma_5 C ]_{\beta\gamma}
[\Lambda_b(p)]_\alpha \psi (k_i,\mu)\;, \label{psii}
\end{eqnarray}
 where $b$, $u$, and $d$ are the quark fields,
$i$, $j$, and $k$ are the color indices, $\alpha$, $\beta$ and
$\gamma$ are the spinor indices, $C$ is the charge conjugation
matrix, $\Lambda_b(p)$ is the $\Lambda_b$ baryon spinor.  The
normalization constant corresponds to the value of the wave function
at the origin in the configuration space. The numerical value
 $f_{\Lambda_b}=4.28^{+0.75}_{-0.64}\times 10^{-3}$ GeV$^2$ used by us
 is determined from the experimental
data on  the semileptonic decay $\Lambda_b \to \Lambda_c l \bar{\nu}_l$
\cite{Amsler:2008zzb}. The quoted value (within the $\pm 1 \sigma$
range) is also in agreement with the ones  estimated in the QCD sum rule
 (QCDSR) approach~\cite{Wang:2008sm,Groote:1997yr}.

The phenomenological model for the  distribution amplitude of the $\Lambda_b$
baryon employed in this work is borrowed from \cite{Schlumpf:1992ce}
\begin{eqnarray}
\psi(x_1,x_2,x_3)=N x_1 x_2 x_3 {\rm{exp}} \bigg[ -{M_{\Lambda_b}^2
\over 2\beta^2 x_1} -{m_l^2 \over 2\beta^2 x_2} -{m_l^2 \over
2\beta^2 x_3}\bigg] \label{wf-lambda_b-CQM}
\end{eqnarray}
with the shape parameter $\beta =1.0 \pm 0.2$ GeV and the mass of
the light degrees of freedom in the $\Lambda_b$ baryon being
$m_l=0.3$ GeV. The normalization
\begin{eqnarray}
\int [dx] \psi (x_1,x_2,x_3) = 1 \;,
\end{eqnarray}
leads to the constant $N=6.67\times 10^{12}$.  We point out
that the complete set of three-quark distributions amplitudes of the
$\Lambda_b$ baryon has been investigated in Ref. \cite{Ball:2008fw}
in the heavy quark limit and the renormalization-group
equation governing the scale-dependence of the leading twist
distribution amplitude is also derived there. It is shown that the
evolution equation for the leading twist distribution amplitude
includes a piece associated with the Lange-Neubert kernel
\cite{Lange:2003ff} which generates a radiative tail extending to
high energies, and a piece relevant to the Brodsky-Lepage kernel
\cite{Lepage:1979zb}, which redistributes the momentum within the
spectator diquark system. It is sufficient to limit the accuracy of the
current pQCD analysis to the leading twist approximation due to the still large
errors of the experimental data.

The model for the twist-2 distribution amplitude for the $\Lambda_b$
baryon proposed in \cite{Ball:2008fw} is:
\begin{eqnarray}
\psi^{\rm QCD}(\omega, u) = \omega^2  u (1-u) \bigg[{1 \over
\epsilon_0^4 } e^{-{\omega \over \epsilon_0}} + a_2 C_2^{3/2} (2 u
-1) {1 \over \epsilon_1^4} e^{-{\omega \over \epsilon_1}} \bigg]
\label{wf2-lambda_b-QCD}
\end{eqnarray}
with $\epsilon_0 = 200 ^{+130}_{-60} {\rm MeV}$, $\epsilon_1 = 650
^{+650}_{-300} {\rm MeV}$ and $a_2 = 0.333^{+0.250}_{-0.333}$. In
the above representation, $\omega$ is the total energy carried by
the light quarks in the rest frame of $\Lambda_b$ baryon and the
dimensionless parameter $u$ describes the momentum fraction carried
by the $u$ quark in the diquark system. The normalization of
$\psi^{\rm QCD}(\omega,u)$ is
\begin{eqnarray}
\int_0^{\infty} \omega d \omega \int_0^1 d u \psi^{\rm
QCD}(\omega,u)=1 .
\end{eqnarray}

\begin{figure}[tb]
\begin{center}
\begin{tabular}{ccc}
\includegraphics[scale=0.6]{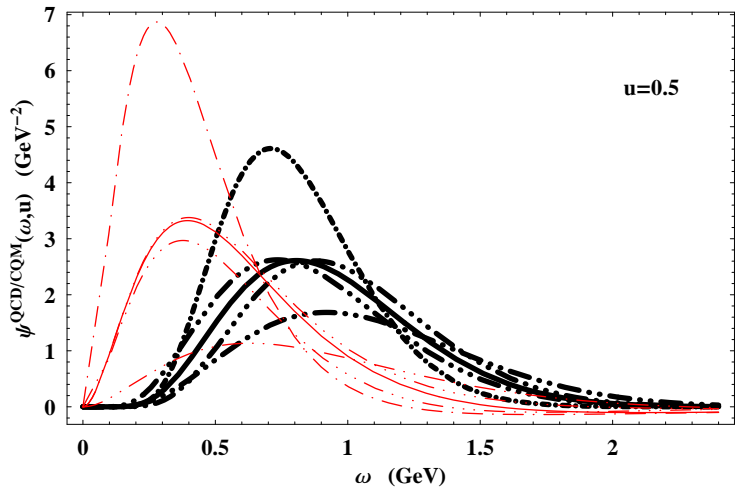}
\end{tabular}
\vspace{-2 cm} \caption{The functions  $\psi^{QCD} (\omega, u)$  and
$\psi^{CQM} (\omega, u)$  plotted against $\omega$ for the fixed
value  $u=0.5$. The solid, dashed-dotted, dashed-double-dotted,
dashed-triple-dotted, and dashed-quartic-dotted curves,  peaking
typically around $\omega =0.8$ GeV, describe the distribution
amplitude $\psi^{CQM} (\omega, u)$ with the values of the parameters
$(\beta=1.0$ GeV, $m_l=0.3$ GeV), $(\beta=0.8$ GeV, $m_l=0.30$ GeV),
$(\beta=1.2$ GeV, $m_l=0.30$ GeV), $(\beta=1.0$ GeV, $m_l=0.24$
GeV), $(\beta=1.0$ GeV, $m_l=0.36$ GeV), respectively. The curves
peaking around $\omega=0.4$ GeV( red curves) correspond to the
distribution amplitudes $\psi^{QCD}(\omega, u)$, where the solid,
dashed-dotted, dashed-double-dotted, dashed-triple-dotted and
dashed-quartic-dotted
 curves correspond to the values of the model parameters
($\epsilon_0=0.20$ GeV, $\epsilon_1=0.65$ GeV), ($\epsilon_0=0.14$
GeV, $\epsilon_1=0.65$ GeV), ($\epsilon_0=0.33$ GeV,
$\epsilon_1=0.65$ GeV), ($\epsilon_0=0.20$ GeV, $\epsilon_1=0.35$
GeV), ($\epsilon_0=0.20$ GeV, $\epsilon_1=1.30$ GeV), respectively.}
\label{fig:wf-lambdab}
\end{center}
\end{figure}

For comparison, we translate  Eq.~(\ref{wf-lambda_b-CQM})
 in terms of the variables $\omega$ and $u$ of Ref. \cite{Ball:2008fw}:
\begin{eqnarray}
\psi^{\rm CQM}(\omega,u) &=&{1 \over M_{\Lambda_b}^4}N \omega^2
u(1-u) \bigg[1- { u \omega \over M_{\Lambda_b}} - { (1-u) \omega
\over M_{\Lambda_b}} \bigg]  \nonumber\\
&&  \times {\rm{exp}} \bigg[ -{M_{\Lambda_b}^2 \over 2\beta^2 (1- {
u \omega \over M_{\Lambda_b}} - { (1-u) \omega \over
M_{\Lambda_b}})} -{m_l^2 \over 2\beta^2  { u \omega \over
M_{\Lambda_b}}} -{m_l^2 \over 2\beta^2 { (1-u) \omega \over
M_{\Lambda_b}}}\bigg]. \label{wf2-lambda_b-CQM}
\end{eqnarray}
The shapes of the LCDAs $\psi^{\rm QCD} (\omega, u)$  and $\psi^{\rm CQM}
(\omega, u)$, given in Eqs.~(\ref{wf2-lambda_b-QCD}) and
(\ref{wf2-lambda_b-CQM}), respectively, are shown in
 Fig.~\ref{fig:wf-lambdab}, and the various curves show the dependence
on the input parameters of the models. The variations of $a_2$
 in $\psi^{\rm QCD}(\omega,u)$  does not play a significant role in the
 behavior of $\psi^{\rm QCD} (\omega, u)$, since the second moment is suppressed
by $\epsilon_0 / \epsilon_1$, and so we have fixed $a_2=0.333$.
At this stage, it is difficult to select one or the other of these LCDAs.
The harder spectrum of $\psi^{CQM} (\omega, u)$ in $\omega$ (the sum of the
energy of the two light quarks in the rest frame of the $\Lambda_b$-baryon)
also reflects in the inverse moments, which are more important for the
dynamics. Following Ref. \cite{Ball:2008fw}, we define the two inverse moments
involving negative powers of the variables $\omega$ and $u$, the
fractional quark momentum
\begin{eqnarray}
\langle (\omega u )^{-1} (\Lambda_{UV})\rangle \equiv
\int_0^{\Lambda_{UV}} d\omega \int_0^1 {du \over u} \psi^{QCD/CQM}
(\omega, u) , \qquad  \langle \omega^{-1}(\Lambda_{UV}) \rangle
\equiv \int_0^{\Lambda_{UV}} d\omega \int_0^1 du \psi^{QCD/CQM}
(\omega, u), \nonumber  \\ \label{lambdab-moments}
\end{eqnarray}
where an additional energy cut $\omega < \Lambda_{UV}$ is introduced
to guarantee that the moments are finite in the presence of a
radiative tail. The values of $\langle (\omega u)^{-1} \rangle $ and
$\langle \omega^{-1} \rangle$ for $\Lambda_{UV}= 2.5 ~{\rm GeV}$ and
$\mu =1 ~{\rm GeV} $ are summarized in Table \ref{number-moments}.
\begin{table}[tb]
\caption{Typical inverse moments defined in Eq.~(\ref{lambdab-moments}) at a
fixed energy cut off  $\Lambda_{UV}= 2.5 ~{\rm GeV}$
and $\mu =1 ~{\rm GeV} $ for the two LCDAs $\psi^{QCD} (\omega, u)$ and
 $\psi^{CQM} (\omega, u)$ discussed in the text. }
\begin{center}
\begin{tabular}{c | c c}
  \hline
    \hline
   & \hspace{2 cm} $\langle \omega^{-1} \rangle  [\rm GeV^{-1}]$ & \hspace{2 cm} $\langle (\omega u)^{-1}  \rangle [\rm GeV^{-1}]$ \\
     \hline
 $\psi^{\rm QCD} (\omega, u)$ & \hspace{2 cm} $1.66^{+0.72}_{-0.67}$ & \hspace{2 cm} $5.38^{+2.22}_{-2.07}$ \\
 $\psi^{\rm CQM}(\omega, u)$ &\hspace{2 cm}  $1.07^{+1.21}_{-0.15}$ & \hspace{2 cm}  $2.53^{+0.42}_{-0.33}$ \\
  \hline
    \hline
\end{tabular}
\label{number-moments}
\end{center}
\end{table}
We note from this table that the moments of the two distribution
amplitudes $\psi^{\rm QCD} (\omega, u)$  and $\psi^{\rm CQM}
(\omega, u)$ are compatible with each other within the errors on the
model parameters (which are large), with the central values of these moments
shifted to lower values for the $\psi^{\rm CQM}(\omega, u)$ LCDA. For the
numerical calculations presented here we use  $\psi^{\rm CQM}(\omega, u)$
with the quoted errors on the model parameters.

\subsubsection{Distribution amplitudes of the proton}

Similarly, the wave functions of the final state proton has
in leading twist the following form \cite{h.n.li_proton}
\begin{eqnarray}
(\bar{Y})_{\alpha\beta\gamma}(k'_i,\mu) &=&-\frac{f_N}{8 \sqrt
{2}N_{c}} \bigg\{[\bar{N}(p')\gamma_{5}]_{\gamma}( C^{-1}
{p}^\prime\!\!\!\slash )_{\alpha\beta} \phi^V(k'_{i},\mu)
+[\bar{N}(p')]_{\gamma}(C^{-1} \gamma_{5} \not{p}^\prime
)_{\alpha\beta} \phi^A(k'_{i},\mu)
\nonumber \\
& & -[\bar{N}(p')\gamma_{5}\gamma^{\mu}]_{\gamma} (C^{-1}\sigma_{\nu
\mu}{p'}^\nu )_{\alpha\beta} \phi^T(k'_{i},\mu) \bigg\}\; ,
\label{4}
\end{eqnarray}
Keeping next-to-leading conformal spin, one obtains the following twist-3
distribution amplitudes \cite{DAs-nucleon-1,DAs-nucleon-2}:
\begin{eqnarray}
\label{All-twist-3} \phi^V(x_i,\mu) &=& 120 x_1 x_2 x_3
\left[\phi_3^0(\mu) + \phi_3^+(\mu) (1- 3 x_3)\right]\,,
\nonumber \\
 \phi^A(x_i,\mu) &=& 120 x_1 x_2 x_3 (x_2 - x_1) \phi_3^-(\mu) \,,
\nonumber \\
 \phi^T(x_i,\mu) &=& 120 x_1 x_2 x_3 \Big[\phi_3^0(\mu) \nonumber-
\frac12\left(\phi_3^+ - \phi_3^-\right)(\mu) (1- 3 x_3)\Big]\,.
\end{eqnarray}
Here the moments of the distribution amplitudes for the proton are
determined by
\begin{eqnarray}
\phi_3^0 = f_N, \,\, \, \, \phi_3^- = \frac{21}{2} f_N \,A_1^u,
\,\,\, \, \phi_3^+ = \frac{7}{2}  f_N \, (1 - 3 V_1^d),
\end{eqnarray}
with all the parameters fixed at the scale $ \mu= 1 $ GeV as
\begin{eqnarray}
 |f_N | &=&         (5.0 \pm 0.5) \times  10^{-3} \mbox{GeV}^2 , \; %\cite{Che84}
\nonumber \\
A_1^u   &=&   0.38 \pm 0.15\,,  %\; \cite{COZ88}
\nonumber \\
V_1^d   &=&    0.23 \pm 0.03\,.  % \; \cite{COZ88}
\end{eqnarray}
It is easy to see that the above proton distribution amplitudes
satisfy the following relations
\begin{eqnarray}
& &\phi^V(x_1,x_2,x_3)=\phi^V(x_2,x_1,x_3)\;,
\nonumber \\
& &\phi^A(x_1,x_2,x_3)=-\phi^A(x_2,x_1,x_3)\;,
\nonumber \\
& &\phi^T(x_1,x_2,x_3)=\phi^T(x_2,x_1,x_3)\;. \label{rel}
\end{eqnarray}

\subsection{A brief review of the conventional pQCD approach}
\label{review of PQCD}

Factorization of amplitudes is a fundamental tool of QCD perturbative
theory to deal with processes involving different energy scales.
Based on the $k_{T}$ factorization, the pQCD approach provides a framework
which has been applied to hard exclusive processes.
In this approach, hard gluon(s) exchange is
essential to ensure the applicability  of the twist expansion, and soft
contributions are expected to be less important owing to the
suppression by the Sudakov factor. This is the case for the transition form factors involving
 mesons. We would like to take the $\Lambda_b
\to p$ transition form factors as an example, first to illustrate the pQCD
factorization theorem, and then offer quantitative estimates for this form factor to check if the
soft contributions remain sub-dominant or not in the baryonic transitions.

The factorization theorem states that the transition form factor can be
expressed as the convolution of hadronic wave functions
$\psi_{\Lambda_b}$, $\psi_p$ and the hard scattering amplitude $T_H$
\begin{eqnarray}
F=\int^1_0 [dx] [dx^{\prime}] \int [d^2{\bf{k}}_{T}] \int
[d^2{\bf{k}}_{T}^{\prime}]
\psi_p(x^{\prime},{\bf{k}}_{T}^{\prime},p^{\prime},\mu)
T_H(x,x^{\prime},M_{\Lambda_b},{\bf{k}}_{T},{\bf{k}}_{T}^{\prime},\mu)
\psi_{\Lambda_b}(x,{\bf{k}}_{T},p,\mu),
\end{eqnarray}
which is usually transformed to the impact parameter $b$ space to
perform the Sudakov resummation of the double logarithms involved in
the radiative corrections to the hadronic wave functions
\begin{eqnarray}
F=\int^1_0 [dx] [dx^{\prime}]\int [{d^2{\bf{b}}}] \int
[d^2{\bf{b}}^{\prime}] {\cal{P}}_p
(x^{\prime},{\bf{b}}^{\prime},p^{\prime},\mu)
T_H(x,x^{\prime},M_{\Lambda_b},{\bf{b}},{\bf{b}}^{\prime},\mu){\cal
P}_{\Lambda_b}(x,{\bf{b}},p,\mu).
\end{eqnarray}
Here ${\cal P}_{\Lambda_b}(x,{\bf{b}},p,\mu)$ and ${\cal{P}}_p
(x^{\prime},{\bf{b}}^{\prime},p^{\prime},\mu)$ are the Fourier
transforms of the $\psi_{\Lambda_b}(x,{\bf{k}}_{T},p,\mu)$ and
$\bar{\psi}_p(x^{\prime},{\bf{k}}_{T}^{\prime},p^{\prime},\mu)$,
respectively. Radiative corrections to the hadronic
wave function can generate a soft logarithm $\alpha_s {\rm ln}\, (Qb)$,
whose overlap with the original collinear logarithm leads to a
double logarithm $\alpha_s {\rm ln}^2(Q b)$. This type of large
logarithm must be organized in order to ensure the validity of the
perturbative expansion. Resummation techniques have been developed to
deal with such double logarithms. The result is a Sudakov exponential
${\rm exp} [-s(Q, b)]$, which decreases fast with  increasing $b$ and vanishes
at $b={1 / \Lambda_{QCD}}$.

The expressions for the Sudakov evolution of the hadronic wave
functions ${\cal P}_{\Lambda_b}(x,{\bf{b}},p,\mu)$ and ${\cal{P}}_p
(x^{\prime},{\bf{b}}^{\prime},p^{\prime},\mu)$ can be expressed as
products of the Sudakov exponents $s(b,Q)$ and reduced hadronic wave
functions, denoted by ${\tilde{\cal
P}}_{\Lambda_b}(x,{\bf{b}},p,\mu) $ and
$\tilde{{\cal{P}}}_p(x^{\prime},{\bf{b}}^{\prime},p^{\prime},\mu )$:
\begin{eqnarray}
{\cal P}_{\Lambda_b}(x,{\bf{b}},p,\mu)=\exp\left[-\sum_{i=2}^{3}
s(w,k_i^{+})\right] {\tilde{\cal P}}_{\Lambda_b}(x,{\bf{b}},p,\mu)\;
,\nonumber \\
{\cal{P}}_p
(x^{\prime},{\bf{b}}^{\prime},p^{\prime},\mu)=\exp\left[-\sum_{i=1}^{3}
s(w^{\prime},k_i^{\prime -})\right] \tilde{{\cal{P}}}_p
(x^{\prime},{\bf{b}}^{\prime},p^{\prime},\mu), \label{sp}
\end{eqnarray}
where $s(b,Q)$  is defined as
\begin{eqnarray}
s(b,Q)&=&~~\frac{A^{(1)}}{2\beta_{1}}\hat{q}\ln\left(\frac{\hat{q}}
{\hat{b}}\right)-
\frac{A^{(1)}}{2\beta_{1}}\left(\hat{q}-\hat{b}\right)+
\frac{A^{(2)}}{4\beta_{1}^{2}}\left(\frac{\hat{q}}{\hat{b}}-1\right)
%\nonumber \\
-\left[\frac{A^{(2)}}{4\beta_{1}^{2}}-\frac{A^{(1)}}{4\beta_{1}}
\ln\left(\frac{e^{2\gamma_E-1}}{2}\right)\right]
\ln\left(\frac{\hat{q}}{\hat{b}}\right)
\nonumber \\
&&+\frac{A^{(1)}\beta_{2}}{4\beta_{1}^{3}}\hat{q}\left[
\frac{\ln(2\hat{q})+1}{\hat{q}}-\frac{\ln(2\hat{b})+1}{\hat{b}}\right]
+\frac{A^{(1)}\beta_{2}}{8\beta_{1}^{3}}\left[
\ln^{2}(2\hat{q})-\ln^{2}(2\hat{b})\right],
\end{eqnarray}
with
\begin{eqnarray}
\hat q\equiv \mbox{ln}[Q/(\sqrt 2\Lambda)],~~~ \hat b\equiv
\mbox{ln}[1/(b\Lambda)],
\end{eqnarray}
and the coefficients $A^{(i)}$ and $\beta_i$ are
\begin{eqnarray}
\beta_1=\frac{33-2n_f}{12},~~\beta_2=\frac{153-19n_f}{24},\nonumber\\
A^{(1)}=\frac{4}{3},~~A^{(2)}=\frac{67}{9}
-\frac{\pi^2}{3}-\frac{10}{27}n_f+\frac{8}{3}\beta_1\mbox{ln}(\frac{1}{2}e^{\gamma_E}),
\end{eqnarray}
$n_f$ is the number of quark flavors and $\gamma_E$ is the Euler
constant. We will use the one-loop running coupling constant, i.e.
we pick up the first four terms in the expression for
the function $s(Q,b)$.

Apart from the double logarithms due to the inclusion of the
transverse momentum, large single logarithms from ultraviolet
divergences can also emerge in the radiative corrections to both the
hadronic wave functions and the hard kernels, which are summed by the
renormalization group (RG) method
\begin{eqnarray}
\left[\mu \frac{\partial}{ \partial \mu}+ \beta(g) \frac{\partial} {
\partial g}\right] \tilde{{\cal P}}_{\Lambda_b}(x,{\bf{b}},p,\mu) &=& -{8 \over 3} \gamma_{q}
\tilde{{\cal P}}_{\Lambda_b}(x,{\bf{b}},p,\mu),
\\
\left[\mu \frac{\partial}{
\partial \mu}+ \beta(g) \frac{\partial} {
\partial g}\right] \tilde{{\cal{P}}}_p (x^{\prime},{\bf{b}}^{\prime},p^{\prime},\mu) &=& -3 \gamma_{q}
\tilde{{\cal{P}}}_p (x^{\prime},{\bf{b}}^{\prime},p^{\prime},\mu),
\\
\left[ \mu \frac{\partial} { \partial \mu}+ \beta(g) \frac{\partial}
{\partial g}
\right]T_H(x,x^{\prime},M_{\Lambda_b},{\bf{b}},{\bf{b}}^{\prime},\mu)
&=& {17 \over 3}
\gamma_{q}T_H(x,x^{\prime},M_{\Lambda_b},{\bf{b}},{\bf{b}}^{\prime},\mu).
\end{eqnarray}
Here the quark anomalous dimension in the axial gauge is
$\gamma_q=-\alpha_s/\pi$. In terms of the above equations, we can
get the RG evolution of the hadronic wave functions and hard
scattering amplitude as
\begin{eqnarray}
\tilde{{\cal P}}_{\Lambda_b}(x,{\bf{b}},p,\mu)&=&\exp\left[-{8 \over
3}\int_{\kappa w}^{\mu} \frac{d\bar{\mu}}{\bar{\mu}}\gamma
_q(\alpha_s(\bar{\mu}))\right]
\times \tilde{{\cal P}}_{\Lambda_b}(x,{\bf{b}},p,w), \nonumber \\
\tilde{{\cal
P}}_{p}(x^{\prime},{\bf{b}^{\prime}},p^{\prime},\mu)&=&\exp\left[-3\int_{\kappa
w^{\prime}}^{\mu} \frac{d\bar{\mu}}{\bar{\mu}}\gamma
_q(\alpha_s(\bar{\mu}))\right] \times \tilde{{\cal
P}_{p}}(x^{\prime},{\bf{b}^{\prime}},p^{\prime},w^{\prime}), \nonumber \\
T_H(x,x^{\prime},M_{\Lambda_b},{\bf{b}},{\bf{b}}^{\prime},\mu)
&=&\exp\left[-{17 \over
3}\,\int_{\mu}^{t}\frac{d\bar{\mu}}{\bar{\mu}}
\gamma_q(\alpha_s(\bar{\mu}))\right]\times
T_H(x,x^{\prime},M_{\Lambda_b},{\bf{b}},{\bf{b}}^{\prime},t)\;,
\end{eqnarray}
The factorization scales $w$ and $w^{\prime}$ represent the
inverse of a typical transverse distance among the three valence
quarks of the initial and final states. The choices of $w$ and
$w^{\prime}$ are
\begin{eqnarray}
w={\rm min} ({1 \over b_1}, {1 \over b_2}, {1\over b_3}), \qquad
w^{\prime}={\rm min} ({1 \over b_1^{\prime}}, {1 \over
b_2^{\prime}}, {1\over b_3^{\prime}}), \label{factorizable scale}
\end{eqnarray}
with the variables $b_1$ and $b_1^{\prime}$ defined as
\begin{eqnarray}
b_1= |{\bf b_2}-{\bf b_3}|, \qquad b_1^{\prime}= |{\bf
b_2}^{\prime}-{\bf b_3}^{\prime}|,
\end{eqnarray}
with the other $b_i$s and $b_i^{\prime}$s defined by permutation.
The introduction of the parameter $\kappa$ is done from the
viewpoint of the resummation, since the scale $\kappa w^{(\prime)}$,
with $\kappa$ of order unity, is equivalent to $w^{(\prime)}$ within
the accuracy of the next-to-leading logarithms \cite{Botts:1989kf}.
The variation of $\kappa$  represents different partitions of the
radiative corrections to the perturbative Sudakov factor and the
non-perturbative wave function. The best fit to the experimental data
of the proton form factor determines the parameter as $\kappa=1.14$
\cite{Kundu:1998gv}.

Furthermore, loop corrections for the weak vertex can also give rise
to another type of double logarithms $\alpha_s {\rm{ln}}^2 x_i$,
which are usually factorized from the hard amplitude and resummed
into the jet function $S_t (x_i)$ to smear the end-point
singularity.
It should be pointed out that the Sudakov factor from
threshold resummation is process-independent, and hence universal~\cite{Kurimoto:2001zj}.
 The following approximate parametrization is proposed in Ref.
\cite{Li:2001ay} for  phenomenological applications
\begin{eqnarray}
S_t(x)=\frac{2^{1+2c}\Gamma(3/2+c)}{\sqrt{\pi}\Gamma(1+c)}
[x(1-x)]^c\;,
\end{eqnarray}
with the parameter $c \simeq 0.3$ determined from the best
fit to the next-to-leading-logarithm threshold resummation in moment
space. The threshold factor modifies  the
end-point behavior of the hadronic distribution amplitudes and forces them to
vanish faster as $x \to 0$. Collecting everything together, we arrive at the
typical expression for the factorization formula of the form factor in the
pQCD approach
\begin{eqnarray}
F&=&\int^1_0 [dx] \int [dx^{\prime}] \int [d^2{\bf{b}}]\int
[d^2{\bf{b}}^{\prime}]\bar{\tilde{{\cal{P}}}}_p
(x^{\prime},{\bf{b}}^{\prime},p^{\prime},w^{\prime})
T_H(x,x^{\prime},M_{\Lambda_b},{\bf{b}},{\bf{b}}^{\prime},t)
{\tilde{\cal P}}_{\Lambda_b}(x,{\bf{b}},p,w) S_t(x^{(\prime)}) \nonumber \\
&&\times \exp\left[-\sum_{i=2}^{3} s(w,k_i^{+}) -{8 \over
3}\int_{\kappa w}^{t} \frac{d\bar{\mu}}{\bar{\mu}}\gamma
_q(\alpha_s(\bar{\mu}))-\sum_{i=1}^{3} s(w^{\prime},k_i^{\prime
-})-{3}\int_{\kappa w^{\prime}}^{t}
\frac{d\bar{\mu}}{\bar{\mu}}\gamma _q(\alpha_s(\bar{\mu}))\right] .
\label{factorization-formulae}
\end{eqnarray}
Apart from the hard perturbative kernel $T_H(x,x^\prime,...)$, the
same expression holds for the mesonic and baryonic transition form
factors. As we shall see quantitatively below, the hard perturbative
kernels entering the latter are parametrically suppressed compared
to the former. Physical interpretation of the Sudakov factor is well
known~\cite{Keum:2004is}, namely it is a probability distribution
function for emitting no soft gluons. When a quark is accelerated in
QCD, infinitely many gluons are emitted. Hence, we may observe many
hadrons (or jets) at the end if gluonic bremsstrahlung occurs.
Therefore, the amplitude for an exclusive decay $\Lambda_b$ to a
light baryon and a light  meson is proportional to the probability
that no bremsstrahlung gluon is emitted. This is just the role of
the Sudakov factor in the $k_{T}$ factorization. It is known that
the Sudakov factor is large only for small transverse intervals
between the quarks in the hadron. A large transverse interval
implies that the quarks in the hadron are separated and hence less
color shielded. Thus the Sudakov factor suppresses the long distance
contributions for the decay amplitude.

\section{Calculations of baryonic decays $\Lambda_b \to p \pi,~ p K$  in the
pQCD  approach}
\label{ff-Feynman-diagrams}

\begin{figure}[tb]
\begin{center}
\begin{tabular}{ccc}
\hspace{-2 cm}
\includegraphics[scale=0.6]{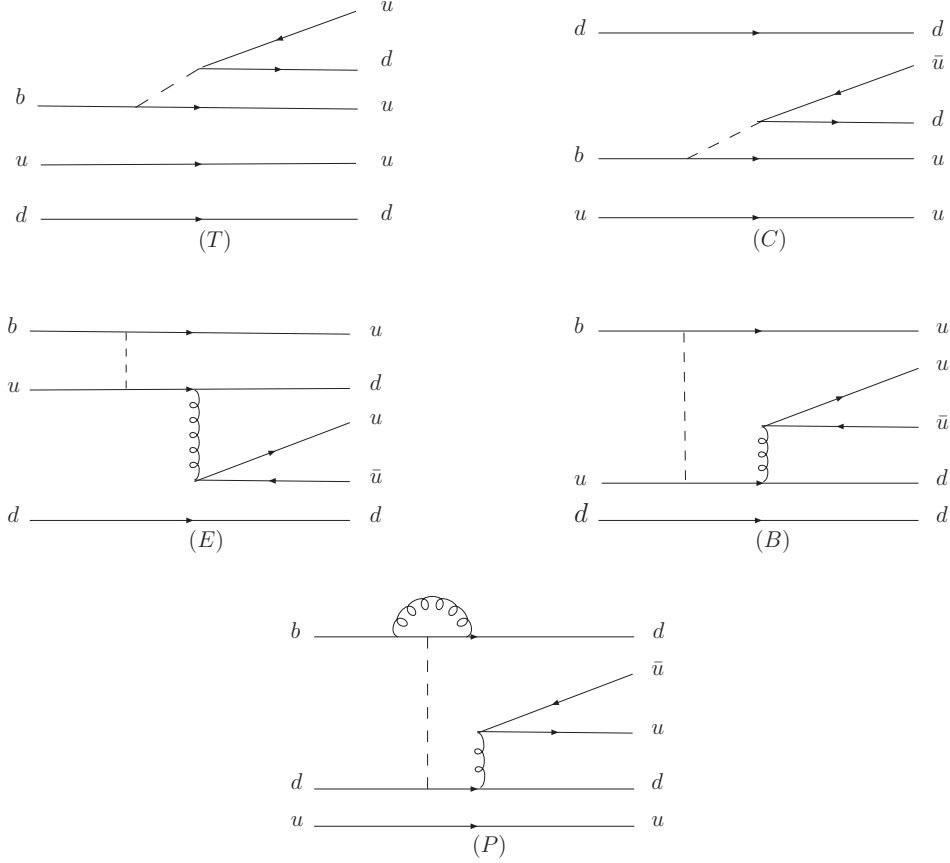}
\end{tabular}
\caption{Topological diagrams responsible for the decay 
$\Lambda_b \to p \pi$, where $T$ denotes the external $W$ emission
diagram; $C$ represents the internal $W$ emission diagram; $E$ labels
$W$ exchange diagram; $B$ denotes the diagram that can be obtained
from the $E$ type diagram by exchanging the two identical down
quarks in the final states; and $P$ represents the diagram that can
only be induced by the penguin operators.}\label{all-Feynman-diagrams}
\end{center}
\end{figure}

Topological diagrams responsible for the decay of
$\Lambda_b$ to a light baryon and a light meson are presented in
Fig. \ref{all-Feynman-diagrams}. In terms of the hard-scattering
mechanism, the exchange of two hard gluons is needed to ensure that
the light spectator quarks in the initial states turn out as
collinear objects in the final state. With this, the various
diagrams for the $\Lambda_b \to p \pi$ decays in the pQCD approach
in the lowest order are displayed in Appendix B. Fig.
\ref{color-allowed-diagrams} shows the external $W$ emission
diagrams, Fig. \ref{color-suppressed-diagrams} the internal $W$
emission diagrams, Fig. \ref{W-exchange-diagrams}
 the $W$ exchange diagrams, Fig. \ref{Bow-tie-diagrams}
 the bow-tie diagrams and
Fig. \ref{penguin-annihilation-diagrams} the penguin diagrams.
 We also include diagrams containing the
three-gluon-vertex displayed in Fig. \ref{three-gluon-diagrams}.
Their contribution is, however, about an order of magnitude smaller
than that from the external $W$ emission ($T$) diagrams, but it can be
comparable to that of the internal $W$ emission ($C$) diagrams. As
for $\Lambda_b \to p K$ decay, only
 Figs.~\ref{color-allowed-diagrams},
 \ref{Bow-tie-diagrams}, \ref{penguin-annihilation-diagrams}, and
\ref{three-gluon-diagrams} contribute to the
decay amplitude.

\subsection{General factorization formulae for $\Lambda_b \to p \pi, ~p K$
decays}

The $\Lambda_b \to p \pi, ~ p K$  decay amplitude $\mathcal{M}$ is
decomposed into two different structures with the corresponding
coefficients $f_1$ and $f_2$:
\begin{eqnarray}
\mathcal{M}=\bar{p}(p^{\prime}) [f_1 + f_2 \gamma_5] \Lambda_b (p).
\label{ff-definition}
\end{eqnarray}
using the equation of motion for a free Dirac particle.
Similar to the factorization formula for the form factors of the $\Lambda_b
\to p$ transition, the coefficients $f_i(i=1,2)$ can be expressed as
\begin{eqnarray}
f_i^j&=& G_F  { \pi^2 \over 144 \sqrt{3}} f_{\Lambda_b} f_p
\sum_{m=V,A,T}^{n=A,P,T} \int [\mathcal{D}x] \int [\mathcal{D}b]^{j}
[\alpha_s(t^j)]^2 a^j(t^j) \psi_{\Lambda_b}(x)
\psi_{p}^{m}(x^{\prime}) \phi^{n}_M(y) H^{mnj}_{i}(x, x^{\prime}, y)
\nonumber \\&& \times \Omega^{j}(b, b^{\prime},b_q) \, {\rm
exp}[-S^j] .
\end{eqnarray}
Here, $f_i^j$ $(i=1,2)$ denotes the contribution to the coefficient $f_i$ by the
``$j_{th}$" diagram displayed in
 Fig.~\ref{color-allowed-diagrams}-\ref{three-gluon-diagrams}, and $a^j$ are the
corresponding Wilson coefficients. The hard function $\Omega^{j}(b, b^{\prime},b_q)$
arises from the Fourier transformation of the denominators of the
internal particle propagators in the $j$th diagram. The hard amplitudes
$H^{mnj}_{i}(x, x^{\prime},y)$  depend on the spin
structures of the three valence quarks in the proton and the form factors
$f_{1,2}$. The integration measure involving the momentum fractions can be written as
\begin{eqnarray}
[\mathcal{D} x]=[d x ][d x^{\prime}] dy, \hspace{0.5 cm} [d x]=dx_1
dx_2 dx_3 \delta(1-\sum_{i=1}^3 x_i), \hspace{0.5 cm} [d
x^{\prime}]=dx^{\prime}_1 dx^{\prime}_2 dx^{\prime}_3
\delta(1-\sum_{i=1}^3 x^{\prime}_i),
\end{eqnarray}
and the expressions for the measure of the transverse extent
$[\mathcal{D} b]$ will be shown in the factorization formulae given in
 Appendix B.

The exponents $S^j$ in the Sudakov factor are determined for the factorizable
diagrams by
\begin{eqnarray}
S^j(x,x^{\prime},b, b^{\prime})=\sum_{i=2}^{3} s(w,k_i^{+})+{8 \over
3}\int_{\kappa w}^{t_j} \frac{d\bar{\mu}}{\bar{\mu}}\gamma
_q(\alpha_s(\bar{\mu}))+\sum_{i=1}^{3} s(w^{\prime},k_i^{\prime
-})+{3}\int_{\kappa w^{\prime}}^{t_j}
\frac{d\bar{\mu}}{\bar{\mu}}\gamma _q(\alpha_s(\bar{\mu}))~,
\end{eqnarray}
and for the non-factorizable diagrams by
\begin{eqnarray}
S^j(x,x^{\prime},y, b, b^{\prime}, b_q)&=&\sum_{i=2}^{3}
s(w,k_i^{+})+{8 \over 3}\int_{\kappa w}^{t_j}
\frac{d\bar{\mu}}{\bar{\mu}}\gamma
_q(\alpha_s(\bar{\mu}))+\sum_{i=1}^{3} s(w^{\prime},k_i^{\prime
-})+{3}\int_{\kappa w^{\prime}}^{t_j}
\frac{d\bar{\mu}}{\bar{\mu}}\gamma _q(\alpha_s(\bar{\mu})) \nonumber
\\ && +\sum_{i=1}^{2} s(w_q,q_i^{+}) +{2}\int_{w_{q}}^{t_j}
\frac{d\bar{\mu}}{\bar{\mu}}\gamma _q(\alpha_s(\bar{\mu}))~,
\end{eqnarray}
where $t^j$ is the typical energy scale of the ``$j_{th}$" diagram and is chosen as
\begin{eqnarray}
t^j= max (t_1^j, t_2^j, t_3^j, t_4^j, w, w^{\prime}, w_q),
\label{hard scale}
\end{eqnarray}
where the hard scales $t_1^j$, $t_2^j$ are relevant to the two
virtual quarks, and $t_3^j$, $t_4^j$ are associated with the two hard
gluons. $w$ and $w^{\prime}$ have been given in Eq.
(\ref{factorizable scale}) and $w_q={1 / b_q}$. The maximum in the
above choice simply indicates that the hard scales should be larger
than the factorization scales.

The factorization formulae for some typical diagrams corresponding
to different topologies in the $\Lambda_b \to p \pi$ decay are given
in Appendix B. The corresponding factorization formulae for
$\Lambda_b \to p K$ decay can be obtained directly following the
same rules.

\subsection{Numerical results for $\Lambda_b \to p \pi, ~p K$ decays}

For the CKM matrix elements, we use as input the updated results  from
\cite{CKMfitter:2006} and drop the (small) errors on $V_{ud}$,
$V_{us}$ and $V_{tb}$:
\begin{equation}
 \begin{array}{lll}
 |V_{ud}|=0.974,   &|V_{us}|=0.225,    &|V_{ub}|=(3.50^{+0.15}_{-0.14})\times 10^{-3},
\\
  |V_{td}|=(8.59^{+0.27}_{-0.29})\times 10^{-3}, &|V_{ts}|=(40.41^{+0.38}_{-1.15})\times 10^{-3},
  &|V_{tb}|=0.999,\\
\beta={(21.58^{+0.91}_{-0.81})}^\circ,&
 \gamma=(67.8^{+4.2}_{-3.9})^\circ.
\end{array}
\end{equation}

It will be shown that the CKM factors mostly yield an overall factor
for the branching ratios and do not introduce large uncertainties to the
numerical results.

\begin{table}[tb]
\caption{The coefficients $f_1$ and $f_2$
contributed by the
Feynman diagrams with definite topologies in the $\Lambda_b \to p \pi$
decay based on the conventional pQCD approach.}
\begin{center}
\begin{tabular}{ccc}
  \hline
  \hline
    & $f_1$ & $f_2$ \\
  \hline
  $T$ & $-2.42 \times 10^{-9} - i 2.07 \times 10^{-9}$ & $-1.74  \times 10^{-9} - i 1.22 \times 10^{-9}$  \\
 $C$ & $2.05 \times 10^{-10} - i 4.60 \times 10^{-10}$ & $-2.35 \times 10^{-10} + i 4.77 \times 10^{-10}$ \\
  $E$ & $2.89 \times 10^{-11} - i 8.95 \times 10^{-12}$ & $1.11 \times 10^{-11} - i 4.36 \times 10^{-12}$ \\
  $B$ & $-7.00 \times 10^{-11} + i 3.33 \times 10^{-10}$ & $2.21 \times 10^{-10} - i 4.04 \times 10^{-11}$  \\
  $P$& $-6.84 \times 10^{-12} + i 4.85 \times 10^{-11}$  & $7.00 \times 10^{-12} - i 4.75 \times 10^{-11}$ \\
  $G$ & $1.37 \times 10^{-10} + i 1.71 \times 10^{-11}$ & $-1.60 \times 10^{-10} + i 2.01 \times 10^{-10}$  \\
  \hline
  \hline
\end{tabular}
\label{results-different-topologies}
\end{center}
 \end{table}

We start by discussing the numerical results in the conventional pQCD approach.
To that end,  we list the coefficients $f_1$ and $f_2$ defined in Eq.
(\ref{ff-definition}) contributed by the Feynman diagrams
with different topologies in the $\Lambda_b \to p \pi$ decay in
Table~\ref{results-different-topologies}. From this table, we
observe that the amplitudes satisfy the relations $T \gg C  \gg E$.

As mentioned earlier, the $T$ type diagrams dominate the $\Lambda_b \to p
\pi$ decays. For this case we present the factorizable and
non-factorizable contributions in the  $\Lambda_b
\to p \pi$ decays in Table \ref{fnf-effects}.
We observe that the factorizable contribution is approximately two orders
of magnitude smaller than the non-factorizable contribution. This is also the reason that
the the conventional pQCD predictions for the semileptonic decay $\Lambda_b \to p l \bar{\nu}$
\cite{h.n.li_proton} and the radiative decay $\Lambda_b \to \Lambda
\gamma$ \cite{He:2006ud} are much smaller than those evaluated in
other theoretical frameworks (such as the constituent quark model or the QCD
sum rules).

\begin{table}[tb]
\caption{The coefficients $f_1$ and $f_2$ in the $\Lambda_b \to p \pi, ~p
K$ decays from the factorizable and non-factorizable external $W$
emission ($T$) diagrams in the conventional pQCD approach.}
\begin{center}
\begin{tabular}{ccc}
  \hline
  \hline
   & factorizable & non-factorizable \\
   \hline
  $f_1 (\Lambda_b \to p \pi )$ & $1.47 \times 10^{-11} -i 1.97 \times 10^{-11}$ & $-2.43 \times 10^{-9} -i 2.05 \times 10^{-9}$ \\
  $f_2 (\Lambda_b \to  p \pi)$ & $1.26 \times 10^{-11} -i 1.94 \times 10^{-11}$& $-1.75 \times 10^{-9} -i 1.20 \times 10^{-9}$ \\
  $f_1 (\Lambda_b \to  p K )$ & $-1.52 \times 10^{-11} -i 0.62 \times 10^{-11}$ & $-0.88 \times 10^{-9} + i 0.54 \times 10^{-10}$ \\
  $f_2 (\Lambda_b \to  p K)$ & $0.17 \times 10^{-11} -i 0.60 \times 10^{-11}$& $-1.06 \times 10^{-9} + i 1.67 \times 10^{-9}$ \\
  \hline
  \hline
\end{tabular}
\label{fnf-effects}
\end{center}
 \end{table}

The suppression of the factorizable contributions in the conventional pQCD
approach has been observed also in the analysis of the
$\Lambda_b \to \Lambda J/\psi$  decays
\cite{h.n.li_lambda}, where the non-factorizable contributions are also found almost
two orders of magnitude larger than those from the factorizable diagrams.
In order to understand the large contribution of the
non-factorizable diagrams in $\Lambda_b$ decays, it is necessary to recall the
role of the Sudakov factor in the $k_T$ factorization approach. As stated in
section II, the Sudakov factor can only suppress the region
with large $b$'s corresponding to small $k_{T}$'s, and has almost no
effect in the region where the transverse momentum $k_{T}$ is
large. Taking the non-factorizable diagram $T_{25}$ as an example,
the two virtual quarks can be on the mass shell even in the region
with large $k_{T}$. Therefore, this diagram is not subjected to the
suppression from the Sudakov factor. It is then expected that the
amplitudes for the non-factorizable diagrams should be much larger than
those from the factorizable diagrams, where the two virtual quarks
can be on the mass shell only in the small $k_T$ region. Actually,
a similar case also occurs in the hadronic $B$ meson decays. There, the
annihilation diagrams contributing to the  $B \to M_1 M_2$ decays in
the pQCD approach is very important, which is responsible for the
large CP violation and the enhancement of the transverse polarization
fractions predicted in the $k_{T}$ factorization. The large
contribution from the annihilation diagrams in the pQCD approach is
due to the fact that the inner quark can be on the mass shell
in the region of large $k_{T}$. The numerical analysis also shows that
the six non-factorizable diagrams $T_{19}, T_{20}, T_{21}, T_{25},
T_{31}, T_{32}$ play the most significant role in the decay
amplitude for the $\Lambda_b \to p \pi$ transition.

We consider the smallness of the factorizable contributions in the conventional
pQCD approach as unrealistic. Consequently, we argue that
the $\Lambda_b \to p$ transition form factors can not be reliably calculated
in the perturbative $k_T$ scheme, i.e.\ these form factors are dominated
by non-perturbative soft contributions, which can not be estimated in
the pQCD approach. Of course, this could easily be checked by measuring
the semileptonic $\Lambda_b$ decays $\Lambda_b \to p \ell \bar{\nu}_\ell$,
 which depend only on the
 factrorizable diagrams. Pending this determination, we consider it as a more reasonable approach
to calculate the $\Lambda_b \to p$ transition form factor by means of some
non-perturbative method.

The form factors of $\Lambda_b \to p$ transition are defined as
\begin{eqnarray}
\langle p (p^{\prime}) |\bar{u}\gamma_{\mu} b|\Lambda_{b}(p)\rangle
&=&\overline{p}(p^{\prime})(g_1 \gamma_{\mu}+g_2 i \sigma_{\mu \nu}
q^{\nu}+g_3 q_{\mu})\Lambda_b(p), \,\, \label{vector matrix element}
\end{eqnarray}
where all the form factors $g_i$ are functions of the square of
momentum transfer $q^2$. We show in Table
 \ref{ff-different-approaches} numerical values for the vector transition
 form factor $g_1$ for the $\Lambda_b \to p$ transition.
These results are obtained in the non-relativistic quark model (NRQM)
\cite{Mohanta:2000nk}, LCSR~\cite{Huang:2004vf}, an  earlier pQCD
calculation \cite{Shih:1998pb}, and this work (also a pQCD
calculation) for comparison. From this Table
we  see that the predictions for the transition form
factor $g_1$ are scattered, with the NRQM~\cite{Mohanta:2000nk} and the
 LCSR~\cite{Huang:2004vf} values differing by a factor 2,
 but the two conventional pQCD results shown, while consistent with each other, are smaller from
 those obtained using the non-perturbative methods typically by an order of magnitude.

\begin{table}[tb]
\caption{The form factor $g_1$  responsible for the $\Lambda_b \to p $
transition at zero momentum transfer,  calculated by us (this work) and 
in the non-relativistic quark model (NRQM),
LCSR, and in another pQCD approach. The uncertainties from the
variations of the hard scale, $\Lambda_{QCD}$ and the shape parameter
$\beta$ in the $\Lambda_b$ wave functions have been combined together
in our work.}
\begin{center}
\begin{tabular}{cccccc}
  \hline
  \hline
    & NRQM \cite{Mohanta:2000nk} & LCSR (full QCD)\cite{Huang:2004vf}
   & pQCD \cite{Shih:1998pb} & pQCD (this work)\\
  $g_1$ & 0.043  & 0.018 & $2.3 \times 10^{-3}$ &   $2.2^{+0.8}_{-0.5} \times 10^{-3}$ \\
  \hline
  \hline
\end{tabular}
\label{ff-different-approaches}
\end{center}
\end{table}

To understand the marked difference of the form factor $g_1$
predicted in the pQCD approach and in the other frameworks, we recall that the
hard dynamics is assumed to be dominant in the heavy-to-light
transition form factors in the former and the
soft contribution, which is not calculable, is assumed to be less important
 due to the Sudakov resummation.
Table \ref{ff-different-approaches} suggests that
the soft dynamics in the heavy-to-light transition form factors is
the dominant effect, in all likelihood
overwhelming the mechanism of the hard
gluon  exchange for the baryonic transitions. Similar
large soft contributions have been also observed in the nonleptonic
charmed meson decays \cite{Li:1997un} as well as in the semileptonic
$\Lambda_b \to \Lambda \gamma, ~\Lambda l^+ l^-$ decays
\cite{He:2006ud, Wang:2008sm}. It is found in \cite{Wang:2008sm}
that the hard contributions to the $\Lambda_b \to \Lambda$ form
factors are almost an order of magnitude smaller than that those from the soft
contributions.

 In the modified version of the pQCD approach, which we call
{\it hybrid } pQCD, the form factors are taken as external inputs. The
perturbative correction to the factorizable amplitude will then enter through
the Wilson coefficients, which are known in next-to-next-to-leading order (NNLO), and the
 vertex corrections, which have been
recently calculated for the tree diagrams in the charmless hadronic
$B$ decays in NNLO~\cite{Bell:2007tv,Bell:2009nk}. As the complete
NNLO corrections, including the QCD penguin amplitudes, are still
not yet at hand,
 we follow the approximate (and less precise) approach proposed in Ref.~\cite{Chen:2005ht}
to neglect the vertex corrections and
vary the renormalization scale $\mu$ of the Wilson coefficients
between $0.5 m_b$ and $1.5 m_b$. Surely, this step of the calculation can be
systematically improved once the complete NNLO virtual corrections are available.
 The non-factorizable contributions will be evaluated as already discussed
in the conventional pQCD approach.

Following the above procedure, we write the complete decay
amplitude for $\Lambda_b \to p \pi, ~p K$ as
\begin{eqnarray}
\mathcal{M}(\Lambda_b \to p \pi)&=&\mathcal{M}_{f}(\Lambda_b \to p
\pi) +
\mathcal{M}_{nf}(\Lambda_b \to p \pi) \, \nonumber \\
\mathcal{M}(\Lambda_b \to p K)&=& \mathcal{M}_{f}(\Lambda_b \to p K)
+ \mathcal{M}_{nf}(\Lambda_b \to p K) ,
\end{eqnarray}
where $\mathcal{M}_{nf}(\Lambda_b \to p \pi)$ and
$\mathcal{M}_{nf}(\Lambda_b \to p K) $ denote the contributions from
the non-factorizable diagrams and have been  computed in the conventional pQCD
approach. To calculate the factorizable  amplitudes
$\mathcal{M}_{f}(\Lambda_b \to p \pi)$ and
$\mathcal{M}_{f}(\Lambda_b \to p K) $, we first need to  deal with
the hadronic matrix elements with the insertion of $(V-A) \otimes
(V+A)$ operators, i.e.,the $O_5 - O_8$ penguin operators. Making use
of the Fierz identity, the factorization assumption and the Dirac
equation, the matrix element of the operator $O_6$ can be written as
\begin{eqnarray}
\label{O6 penguin}
 \langle p M|O_6| \Lambda_b \rangle = [R_1^M
\langle p | \bar{q^{\prime}} \gamma_{\mu}  b| \Lambda_b \rangle +
R_2^M
 \langle p | \bar{q^{\prime}} \gamma_{\mu} \gamma_5 b| \Lambda_b \rangle]
 \langle M |\bar{q} \gamma_{\mu} (1-\gamma_5) q^{\prime}| 0\rangle
\end{eqnarray}
with
\begin{eqnarray}
R_{1}^M = {2 m_{M}^2 \over (m_b - m_u) (m_u +m_q)}, \qquad R_{2}^M =
{2 m_{M}^2 \over (m_b + m_u) (m_u +m_q)},
\end{eqnarray}
where the quark masses are the current quark masses. In addition to the
form factors defined in Eq. (\ref{vector matrix element}), we need the
matrix element describing the $\Lambda_b \to
p$ transition induced by the axial-vector current
\begin{eqnarray}
\langle \Lambda(P)|\bar{s}\gamma_{\mu}\gamma_5
b|\Lambda_{b}(P+q)\rangle &=&\overline{\Lambda}(P)(G_1
\gamma_{\mu}+G_2 i\sigma_{\mu \nu} q^{\nu}+G_3
q_{\mu})\gamma_{5}\Lambda_b(P+q). \,\, \label{axial-vector matrix
element}
\end{eqnarray}
It is then straightforward to write the factorizable amplitudes
$\mathcal{M}_{f}(\Lambda_b \to p \pi)$ and
$\mathcal{M}_{f}(\Lambda_b \to p K) $ as
\begin{eqnarray}
\label{fC-NF-pion}
&&\mathcal{M}_{f}(\Lambda_b \to p \pi) \nonumber \\
&&={G_F \over \sqrt{2}} f_{\pi} \bar{p}(p^{\prime})\bigg  \{ \bigg
[V_{ub}V_{ud}^{\ast} a_1 - V_{tb}V_{td}^{\ast}(a_4+a_{10}+R_1^{\pi}
(a_6+a_8)) \bigg] \bigg [g_1(m_{\pi}^2)(M_{\Lambda_b}-M_p)+
g_3(m_{\pi}^2)m_{\pi}^2\bigg]
\nonumber \\
&& + \bigg [V_{ub}V_{ud}^{\ast} a_1 -
V_{tb}V_{td}^{\ast}(a_4+a_{10}-R_2^{\pi} (a_6+a_8)) \bigg] \bigg
[G_1(m_{\pi}^2)(M_{\Lambda_b}+M_p)- G_3(m_{\pi}^2)m_{\pi}^2\bigg]
\gamma_5 \bigg\} \Lambda_b(p),   \\
&&\mathcal{M}_{f}(\Lambda_b \to p K) \nonumber \\
&&= {G_F \over \sqrt{2}} f_{K} \bar{p}(p^{\prime})\bigg  \{ \bigg
[V_{ub}V_{us}^{\ast} a_1 - V_{tb}V_{ts}^{\ast}(a_4+a_{10}+R_1^{K}
(a_6+a_8)) \bigg] \bigg [g_1(m_{K}^2)(M_{\Lambda_b}-M_p)+
g_3(m_{K}^2)m_{K}^2\bigg]
\nonumber \\
&& + \bigg [V_{ub}V_{us}^{\ast} a_1 -
V_{tb}V_{ts}^{\ast}(a_4+a_{10}-R_2^{K} (a_6+a_8)) \bigg] \bigg
[G_1(m_{K}^2)(M_{\Lambda_b}+M_p)- G_3(m_{K}^2)m_{K}^2\bigg] \gamma_5
\bigg\} \Lambda_b(p).  \hspace{1.0 cm}
 \label{fC-NF-kaon}
\end{eqnarray}
The masses of the pseudoscalar mesons of $\pi$ and $K$ can safely be
neglected, therefore only the form factors at the zero-momentum
transfer will be involved in the numerical computations.

To evaluate the  $\Lambda_b \to p \pi, ~p K$ decays numerically, we
need to specify the form factors responsible for the
$\Lambda_b \to p$ transition.
\begin{table}[tb]
\caption{Numerical values of the form factors $g_1$ and
$m_{\Lambda_b}g_2$ at zero momentum transfer, responsible for the
$\Lambda_b \to p$ transition, estimated in the LCSR approach
\cite{Huang:2004vf}.}
\begin{center}
\begin{tabular}{ccc}
  \hline
  \hline
   form factors & $ \hspace{2.5 cm} g_1$ & $ \hspace{2.5 cm} m_{\Lambda_b} g_2$  \\
  $\Lambda_b \to p$ &  \hspace{2.5 cm} 0.018 & $ \hspace{2.5 cm} -0.159$  \\
  \hline
  \hline
\end{tabular}
\label{ff-models}
\end{center}
\end{table}
As can be seen from Eqs. (\ref{fC-NF-pion}-\ref{fC-NF-kaon}), the
form factors $g_3$ and $G_3$, whose contributions are proportional
to the mass of the corresponding meson, are inessential for the calculation
of the decay amplitudes. In view of the minor effects of these two form
factors, it is quite adequate to determine them in terms of the relations
derived in the heavy quark limit. As is well known, the form factors
$g_i$ and  $G_i$ satisfy
\begin{eqnarray}
g_1&=&G_1= \xi_1 + { m_{\Lambda} \over  m_{\Lambda_b}}\xi_2,
\label{relatins of from factors in HQET 1} \\
g_2&=&G_2=g_3=G_3={ \xi_2 \over  m_{\Lambda_b}}. \label{relatins of
from factors in HQET 2}
\end{eqnarray}
in the heavy quark effective theory (HQET), where the   two
independent form factors $\xi_1$ and  $\xi_2$  are defined as
\begin{eqnarray}
\langle\Lambda(P)|\bar{b}\Gamma s|\Lambda_{b}(P+q)\rangle
&=&\overline{\Lambda}(P)[\xi_1(q^2)+\not \! v \xi_2(q^2)] \Gamma
\Lambda_b(P+q), \label{ff-HQET}
\end{eqnarray}
with $\Gamma$ being an arbitrary Lorentz structure and $v_{\mu}$
being the four-velocity of the $\Lambda_b$ baryon.
 An analysis of the form factors
$g_i$ and $G_i$ has been performed in the LCSR \cite{Huang:2004vf},
which we shall use here. The numerical  results for $g_1$ and
$m_{\Lambda_b} g_2$ needed for our numerical calculations are
grouped in Table \ref{ff-models}, which correspond to $\xi_1=0.050$ and
 $\xi_2=-0.16$.

Utilizing the Wilson coefficients, the input form factors just discussed and the CKM factors
given earlier, we can now compute the factorizable contributions to $f_1$ and
$f_2$ in the hybrid pQCD approach and compare them to the corresponding non-factorizable
contributions, which have been given already earlier. The results are given in
Table \ref{fnf-mixed-scheme}. From this table we see that the factorizable contributions
are now much larger than in the conventional pQCD approach, though they are still smaller than
the corresponding non-factorizable contributions.

\begin{table}[tb]
\caption{The coefficients $f_1$ and $f_2$ in the decays $\Lambda_b \to p \pi, ~p
K$ contributed by the factorizable and non-factorizable external $W$
emission ($T$) diagrams in the hybrid pQCD scheme.}
\begin{center}
\begin{tabular}{ccc}
  \hline
  \hline
    & factorizable & non-factorizable \\
   \hline
  $f_1 (\Lambda_b \to p \pi )$ & $2.43 \times 10^{-10} -i 4.39 \times 10^{-10}$ & $-2.43 \times 10^{-9} -i 2.05 \times 10^{-9}$ \\
  $f_2 (\Lambda_b \to  p \pi)$ & $2.64 \times 10^{-10} -i 6.54 \times 10^{-10}$& $-1.75 \times 10^{-9} -i 1.20 \times 10^{-9}$ \\
  $f_1 (\Lambda_b \to  p K )$ & $-3.17 \times 10^{-10} -i 1.22 \times 10^{-10}$ & $-0.88 \times 10^{-9} + i 0.54 \times 10^{-10}$ \\
  $f_2 (\Lambda_b \to  p K)$ & $1.74 \times 10^{-10} -i 1.96 \times 10^{-10}$& $-1.06 \times 10^{-9} + i 1.67 \times 10^{-9}$ \\
  \hline
  \hline
\end{tabular}
\label{fnf-mixed-scheme}
\end{center}
 \end{table}

We are now in a position to present our final results concerning the branching
ratios, direct CP asymmetries and the polarization asymmetry parameter
$\alpha $ for the two decay channels in the conventional pQCD and in the hybrid
pQCD approach. The CP-asymmetry $A_{\rm CP}(\Lambda_b^0 \to p\pi^-)$ is defined
 as follows:
\begin{eqnarray}
A_{\rm CP}(\Lambda_b^0 \to p\pi^-) &\equiv&
 \frac{{\cal B}(\bar{\Lambda}_b^0 \to \bar{p}\pi^+) -
 {\cal B}(\Lambda_b^0 \to p \pi^-)}
{{\cal B}(\bar{\Lambda}_b^0 \to \bar{p}\pi^+) +
 {\cal B}(\Lambda_b^0 \to p \pi^-)}~,
\label{CP-def}
\end{eqnarray}
with $A_{\rm CP}(\Lambda_b^0 \to p K^-)$ defined similarly.
The  asymmetry parameter $\alpha$
associated with the anisotropic angular distribution of the proton
emitted in the polarized $\Lambda_b$ baryon decays is defined
as follows:
\begin{eqnarray}
\Gamma = \Gamma_0 (1+ \alpha {\bf p} \cdot {\bf s_{\Lambda_b}}) \,
\end{eqnarray}
with ${\bf p}$, ${\bf s_{\Lambda_b}}$ being the three-momentum and
spin vector of the proton in the rest frame of the $\Lambda_b$ baryon. The
explicit expression of $\alpha$ can be written as
\cite{Cheng:1996cs}
\begin{eqnarray}
\alpha = - {2 \tilde{\kappa} Re(f_1^{\ast} f_2) \over (|f_1|^2+\tilde{\kappa}^2
|f_2|^2) } \, ,
\end{eqnarray}
with $\tilde{\kappa} = { |{\bf p}| / (E_p + m_p)} = \sqrt{(E_p + m_p) /
(E_p- m_p)}$.

\begin{table}[htbp]
\caption{The CP-averaged branching ratios, direct CP asymmetries and
the polarization asymmetry parameter $\alpha$ for the $\Lambda_b \to p \pi, p
K$ decays obtained in the conventional and the hybrid pQCD approaches.
The errors for
these entries correspond to the uncertainties in the input hadronic
quantities, the scale-dependence, and the CKM matrix elements,
respectively.  Current experimental measurements at the Tevatron
\cite{ Morello:2008gy} are also listed. }
\begin{center}
\begin{tabular}{cccc}
  \hline
  \hline
      & pQCD  (conventional) & pQCD  (hybrid scheme) & Exp. \\
  \hline
 ${\cal B}(\Lambda_b \to p \pi)$  & $4.66^{+2.08+0.70+0.35}_{-1.74-0.35-0.35} \times 10^{-6}$ & $5.21 ^{+2.42 +0.30+0.42}_{-1.89-0.10-0.37} \times 10^{-6}$& $3.5 \pm 0.6 \pm 0.9 \times 10^{-6}$ \\
 ${\cal B}(\Lambda_b \to p K)$  & $1.82^{+0.74+0.62+0.07}_{-0.71-0.80-0.05} \times 10^{-6}$ & $2.02^{+0.78+0.55+0.10}_{-0.86-0.90-0.05} \times 10^{-6}$& $5.6 \pm 0.8 \pm 1.5 \times 10^{-6}$ \\
 $A_{\rm CP}(\Lambda_b \to p \pi)$  & $-0.32^{+0.27+0.41+0.01}_{-0.00-0.00-0.01}$ & $-0.31^{+0.28+0.32+0.01}_{-0.00-0.00-0.01}$ & $-0.03 \pm 0.17 \pm 0.05$ \\
 $A_{\rm CP}(\Lambda_b \to p K)$  & $-0.03^{+0.21+0.13+0.00}_{-0.00-0.04-0.00}$ & $-0.05 ^{+0.26+0.03+0.01}_{-0.00-0.05-0.00}$& $-0.37 \pm 0.17 \pm 0.03$ \\
 $\alpha(\Lambda_b \to p \pi)$  & $-0.83^{+0.03+0.03+0.01}_{-0.01-0.07-0.01}$  & $-0.84^{+0.03+0.00+0.01}_{-0.00-0.00-0.01}$ &  ---  \\
 $\alpha(\Lambda_b \to p K)$  & $0.03^{+0.00+0.00+0.03}_{-0.36-0.07-0.05}$ &$0.08^{+0.00+0.05+0.04}_{-0.38-0.42-0.04}$&  ---  \\
 \hline
  \hline
\end{tabular}
\label{BR-CP}
\end{center}
 \end{table}

We present our results in the two pQCD approaches  and compare them with
 the current
 experimental data from the Tevatron~\cite{Morello:2008gy} in Table~\ref{BR-CP}.
The first error in the pQCD-based entries arises from the input hadronic
parameters, which is dominated by the errors on the 
normalization constant of the
$\Lambda_b$ baryon (taken as $f_{\Lambda_b}=4.28 ^{+0.75}_{-0.64}
\times 10^{-3}{\rm GeV}^2$) and the $\Lambda_b$ baryon wave function
shape parameter (taken as $\beta= 1.0 \pm 0.2$ GeV). The second
error is the combined error from the hard scale $t$,
 defined in Eq.~(\ref{hard scale}), which is varied from $0.75t$ to $1.25t$,
 and the renormalization scale of the Wilson coefficients, given in
Table I.  The third error is the combined uncertainty due to
the CKM matrix elements.

We observe from Table \ref{BR-CP} that the results for the
conventional pQCD and the hybrid pQCD approaches do not differ very
much, although in the hybrid approach the factorizable contributions
have increased by almost an order of magnitude as compared to the
conventional pQCD approach. The reason for this is that in the
hybrid approach the factorizable amplitudes $f_i$ are still only a
fraction of the non-factorizable amplitudes, as is apparent by
comparing the results in Table \ref{fnf-mixed-scheme}. Of course, it
remains to be checked if the non-factorizable amplitude is correctly
estimated in the pQCD approach for the $b$-baryonic decays due to
the exchange of two gluons. This involves, among other diagrams,
those where both the gluons are attached to the same outgoing quark
line (see, for example, the diagrams in the fourth row in
Fig.~\ref{color-allowed-diagrams}). These contributions are more
sharply peaked, compared to the others encountered here or in the
decays of $B$-mesons, which involve single gluon attachments on a
quark line.

The ratio of the decay rates for the $\Lambda_b \to p \pi$ and
$\Lambda_b \to p K$ decays, called below $R_{\pi K}(\Lambda_b)$,
 can be calculated from Table \ref{BR-CP}, and is estimated by us as
 \begin{eqnarray}
 R_{\pi K}(\Lambda_b) \equiv  { BR(\Lambda_b \to p \pi) \over BR(\Lambda_b
 \to p K)} = 2.6^{+2.0}_{-0.5}
\end{eqnarray}
in the hybrid pQCD approach. This  can be
understood from Eqs.~(\ref{fC-NF-pion}-\ref{fC-NF-kaon}), which show that
 the QCD penguin operators contribute to the coefficients $f_1$ and $f_2$
(defined in Eq. (\ref{ff-definition})) in the combination
$a_4 + R_1^K a_6 $ and $a_4 - R_2^K a_6 $, respectively. This is
quite different from the two-body hadronic decays of the $B$ mesons,
$B \to PP$ or $B \to PV$, where $P(V)$ is a light pseudoscalar (vector) meson.
 The key
point is that both  the hadronic matrix elements $\langle
\Lambda(P)|\bar{s}\gamma_{\mu} b|\Lambda_{b}(P+q)\rangle$ and
$\langle \Lambda(P)|\bar{s}\gamma_{\mu}\gamma_5
b|\Lambda_{b}(P+q)\rangle$  contribute to the baryonic decays.
Theoretical predictions presented here deviate from the experimental
data $R_{\pi K}(\Lambda_b)= 0.66 \pm 0.14 \pm 0.08
$~\cite{Morello:2008gy}. Whether this discrepancy reflects the
inadequacy of the current theoretical formalism embedded in the
standard model, or the standard model itself, or requires improved data remains to be seen.
We note
{\it en passant} that the estimates of the branching ratios for the
decays $\Lambda_b \to p\pi$ and $\Lambda_b \to p K$, and hence of
the quantity $R_{\pi K}(\Lambda_b)$, reported in
~\cite{Mohanta:2000nk} in the generalized factorization
approximation, are in error due to the incorrect relative sign of
the two terms in Eq.~(18) in that paper. We are convinced that
the correct relative sign in question is given in our Eq.~(\ref{O6 penguin}).

As for the direct CP asymmetries, theoretical predictions suffer
from large uncertainties due to the hadronic distributions, the hard
scattering and the renormalization scales in the factorizable
amplitudes. For the CP asymmetries, one needs the complete NNLO
vertex corrections, as only with this input will it be possible to
make quantitative predictions.
As can be seen from Table \ref{BR-CP}, theoretical
estimates for the parameter $\alpha$ for the decay $\Lambda_b \to p \pi$
have  negative values in both the pQCD approaches, reflecting the
$(V-A)$ structure of the weak current \cite{Cheng:1995fe}. It is
pointed out in \cite{Mannel:1991fg} that the parameter $\alpha$ in $B_i
({1 \over 2}^{+}) \to B_f ({1 \over 2}^{+}) P(V)$ decays
approaches  $-1$ in the soft pseudoscalar meson or vector meson limit,
i.e., for $m_P \to 0$ or $m_V \to 0$. This argument, however,  is only valid for
the tree-dominated processes. As for the $\Lambda_b \to p K$ decay,
the contributions from the QCD penguin operators are comparable to that
of the tree amplitude. The operator $O_6$ contributes to the
$\Lambda_b \to p$ transition via the $(V+A)$ current (see Eq.
(\ref{O6 penguin})) and the Wilson coefficient $a_6$ is very
sensitive to the energy scale as can be seen from Table
\ref{Wilson-coefficients}. Hence, the asymmetry
parameter $\alpha$ can flip its sign for the $\Lambda_b \to p K$
decay due to the large penguin contributions. As a final remark, we find
that the predictions for the parameter $\alpha$ in the $\Lambda_b
\to p \pi$ decay  are relatively stable with respect to the
variations of hadronic parameters, the CKM matrix elements and the hard
scale, and therefore it serves as a good quantity to test the
standard model \cite{h.n.li_lambda}.

\section{Discussions and conclusions}

Thanks to the current and impending experimental programs at the Tevatron and the LHC,
dedicated studies of the decays of the $\Lambda_b$ baryon (and other heavy baryons) will be
carried out, following the first measurements of the decays
$\Lambda_b \to p \pi, ~p K$, performed at the Tevatron.
Baryonic decays are flavor self-tagging
processes. Therefore, they should be easier to
reconstruct experimentally. In particular, the CP-asymmetry measurements amount to
counting these self-tagged decay modes and their CP-conjugates.
From the theoretical viewpoint, however, $b$-baryon decays are
less tractable as the underlying QCD dynamics is more involved. Hence, it is
far from being obvious if the theoretical approaches developed for the
quantitative studies of the two-body non-leptonic decays of the $B$-mesons will
work also for the corresponding $b$-baryon decays.
We have carried out an exploratory study of the charmless
hadronic decays $\Lambda_b \to p \pi, ~p K$
in the pQCD approach and
find that the factorizable diagrams in the conventional pQCD approach
contribute very little to the branching ratios, as the hard (perturbative) contributions to
the baryonic transition form factors in this case turn out to be quite
small compared to the estimates dominated by the soft dynamics.
As an alternative, we adopted a hybrid approach, in spirit similar
to the one advocated in Ref.~\cite{Chen:2005ht} for the analysis of the color-suppressed
decays, such as $B^0 \to J/\psi K^0$.
An essential characteristic of this hybrid scheme is that
the transition form
factors are treated as non-perturbative objects, i.e., they are input in the theoretical
 analysis and
are not computed perturbatively, as in the conventional pQCD approach.
 Employing the form factors
estimated in the LCSR approach, we find that the factorizable
contributions are no longer negligible, though for the two decays
worked out here, the amplitudes are still dominated by the
non-factorizing contributions.

Our predictions for the branching fraction for the decay $\Lambda_b
\to p \pi$, which is dominated by the tree diagrams, are essentially
in agreement with the current data, whereas estimates of the
branching ratio for the $\Lambda_b \to p K$ decay, dominated by the
penguin-amplitude, are found to be smaller typically by a factor 2.
This deserves an improved theoretical analysis, as the data gets
consolidated.
 The asymmetry parameter
$\alpha$ associated with the anisotropic angular distribution of the
proton produced in the polarized $\Lambda_b$ baryon decays is also
derived and is found to be relatively stable with respect to
variations of hadronic inputs and higher-order corrections in
$\Lambda_b \to p \pi$ decay. The asymmetry parameter $\alpha$ in the $\Lambda_b \to p K$
decay, however,  can flip its sign due to the large
penguin contributions and the sensitive scale dependence of the effective
Wilson coefficient $a_6(\mu)$. The Feynman diagrams (G) with the
three-gluon-vertices present in the perturbative amplitudes included
in this work are found to be less important compared with the
$T$ diagrams. However, these three-gluon-vertex diagrams are
comparable to the $C$ diagrams, as can be seen from Table
 \ref{results-different-topologies}, and hence they may induce significant
corrections to the color suppressed modes, such as the $\Lambda_b \to
\Lambda J/\psi$ decay. Finally, quantitative estimates of the CP asymmetries
presented here show large scale uncertainties and require
NNLO vertex corrections to be firmed up, which are not yet available completely.

\section*{Acknowledgments}

This work is partly supported by the National Science Foundation of
China under Grant No.10735080 and 10625525 and by the German Academic
 Exchange Program (DAAD). The authors would like
to thank Hai-Yang Cheng, Hsiang-nan Li, Run-Hui Li and Yue-Long Shen
for valuable  comments. Y.~M.~W.  would like to
acknowledge Lei Dang, Cheng Li, Ping Ren, Qian Wang,  Xiao-Xia Wang
and Yu-Min Wang for allowing us to share the computing resources.
H.~Z. is grateful to  Marco Drewes, Christian Hambrock, Sebastian Mendizabal,
Satoshi Mishima and Alexander Parkhomenko for helpful discussions
during his stay at DESY.

\appendix

\section{Fourier integrations and $b$-space measures}

We list below the Fourier integration formulae which have been
employed in the derivation of the hard amplitudes in the impact
parameter (or $b$) space. The symbols
$J_1$, $N_1$, $K_0$ and $K_1$ are the various Bessel functions; $z_i$
are the Feynman parameters; and the relation
\begin{eqnarray}
K_n(-i z)={\pi i \over 2}e^{i n \pi \over 2}[J_n(z)+iN_n(z)]
\end{eqnarray}
has been used used in the derivation of the Fourier transformation.
With this, we get:

\begin{eqnarray} &
&\int d^2k \frac{e^{i {\bf k}\cdot {\bf b}}}{k^2+A} = 2\pi
\{K_0(\sqrt{A}b)\theta(A)+{\pi i \over 2}  [J_0(\sqrt{|A|}b)+ i
N_0(\sqrt{|A|}b)]\theta(-A)\},
\\
& &\nonumber\\
& &\int d^2k \frac{e^{i {\bf k}\cdot {\bf b}}}{(k^2+A)(k^2+B)}
=\pi\int_0^1 dz\frac{b}{\sqrt{|Z_1|}} \{K_1(\sqrt{Z_1}b)
\theta(Z_1)+{\pi \over 2}  [N_1(\sqrt{|Z_1|}b)- i
J_1(\sqrt{|Z_1|}b)]\theta(-Z_1) \},
\\
& &\nonumber\\
& &\int d^2k_1d^2k_2 \frac{e^{i ({\bf k}_1\cdot {\bf b}_1 +{\bf
k}_2\cdot {\bf b}_2)}}{(k_1^2+A)(k_2^2+B)[(k_1+k_2)^2+C]}
\nonumber \\
&&= \pi^2\int_0^1\frac{dz_1dz_2}{z_1(1-z_1)}
\frac{\sqrt{X_2}}{\sqrt{|Z_2|}} \Bigg\{K_1(\sqrt{X_2Z_2})\theta(Z_2)
+\frac{\pi}{2}\left[N_1(\sqrt{X_2|Z_2|}) -i
J_1(\sqrt{X_2|Z_2|})\right]\theta(-Z_2) \Bigg\}\;,
\nonumber\\
& &\;\;\;\; {\rm where}\; A\, >\, 0\;,\; {\rm and}\;B, C \;
 {\rm are~arbitrary}\;,
\\
& &\nonumber\\
& &\int d^2k_1d^2k_2d^2k_3 \frac{e^{i ({\bf k}_1\cdot {\bf b}_1
+{\bf k}_2\cdot {\bf b}_2 +{\bf k}_3\cdot {\bf b}_3)}}
{(k_1^2+A)(k_2^2+B)(k_3^2+C)[(k_1+k_2+k_3)^2+D]}
\nonumber  \\
&&=\pi^3\int_0^1\frac{dz_1dz_2dz_3}{z_1(1-z_1)z_2(1-z_2)}
\frac{\sqrt{X_3}}{\sqrt{|Z_3|}} \Bigg\{K_1(\sqrt{X_3Z_3})\theta(Z_3)
+\frac{\pi}{2}\left[N_1(\sqrt{X_3|Z_3|}) - i
J_1(\sqrt{X_3|Z_3|})\right]\theta(-Z_3) \Bigg\}\;,
\nonumber\\
& &\;\;\;\;A, B\;>\;0\;, \;{\rm and}\;C, D\;{\rm arbitrary}\;,
\end{eqnarray}
with the variables,
\begin{eqnarray} Z_1&=&A\;z+B\;(1-z)\;,
\\
Z_2 &=& A\;(1-z_2)+\frac{z_2}{z_1(1-z_1)}[B\;(1-z_1)+C\;z_1] \;,
\nonumber \\
X_2 &=&(b_1-z_1b_2)^2+\frac{z_1 (1-z_1)}{z_2}b_2^2 \;,
\\
Z_3 &=& A\;(1-z_3)+\frac{z_3}{z_2(1-z_2)}\left\{B\;(1-z_2)
+\frac{z_2}{z_1(1-z_1)}[C\;(1-z_1)+D\;z_1]\right\} \;,
\nonumber\\
X_3 &=& [b_1-b_2z_2-b_3z_1(1-z_2)]^2+\frac{z_2 (1-z_2)}{z_3}
(b_2-b_3 z_1)^2  +\frac{z_1 (1-z_1) z_2 (1-z_2)}{z_2 z_3}b_3^2
\;.\end{eqnarray}

\section{Factorization formulae for the Feynman diagrams with various topologies}
 \label{appendix: fformulae }
 In this appendix, we would like to collect the
factorizable formulae for typical diagrams corresponding to
different topologies in the $\Lambda_b \to p \pi$ decays.
In doing so, we give the expressions only for
a certain representative set of diagrams in each class,  with the rest following
from appropriate substitutions.

\subsection{Factorization formulae for the color allowed emission diagrams}

\begin{figure}[tb]
\begin{center}
\begin{tabular}{ccc}
\hspace{-2 cm}
\includegraphics[scale=0.70]{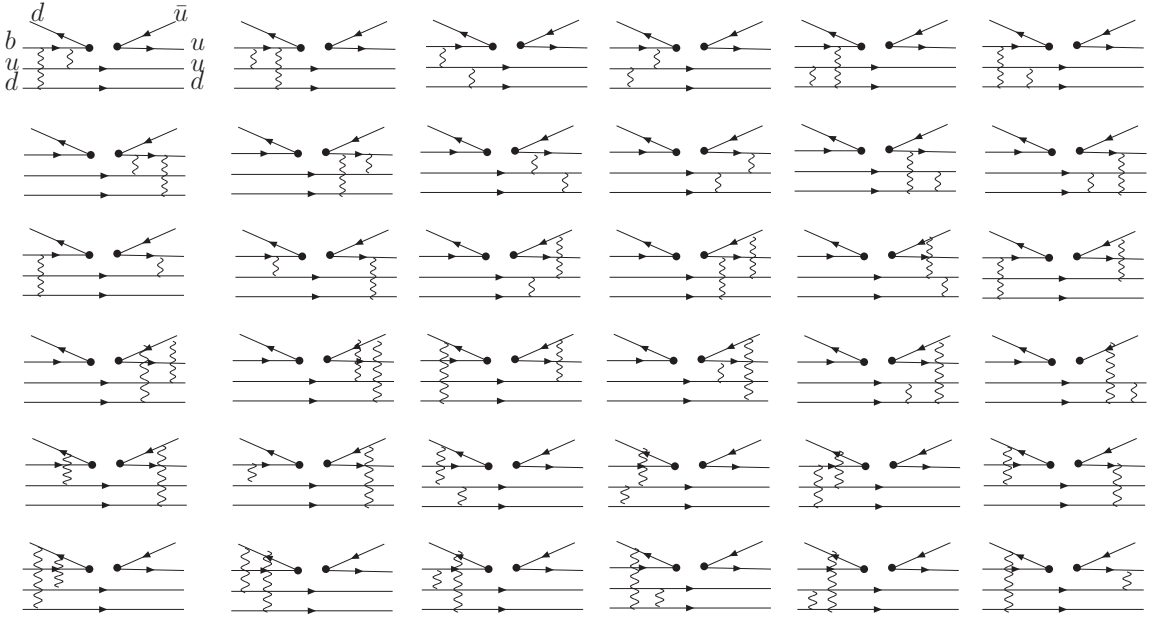}
\end{tabular}
\caption{External $W$ emission ($T$) diagrams for the  $\Lambda_b \to p
\pi$ decay to the lowest order in the pQCD approach where the  dots
denote the weak interactions vertices. The two hard gluons are
needed to transfer the large momentum to the light quarks in the
initial state so that these two light quarks are collinear in the
final state. These diagrams are called $T_1, T_2,...,T_{36}$ in the
text.}\label{color-allowed-diagrams}
\end{center}
\end{figure}

For the first diagram in Fig. \ref{color-allowed-diagrams}
(labeled as figure $T_1$ ), which is a factorizable
diagram and included only in the conventional pQCD approach, we have:
\begin{eqnarray}
f_1^{T_1}&=& G_F  { \pi^2 \over 18 \sqrt{3}} f_{\Lambda_b} f_p \int
[d x ] \int [d x^{\prime}] \int dy \,
[\alpha_s(t^{T_1})]^2 \psi_{\Lambda_b}(x) \nonumber\\
&\times &\bigg\{ \bigg[16 M_{\Lambda_b}^5
[(\frac{1}{3}C_1+C_2)V_{ub}V_{ud}^*+(\frac{1}{3}C_3+
C_4+\frac{1}{3}C_9+C_{10})V_{tb}V_{td}^*](-2x_2+(1-2x_2)x_1'-x_3')\phi_M^A(y)\nonumber\\
&&\;\;\;\;-32m_0 M_{\Lambda_b}^4(\frac{1}{3}C_5+
C_6+\frac{1}{3}C_7+C_8)V_{tb}V_{td}^*(4x_1+4x_3-x_3'-3) \phi_M^P (y) \bigg]\psi_p^V(x^{\prime})\nonumber\\
&&+\bigg[16 M_{\Lambda_b}^5
[(\frac{1}{3}C_1+C_2)V_{ub}V_{ud}^*+(\frac{1}{3}C_3+
C_4+\frac{1}{3}C_9+C_{10})V_{tb}V_{td}^* ](1+x_2')\phi_M^A(y)\nonumber\\
&&\;\;\;\;-32m_0 M_{\Lambda_b}^4(\frac{1}{3}C_5+
C_6+\frac{1}{3}C_7+C_8)V_{tb}V_{td}^*(x_3'-1)\phi_M^P (y) \bigg]
\psi_p^{A}(x^{\prime})\nonumber\\
&&+ \bigg[ 16 M_{\Lambda_b}^5
[(\frac{1}{3}C_1+C_2)V_{ub}V_{ud}^*+(\frac{1}{3}C_3+
C_4+\frac{1}{3}C_9+C_{10})V_{tb}V_{td}^*](2(x_1+x_3))\phi_M^{A}(y)\nonumber\\
&&\;\;\;\;+32m_0 M_{\Lambda_b}^4(\frac{1}{3}C_5+
C_6+\frac{1}{3}C_7+C_8)V_{tb}V_{td}^*(2(x_1+x_3)) \phi_M^{T}(y) \bigg]\psi_p^{T}(x^{\prime}) \bigg \}\nonumber\\
&\times &\frac{1}{16 \pi^2}\int b_1'db_1'\int b_2db_2 \int b_3 db_3
\int d\theta_1 \int d\theta_2 \, {\rm exp} [-S^{T_1} (x, x^{\prime},
b, b^{\prime})] \, K_0(\sqrt{D^{T_1}}|b_3|)
\nonumber\\
&&\int_0^1\frac{dz_1dz_2}{z_1(1-z_1)}\sqrt{\frac{X_2^{T_1}}{|Z_2^{T_1}|}}
\bigg\{ K_1 (\sqrt{X_2^{T_1} Z_2^{T_1}})\Theta(Z_2^{T_1})
+\frac{\pi}{2}[J_1(\sqrt{X_2^{T_1}
|Z_2^{T_1}|})+iN_1(\sqrt{X_2^{T_1} |Z_2^{T_1}|})]\Theta(-Z_2^{T_1})
\bigg \} , \nonumber \\
\end{eqnarray}
where the auxiliary functions in the above expression are defined as
\begin{eqnarray}
&&A^{T_1}=(1-x_1')M_{\Lambda_b}^2,
B^{T_1}=(x_2+x_3'-x_2x_3')M_{\Lambda_b}^2,
C^{T_1}=x_2x_2'M_{\Lambda_b}^2,\,\,
D^{T_1}=x_3x_3'M_{\Lambda_b}^2, \nonumber\\
&&Z_2^{T_1} = A^{T_1}(1-z_2)+\frac{z_2}{z_1(1-z_1)}[B^{T_1}(1-z_1)+C^{T_1}z_1] , \nonumber\\
&&X_2^{T_1} = ({b_1'}-z_1{b_2})^2 + \frac{z_1(1-z_1)}{z_2}b_2^2. \nonumber\\
&&t^{T_{1}} = max(
\sqrt{|A^{T_1}|},\sqrt{|B^{T_1}|},\sqrt{|C^{T_1}|},\sqrt{|D^{T_1}|},\omega,\omega^{\prime})
.
\end{eqnarray}
Similarly, the factorization formula for the form factor $f_2$
contributed by $T_1$ can be written as
\begin{eqnarray}
f_2^{T_1}&=& G_F  { \pi^2 \over 18 \sqrt{3}} f_{\Lambda_b} f_p \int
[d x ] \int [d x^{\prime}] \int dy \,
[\alpha_s(t^{T_1})]^2 \psi_{\Lambda_b}(x) \nonumber\\
&\times &\bigg\{\bigg[16 M_{\Lambda_b}^5
[(\frac{1}{3}C_1+C_2)V_{ub}V_{ud}^*+(\frac{1}{3}C_3+
C_4+\frac{1}{3}C_9+C_{10})V_{tb}V_{td}^* ](1+x_2')\phi_M^A(y)\nonumber\\
&&\;\;\;\;+32m_0 M_{\Lambda_b}^4(\frac{1}{3}C_5+
C_6+\frac{1}{3}C_7+C_8)V_{tb}V_{td}^*(x_3'-1)\phi_M^P (y) \bigg] \psi_p^{V}(x^{\prime})\nonumber\\
&&+ \bigg[16 M_{\Lambda_b}^5
[(\frac{1}{3}C_1+C_2)V_{ub}V_{ud}^*+(\frac{1}{3}C_3+
C_4+\frac{1}{3}C_9+C_{10})V_{tb}V_{td}^*](-2x_2+(1-2x_2)x_1'-x_3')\phi_M^A(y)\nonumber\\
&&\;\;\;\;-32m_0 M_{\Lambda_b}^4(\frac{1}{3}C_5+
C_6+\frac{1}{3}C_7+C_8)V_{tb}V_{td}^*(4x_1+4x_3-x_3'-3) \phi_M^P (y) \bigg]\psi_p^A(x^{\prime})\nonumber\\
&&+ \bigg[ 16 M_{\Lambda_b}^5
[(\frac{1}{3}C_1+C_2)V_{ub}V_{ud}^*+(\frac{1}{3}C_3+
C_4+\frac{1}{3}C_9+C_{10})V_{tb}V_{td}^*](2(x_1+x_3))\phi_M^{A}(y)\nonumber\\
&&\;\;\;\;-32m_0 M_{\Lambda_b}^4(\frac{1}{3}C_5+
C_6+\frac{1}{3}C_7+C_8)V_{tb}V_{td}^*(2(x_1+x_3)) \phi_M^{T}(y) \bigg]\psi_p^{T}(x^{\prime}) \bigg \}\nonumber\\
&\times &\frac{1}{16 \pi^2}\int b_1'db_1'\int b_2db_2 \int b_3 db_3
\int d\theta_1 \int d\theta_2 \, {\rm exp} [-S^{T_1} (x, x^{\prime},
b, b^{\prime})] \, K_0(\sqrt{D^{T_1}}|b_3|)
\nonumber\\
&&\int_0^1\frac{dz_1dz_2}{z_1(1-z_1)}\sqrt{\frac{X_2^{T_1}}{|Z_2^{T_1}|}}
\bigg\{ K_1 (\sqrt{X_2^{T_1} Z_2^{T_1}})\Theta(Z_2^{T_1})
+\frac{\pi}{2}[J_1(\sqrt{X_2^{T_1}
|Z_2^{T_1}|})+iN_1(\sqrt{X_2^{T_1} |Z_2^{T_1}|})]\Theta(-Z_2^{T_1})
\bigg \} . \nonumber \\
\end{eqnarray}

For the 25th diagram in Fig. \ref{color-allowed-diagrams}
(labeled as $T_{25}$), which is a non-factorizable
diagram, we have:
\begin{eqnarray}
f_1^{T_{25}}&=& G_F  { \pi^2 \over 144  \sqrt{3}} f_{\Lambda_b} f_p
\int [d x ] \int [d x^{\prime}] \int dy \,
[\alpha_s(t^{T_{25}})]^2 \psi_{\Lambda_b}(x) \nonumber\\
&\times &\bigg\{ \bigg[16 M_{\Lambda_b}^5
[(\frac{8}{3}C_1-2C_2)V_{ub}V_{ud}^*+(\frac{8}{3}C_3-2
C_4+\frac{8}{3}C_9-2C_{10})V_{tb}V_{td}^*](x_2-y)(-x_3+x_3^{\prime}-y+1)\phi_M^A(y)\nonumber\\
&&\;\;\;\;+16 m_0 M_{\Lambda_b}^4(\frac{8}{3}C_5-
2 C_6+\frac{8}{3}C_7-2C_8)V_{tb}V_{td}^* x_2^{\prime}(1-x_3-y) (\phi_M^P (y)-\phi_M^T (y)) \bigg]\psi_p^V(x^{\prime})\nonumber\\
&&+\bigg[16 M_{\Lambda_b}^5
[(\frac{8}{3}C_1-2C_2)V_{ub}V_{ud}^*+(\frac{8}{3}C_3-2
C_4+\frac{8}{3}C_9-2C_{10})V_{tb}V_{td}^* ](x_2-y)(1-x_3-x_3^{\prime}-y)\phi_M^A(y)\nonumber\\
&&\;\;\;\;+16 m_0 M_{\Lambda_b}^4(\frac{8}{3}C_5-2
C_6+\frac{8}{3}C_7-2C_8)V_{tb}V_{td}^* x_2^{\prime} (1-x_3-y)
(\phi_M^P (y)-\phi_M^T (y)) \bigg]
\psi_p^{A}(x^{\prime})\nonumber\\
&&+ \bigg[32 M_{\Lambda_b}^5
[(\frac{8}{3}C_1-2C_2)V_{ub}V_{ud}^*+(\frac{8}{3}C_3-2
C_4+\frac{8}{3}C_9-2C_{10})V_{tb}V_{td}^* ] (x_2-y) (1-x_3-y) \phi_M^{A}(y)\nonumber\\
&&\;\;\;\;+32m_0 M_{\Lambda_b}^4(\frac{8}{3}C_5-2
C_6+\frac{8}{3}C_7-2C_8)V_{tb}V_{td}^* x_3^{\prime} (y-x_2) \phi_M^{T}(y) \bigg]\psi_p^{T}(x^{\prime}) \bigg \}\nonumber\\
&\times &\frac{1}{32 \pi^2}\int b_2 db_2 \int b_3 db_3  \int b_q
db_q \int d\theta_1 \int d\theta_2 \, {\rm exp} [-S^{T_{25}} (x,
x^{\prime},y, b, b^{\prime}, b_q)] \,  \int_0^1\frac{dz_1dz_2
dz_3}{z_1(1-z_1)z_2(1-z_2)} \nonumber\\
&& \sqrt{\frac{X_3^{T_{25}}}{|Z_3^{T_{25}}|}} \bigg\{ K_1
(\sqrt{X_3^{T_{25}}Z_3^{T_{25}}})\Theta(Z_3^{T_{25}})
+\frac{\pi}{2}[J_1(\sqrt{X_3^{T_{25}}|Z_3^{T_{25}}|})+iN_1(\sqrt{X_3^{T_{25}}|Z_3^{T_{25}}|})]\Theta(-Z_3^{T_{25}})
\bigg \} ,
\end{eqnarray}
where the auxiliary functions in the expressions above are defined as
\begin{eqnarray}
&&A^{T_{25}}=x_3^{\prime}(x_3+y-1)M_{\Lambda_b}^2,
B^{T_{25}}=x_2^{\prime}(x_2-y)M_{\Lambda_b}^2, C^{T_{25}}=x_2
x_2^{\prime}M_{\Lambda_b}^2,\,\,
D^{T_{25}}=x_3 x_3^{\prime}M_{\Lambda_b}^2\nonumber\\
&&X_3^{T_{25}}=(b_2-(b_3-b_q)z_2+b_qz_1(1-z_2))^2+\frac{z_2(1-z_2)}{z_3}(b_3-b_q+b_qz_1)^2
+\frac{z_1(1-z_1)z_2(1-z_2)}{z_2z_3}b_q^2~,\nonumber\\
&&Z_3^{T_{25}}=C(1-z_3)+\frac{z_3}{z_2(1-z_2)}[D(1-z_2)+\frac{z_2}{z_1(1-z_1)}[A(1-z_1)+Bz_1]], \nonumber\\
&&t^{T_{25}} =
max(\sqrt{|A^{T_{25}}|},\sqrt{|B^{T_{25}}|},\sqrt{|C^{T_{25}}|},\sqrt{|D^{T_{25}}|},
\omega,\omega^{\prime},\omega_q). \label{f1 T25}
\end{eqnarray}
Similarly, the factorization formula for the form factor $f_2$
contributed by $T_{25}$ can be written as
\begin{eqnarray}
f_2^{T_{25}}&=& G_F  { \pi^2 \over 144 \sqrt{3}} f_{\Lambda_b} f_p
\int [d x ] \int [d x^{\prime}] \int dy \,
[\alpha_s(t^{T_{25}})]^2 \psi_{\Lambda_b}(x) \nonumber\\
&\times &\bigg\{\bigg[16 M_{\Lambda_b}^5
[(\frac{8}{3}C_1-2C_2)V_{ub}V_{ud}^*+(\frac{8}{3}C_3-2
C_4+\frac{8}{3}C_9-2C_{10})V_{tb}V_{td}^* ](x_2-y)(1-x_3-x_3^{\prime}-y)\phi_M^A(y)\nonumber\\
&&\;\;\;\;-16 m_0 M_{\Lambda_b}^4(\frac{8}{3}C_5-2
C_6+\frac{8}{3}C_7-2C_8)V_{tb}V_{td}^* x_2^{\prime} (1-x_3-y)
(\phi_M^P (y)-\phi_M^T (y)) \bigg]\psi_p^{V}(x^{\prime})\nonumber\\
&&+\bigg[16 M_{\Lambda_b}^5
[(\frac{8}{3}C_1-2C_2)V_{ub}V_{ud}^*+(\frac{8}{3}C_3-2
C_4+\frac{8}{3}C_9-2C_{10})V_{tb}V_{td}^*](x_2-y)(-x_3+x_3^{\prime}-y+1)\phi_M^A(y)\nonumber\\
&&\;\;\;\;-16 m_0 M_{\Lambda_b}^4(\frac{8}{3}C_5-
2 C_6+\frac{8}{3}C_7-2C_8)V_{tb}V_{td}^* x_2^{\prime}(1-x_3-y) (\phi_M^P (y)-\phi_M^T (y)) \bigg]\psi_p^A(x^{\prime})\nonumber\\
&&+ \bigg[32 M_{\Lambda_b}^5
[(\frac{8}{3}C_1-2C_2)V_{ub}V_{ud}^*+(\frac{8}{3}C_3-2
C_4+\frac{8}{3}C_9-2C_{10})V_{tb}V_{td}^* ] (x_2-y) (1-x_3-y) \phi_M^{A}(y)\nonumber\\
&&\;\;\;\;-32m_0 M_{\Lambda_b}^4(\frac{8}{3}C_5-2
C_6+\frac{8}{3}C_7-2C_8)V_{tb}V_{td}^* x_3^{\prime} (y-x_2) \phi_M^{T}(y) \bigg]\psi_p^{T}(x^{\prime}) \bigg \}\nonumber\\
&\times &\frac{1}{32 \pi^2}\int b_2 db_2 \int b_3 db_3  \int b_q
db_q \int d\theta_1 \int d\theta_2 \, {\rm exp} [-S^{T_{25}} (x,
x^{\prime},y, b, b^{\prime}, b_q)] \,  \int_0^1\frac{dz_1dz_2
dz_3}{z_1(1-z_1)z_2(1-z_2)} \nonumber\\
&& \sqrt{\frac{X_3^{T_{25}}}{|Z_3^{T_{25}}|}} \bigg\{ K_1
(\sqrt{X_3^{T_{25}}Z_3^{T_{25}}})\Theta(Z_3^{T_{25}})
+\frac{\pi}{2}[J_1(\sqrt{X_3^{T_{25}}|Z_3^{T_{25}}|})+iN_1(\sqrt{X_3^{T_{25}}|Z_3^{T_{25}}|})]\Theta(-Z_3^{T_{25}})
\bigg \} .
\end{eqnarray}
As can be seen from Eq.~(\ref{f1 T25}), the color structures for
the baryonic decays are quite different from that in the mesonic
decays. Only the operators with the color indices the same as $O_1$
can contribute to the non-factorizable emission diagrams in the
two-body hadronic $B$ meson decays. However, all the operators $O_i
(i=1-10)$ contribute to the non-factorizable emission diagrams in the
non-leptonic two-body bottom baryon $\Lambda_b$ decays. In
particular, the $(V-A)\otimes (V+A)$ type operators have no effect
on the non-factorizable emission diagrams for the hadronic $B \to P
P$ decays, if the emitted meson is a $\pi, \eta$ or $\eta^{\prime}$.
In contrast, both the $(V-A)\otimes (V-A)$ and $(V-A)\otimes (V+A)$
operators contribute to the non-factorizable emission diagrams in
their baryonic counterparts.

\subsection{Factorization formulae for the color suppressed emission
diagrams}

\begin{figure}[tb]
\begin{center}
\begin{tabular}{ccc}
\hspace{-2 cm}
\includegraphics[scale=0.70]{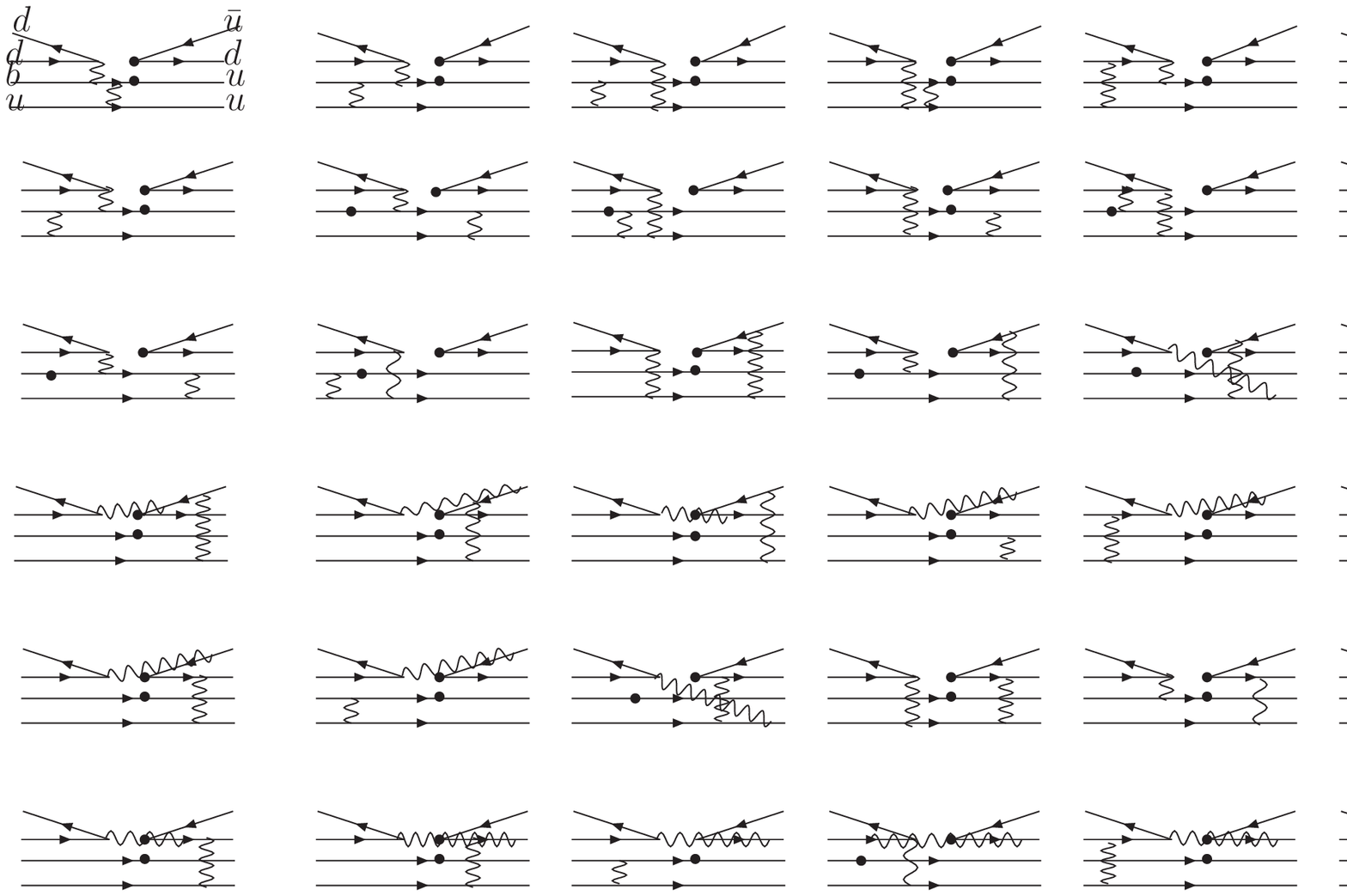}
\end{tabular}
\caption{Internal $W$ emission ($C$) diagrams for  the $\Lambda_b \to p
\pi$ decay to lowest order in the PQCD approach where the  dots
denote the weak interactions vertices. As in the preceding figure, the two hard
 gluons are essential to transfer the large momentum to the light quarks
in the initial state. These diagrams are called $C_1, C_2,...,C_{36}$.
 }\label{color-suppressed-diagrams}
\end{center}
\end{figure}

For the first diagram in Fig. \ref{color-suppressed-diagrams} (labeled as $C_1$),
a factorizable diagram, we have:
\begin{eqnarray}
f_1^{C_1}&=& G_F  { \pi^2 \over 54 \sqrt{3}} f_{\Lambda_b} f_p \int
[d x ] \int [d x^{\prime}] \int dy \,
[\alpha_s(t^{C_1})]^2 \psi_{\Lambda_b}(x) \nonumber\\
&\times &\bigg\{ \bigg[-16  M_{\Lambda_b}^4(-(C_5+
C_7)+(C_6+C_8))V_{tb}V_{td}^* \, [m_0 (x_2 (y-2)+(y-1)y) (\phi_M^P
(y)+\phi_M^T (y)) \nonumber\\&& \hspace{0.5 cm}
-M_{\Lambda_b}(x_2(y-1)+(y-2)y)\phi_M^A(y) \,] \bigg]
(\psi_p^V(x^{\prime})-\psi_p^{A}(x^{\prime}))\nonumber\\
&&+ \bigg[ -32 M_{\Lambda_b}^4(-(C_5+ C_7)+(C_6+C_8))V_{tb}V_{td}^*
\,(y-1) \, [m_0 (x_2+y-1) (\phi_M^P (y)+\phi_M^T (y))
\nonumber\\&& \hspace{0.5 cm} -M_{\Lambda_b}\phi_M^A(y) \,] \bigg ] \psi_p^{T}(x^{\prime}) \bigg \}\nonumber\\
&\times &\frac{1}{16 \pi^2}\int b_2 db_2 \int b_2' db_2' \int b_q
db_q \int d\theta_1 \int d\theta_2 \, {\rm exp} [-S^{C_1} (x,
x^{\prime}, b, b^{\prime})] \,
K_0(\sqrt{D^{C_1}}|b_2+b_2^{\prime}-b_q|)
\nonumber\\
&&\int_0^1\frac{dz_1dz_2}{z_1(1-z_1)}\sqrt{\frac{X_2^{C_1}}{|Z_2^{C_1}|}}
\bigg\{ K_1 (\sqrt{X_2^{C_1} Z_2^{C_1}})\Theta(Z_2^{C_1})
+\frac{\pi}{2}[J_1(\sqrt{X_2^{C_1}
|Z_2^{C_1}|})+iN_1(\sqrt{X_2^{C_1} |Z_2^{C_1}|})]\Theta(-Z_2^{C_1})
\bigg \} , \nonumber \\
\end{eqnarray}
where the auxiliary functions in the expressions above are defined as
\begin{eqnarray}
&&A^{T_1}=(x_2'+y-x_2'y)M_{\Lambda_b}^2,
B^{T_1}=(x_2+y)M_{\Lambda_b}^2,
 C^{T_1}=x_2 x_2'M_{\Lambda_b}^2,\,\,
D^{T_1}=x_3 y M_{\Lambda_b}^2, \nonumber\\
&&Z_2^{T_1} = A^{T_1}(1-z_2)+\frac{z_2}{z_1(1-z_1)}[B^{T_1}(1-z_1)+C^{T_1}z_1] , \nonumber\\
&&X_2^{T_1} = ({b_1'}+z_1{b_2})^2 + \frac{z_1(1-z_1)}{z_2}b_2^2, \nonumber\\
&&t^{C_1} =
max(\sqrt{|A^{C_1}|},\sqrt{|B^{C_1}|},\sqrt{|C^{C_1}|},\sqrt{|D^{C_1}|},
\omega,\omega^{\prime}).
\end{eqnarray}
Similarly, the factorization formula for the form factor $f_2$
contributed by $C_{1}$ can be written as
\begin{eqnarray}
f_2^{C_1}&=& G_F  { \pi^2 \over 54 \sqrt{3}} f_{\Lambda_b} f_p \int
[d x ] \int [d x^{\prime}] \int dy \,
[\alpha_s(t^{C_1})]^2 \psi_{\Lambda_b}(x) \nonumber\\
&\times &\bigg\{ \bigg[-16  M_{\Lambda_b}^4(-(C_5+
C_7)+(C_6+C_8))V_{tb}V_{td}^* \, [m_0 (x_2 (y-2)+(y-1)y) (\phi_M^P
(y)+\phi_M^T (y)) \nonumber\\&& \hspace{0.5 cm}
-M_{\Lambda_b}(x_2(y-1)+(y-2)y)\phi_M^A(y) \,] \bigg]
(\psi_p^V(x^{\prime})-\psi_p^{A}(x^{\prime}))\nonumber\\
&&+ \bigg[ 32 M_{\Lambda_b}^4(-(C_5+ C_7)+(C_6+C_8))V_{tb}V_{td}^*
\,(y-1) \, [m_0 (x_2+y-1) (\phi_M^P (y)+\phi_M^T (y))
\nonumber\\&& \hspace{0.5 cm} -M_{\Lambda_b}\phi_M^A(y) \,] \bigg ] \psi_p^{T}(x^{\prime}) \bigg \}\nonumber\\
&\times &\frac{1}{16 \pi^2}\int b_2 db_2 \int b_2' db_2' \int b_q
db_q \int d\theta_1 \int d\theta_2 \, {\rm exp} [-S^{C_1} (x,
x^{\prime}, b, b^{\prime})] \,
K_0(\sqrt{D^{C_1}}|b_2+b_2^{\prime}-b_q|)
\nonumber\\
&&\int_0^1\frac{dz_1dz_2}{z_1(1-z_1)}\sqrt{\frac{X_2^{C_1}}{|Z_2^{C_1}|}}
\bigg\{ K_1 (\sqrt{X_2^{C_1} Z_2^{C_1}})\Theta(Z_2^{C_1})
+\frac{\pi}{2}[J_1(\sqrt{X_2^{C_1}
|Z_2^{C_1}|})+iN_1(\sqrt{X_2^{C_1} |Z_2^{C_1}|})]\Theta(-Z_2^{C_1})
\bigg \} . \nonumber \\
\end{eqnarray}

For the 20th diagram in Fig. \ref{color-suppressed-diagrams}
(labeled $C_{20}$ ),  a non-factorizable diagram, we have:
\begin{eqnarray}
f_1^{C_{20}}&=& G_F  { \pi^2 \over 27 \sqrt{3}} f_{\Lambda_b} f_p
\int [d x ] \int [d x^{\prime}] \int dy \,
[\alpha_s(t^{C_{20}})]^2 \psi_{\Lambda_b}(x) \nonumber\\
&\times &\bigg\{  -32 M_{\Lambda_b}^4((C_5+
C_7)-(C_6+C_8))V_{tb}V_{td}^* \, [M_{\Lambda_b} x_3 \phi_M^A(y) - 2
m_0 (x_3-1)\phi_M^P (y) \,]  \psi_p^{T}(x^{\prime}) \bigg \}\nonumber\\
&\times &\frac{1}{16 \pi^2}\int b_1 db_1 \int b_3 db_3 \int b_q db_q
\int d\theta_1 \int d\theta_2 \, {\rm exp} [-S^{C_{20}} (x,
x^{\prime}, y, b, b^{\prime}, b_q)] \,
K_0(\sqrt{D^{C_{20}}}|b_2+b_2^{\prime}-b_q|)
\nonumber\\
&&\hspace{-1 cm}
\int_0^1\frac{dz_1dz_2}{z_1(1-z_1)}\sqrt{\frac{X_2^{C_{20}}}{|Z_2^{C_{20}}|}}
\bigg\{ K_1 (\sqrt{X_2^{C_{20}} Z_2^{C_{20}}})\Theta(Z_2^{C_{20}})
+\frac{\pi}{2}[J_1(\sqrt{X_2^{C_{20}}
|Z_2^{C_{20}}|})+iN_1(\sqrt{X_2^{C_{20}}
|Z_2^{C_{20}}|})]\Theta(-Z_2^{C_{20}})
\bigg \} , \nonumber \\
\end{eqnarray}
where the auxiliary functions in the above expression are defined as
\begin{eqnarray}
&&A^{C_{20}}=x_3 M_{\Lambda_b}^2,  B^{C_{20}}=-x_2^{\prime}
M_{\Lambda_b}^2, C^{C_{20}}=x_2 x_2'M_{\Lambda_b}^2,\,\,
D^{C_{20}}=x_3 y M_{\Lambda_b}^2, \nonumber\\
&&Z_2^{C_{20}} = A^{C_{20}}(1-z_2)+\frac{z_2}{z_1(1-z_1)}[B^{C_{20}}(1-z_1)+C^{C_{20}}z_1] , \nonumber\\
&&X_2^{C_{20}} = [(b_3+b_q)-z_1{b_1}]^2 + \frac{z_1(1-z_1)}{z_2}b_1
^2 ,\nonumber\\
&&t^{C_{20}} =
max(\sqrt{|A^{C_{20}}|},\sqrt{|B^{C_{20}}|},\sqrt{|C^{C_{20}}|},\sqrt{|D^{C_{20}}|},
\omega,\omega^{\prime}, \omega_q).
\end{eqnarray}

Similarly, the factorization formula for the form factor $f_2$
contributed by $C_{20}$ can be written as
\begin{eqnarray}
f_1^{C_{20}}=-f_2^{C_{20}}.
\end{eqnarray}

\subsection{Factorization formulae for the exchange diagrams}

\begin{figure}[tb]
\begin{center}
\begin{tabular}{ccc}
\hspace{-2 cm}
\includegraphics[scale=0.70]{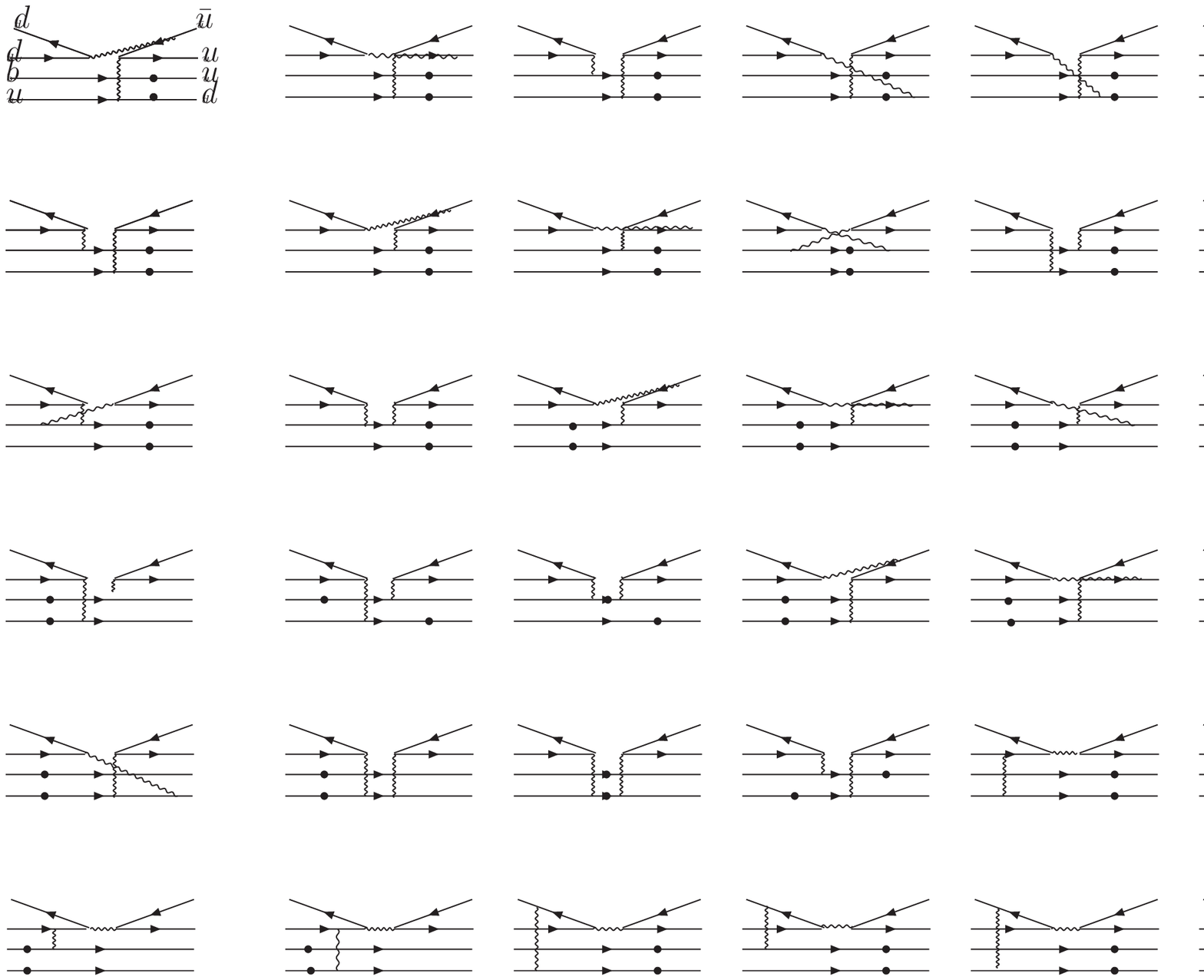}
\end{tabular}
\caption{$W$ exchange ($E$) diagrams for the   $\Lambda_b \to p \pi$
decay to lowest order in the pQCD approach where the  dots denote
the weak interactions vertices. As in the preceding figure, the two hard
 gluons are needed to
transfer the large momentum to the light quarks in the initial
state. These diarams are called $E_1, E_2,...,E_{36}$.
 }\label{W-exchange-diagrams}
\end{center}
\end{figure}

For the 18th diagram in Fig. \ref{W-exchange-diagrams}
(labeled as $E_{18}$), we have:
\begin{eqnarray}
f_1^{E_{18}}&=& G_F  { \pi^2 \over 54 \sqrt{3}} f_{\Lambda_b} f_p
\int [d x ] \int [d x^{\prime}] \int dy \,
[\alpha_s(t^{E_{18}})]^2 \psi_{\Lambda_b}(x) \nonumber\\
&\times &\bigg\{ \bigg[16 m_0 M_{\Lambda_b}^4
[(C_1-C_2)V_{ub}V_{ud}^*+((C_3+
C_9)-(C_4+C_{10})) V_{tb}V_{td}^*](y-1) (\phi_M^P(y)+\phi_M^T(y)) \nonumber\\
&&\;\;\;\;+16 m_0 M_{\Lambda_b}^4((C_5+
C_7)-(C_6+C_8))V_{tb}V_{td}^* (y-1) (\phi_M^P(y)+\phi_M^T(y)) \bigg]\psi_p^V(x^{\prime})\nonumber\\
&&+\bigg[16 m_0 M_{\Lambda_b}^4 [(C_1-C_2)V_{ub}V_{ud}^*+((C_3+
C_9)-(C_4+C_{10})) V_{tb}V_{td}^*](y-1) (\phi_M^P(y)+\phi_M^T(y)) \nonumber\\
&&\;\;\;\;-16 m_0 M_{\Lambda_b}^4((C_5+
C_7)-(C_6+C_8))V_{tb}V_{td}^* (y-1) (\phi_M^P(y)+\phi_M^T(y)) \bigg]
\psi_p^{A}(x^{\prime}) \bigg \}\nonumber\\
&\times &\frac{1}{16 \pi^2}\int b_2db_2 \int b_3 db_3 \int b_q db_q
\int d\theta_1 \int d\theta_2 \, {\rm exp} [-S^{E_{18}} (x,
x^{\prime}, y,  b, b^{\prime}, b_q)] \,  \nonumber \\
&&\{ K_0(\sqrt{C^{E_{18}}}|b_2^{\prime}|) \theta(C^{E_{18}})+ {\pi i
\over 2} [J_0(\sqrt{|C^{E_{18}}|}|b_2^{\prime}|) + i
N_0(\sqrt{|C^{E_{18}}|}|b_2^{\prime}|)] \theta(-C^{E_{18}}) \}
\int_0^1\frac{dz_1dz_2}{z_1(1-z_1)} \nonumber\\
&&\sqrt{\frac{X_2^{E_{18}}}{|Z_2^{E_{18}}|}} \bigg\{ K_1
(\sqrt{X_2^{E_{18}} Z_2^{E_{18}}})\Theta(Z_2^{E_{18}})
+\frac{\pi}{2}[J_1(\sqrt{X_2^{E_{18}}
|Z_2^{E_{18}}|})+iN_1(\sqrt{X_2^{E_{18}}
|Z_2^{E_{18}}|})]\Theta(-Z_2^{E_{18}})
\bigg \} , \nonumber \\
\end{eqnarray}
where the auxiliary functions in the expression above are defined as
\begin{eqnarray}
&&A^{E_{18}}=(x_3^{\prime}-1)M_{\Lambda_b}^2,
B^{E_{18}}=(y-1)(1-x_3^{\prime})M_{\Lambda_b}^2,
C^{E_{18}}=x_2^{\prime} (y-1)M_{\Lambda_b}^2,\,\,
D^{E_{18}}=x_3 yM_{\Lambda_b}^2, \nonumber\\
&&Z_2^{E_{18}} = A^{E_{18}}(1-z_2)+\frac{z_2}{z_1(1-z_1)}[B^{E_{18}}(1-z_1)+D^{E_{18}} z_1] , \nonumber\\
&&X_2^{E_{18}} = [{b_3}+z_1(b_2^{\prime}+b_q)]^2 +
\frac{z_1(1-z_1)}{z_2} (b_2^{\prime}+b_q)^2,
\nonumber\\
&&t^{E_{18}} =
max(\sqrt{|A^{E_{18}}|},\sqrt{|B^{E_{18}}|},\sqrt{|C^{E_{18}}|},\sqrt{|D^{E_{18}}|},
\omega,\omega^{\prime}, \omega_q).
\end{eqnarray}
Similarly, the factorization formula for the form factor $f_2$
contributed from $E_{18}$ can be written as
\begin{eqnarray}
f_2^{E_{18}}&=& f_1^{E_{18}} .
\end{eqnarray}

For the 26th diagram in Fig. \ref{W-exchange-diagrams}
(labeled as $E_{26}$), we have:
\begin{eqnarray}
f_1^{E_{26}}&=& G_F  { \pi^2 \over 54 \sqrt{3}} f_{\Lambda_b} f_p
\int [d x ] \int [d x^{\prime}] \int dy \,
[\alpha_s(t^{E_{26}})]^2 \psi_{\Lambda_b}(x) \nonumber\\
&\times &\bigg\{ \bigg[16 m_0 M_{\Lambda_b}^4
[(C_1-C_2)V_{ub}V_{ud}^*+((C_3+ C_9)-(C_4+C_{10}))
V_{tb}V_{td}^*](1-y) (\phi_M^P(y)+\phi_M^T(y)) \bigg]\psi_p^V(x^{\prime})\nonumber\\
&&+\bigg[16 m_0 M_{\Lambda_b}^4 [(C_1-C_2)V_{ub}V_{ud}^*+((C_3+
C_9)-(C_4+C_{10})) V_{tb}V_{td}^*](1-y) (\phi_M^P(y)+\phi_M^T(y))
\bigg] \psi_p^{A}(x^{\prime}) \nonumber \\
&&+\bigg[32 M_{\Lambda_b}^4((C_5+ C_7)-(C_6+C_8))V_{tb}V_{td}^*
(y-1)[M_{\Lambda_b} (x_1^{\prime}-1) \phi_M^A(y) \nonumber\\
&&\hspace{0.5 cm}-m_0 (x_1^{\prime}-2) (\phi_M^P(y)+\phi_M^T(y))] \bigg] \psi_p^{T}(x^{\prime}) \bigg \}\nonumber\\
&\times &\frac{1}{16 \pi^2}\int b_1^{\prime} db_1^{\prime} \int
b_2^{\prime} db_2^{\prime} \int b_q db_q  \int d\theta_1 \int
d\theta_2 \, {\rm exp} [-S^{E_{26}} (x, x^{\prime}, y,  b,
b^{\prime}, b_q)] \, \nonumber \\
&&\{ K_0(\sqrt{C^{E_{26}}}|b_2^{\prime}|) \theta(C^{E_{26}}) + {i
\pi \over 2} [J_0(\sqrt{C^{E_{26}}}|b_2^{\prime}|) + i
N_0(\sqrt{C^{E_{26}}}|b_2^{\prime}|) ] \theta(-C^{E_{26}}) \}
\int_0^1\frac{dz_1dz_2}{z_1(1-z_1)} \nonumber\\
&& \sqrt{\frac{X_2^{E_{26}}}{|Z_2^{E_{26}}|}} \bigg\{ K_1
(\sqrt{X_2^{E_{26}} Z_2^{E_{26}}})\Theta(Z_2^{E_{26}})
+\frac{\pi}{2}[J_1(\sqrt{X_2^{E_{26}}
|Z_2^{E_{26}}|})+iN_1(\sqrt{X_2^{E_{26}}
|Z_2^{E_{26}}|})]\Theta(-Z_2^{E_{26}})
\bigg \} , \nonumber \\
\end{eqnarray}
where the auxiliary functions above are defined as
\begin{eqnarray}
&&A^{E_{26}}=(y-1)(1-x_1^{\prime})M_{\Lambda_b}^2,
B^{E_{26}}=(x_1^{\prime}-1)M_{\Lambda_b}^2,
C^{E_{26}}=x_2^{\prime}(y-1)M_{\Lambda_b}^2,\,\,
D^{E_{26}}=x_3 y M_{\Lambda_b}^2, \nonumber\\
&&Z_2^{E_{26}} = A^{E_{18}}(1-z_2)+\frac{z_2}{z_1(1-z_1)}[B^{E_{18}}(1-z_1)+D^{E_{18}} z_1] , \nonumber\\
&&X_2^{E_{26}} =[(b_2^{\prime}+b_q)
+z_1(b_1^{\prime}-b_2^{\prime}-b_q)]^2  + \frac{z_1(1-z_1)}{z_2}
(b_1^{\prime}-b_2^{\prime}-b_q)^2 ,
\nonumber\\
&&t^{E_{26}} =
max(\sqrt{|A^{E_{26}}|},\sqrt{|B^{E_{26}}|},\sqrt{|C^{E_{26}}|},\sqrt{|D^{E_{26}}|},
\omega,\omega^{\prime}, \omega_q).
\end{eqnarray}

Similarly, the factorization formula for the form factor $f_2$
contributed by $E_{26}$ can be written as
\begin{eqnarray}
f_2^{E_{26}}&=& G_F  { \pi^2 \over 54 \sqrt{3}} f_{\Lambda_b} f_p
\int [d x ] \int [d x^{\prime}] \int dy \,
[\alpha_s(t^{E_{26}})]^2  \psi_{\Lambda_b}(x) \nonumber\\
&\times &\bigg\{ \bigg[16 m_0 M_{\Lambda_b}^4
[(C_1-C_2)V_{ub}V_{ud}^*+((C_3+ C_9)-(C_4+C_{10}))
V_{tb}V_{td}^*](1-y) (\phi_M^P(y)+\phi_M^T(y)) \bigg]\psi_p^V(x^{\prime})\nonumber\\
&&+\bigg[16 m_0 M_{\Lambda_b}^4 [(C_1-C_2)V_{ub}V_{ud}^*+((C_3+
C_9)-(C_4+C_{10})) V_{tb}V_{td}^*](1-y) (\phi_M^P(y)+\phi_M^T(y))
\bigg] \psi_p^{A}(x^{\prime}) \nonumber \\
&&+\bigg[-32 M_{\Lambda_b}^4((C_5+ C_7)-(C_6+C_8))V_{tb}V_{td}^*
(y-1)[M_{\Lambda_b} (x_1^{\prime}-1) \phi_M^A(y) \nonumber\\
&&\hspace{0.5 cm}-m_0 (x_1^{\prime}-2) (\phi_M^P(y)+\phi_M^T(y))] \bigg] \psi_p^{T}(x^{\prime}) \bigg \}\nonumber\\
&\times &\frac{1}{16 \pi^2}\int b_1^{\prime} db_1^{\prime} \int
b_2^{\prime} db_2^{\prime} \int b_q db_q  \int d\theta_1 \int
d\theta_2 \, {\rm exp} [-S^{E_{26}} (x, x^{\prime}, y,  b,
b^{\prime}, b_q)] \, \nonumber \\
&&\{K_0(\sqrt{C^{E_{26}}}|b_2^{\prime}|) \theta(C^{E_{26}}) + {i \pi
\over 2} [J_0(\sqrt{C^{E_{26}}}|b_2^{\prime}|) + i
N_0(\sqrt{C^{E_{26}}}|b_2^{\prime}|) ] \theta(-C^{E_{26}}) \}
\int_0^1\frac{dz_1dz_2}{z_1(1-z_1)} \nonumber\\
&&\sqrt{\frac{X_2^{E_{26}}}{|Z_2^{E_{26}}|}} \bigg\{ K_1
(\sqrt{X_2^{E_{26}} Z_2^{E_{26}}})\Theta(Z_2^{E_{26}})
+\frac{\pi}{2}[J_1(\sqrt{X_2^{E_{26}}
|Z_2^{E_{26}}|})+iN_1(\sqrt{X_2^{E_{26}}
|Z_2^{E_{26}}|})]\Theta(-Z_2^{E_{26}})
\bigg \} . \nonumber \\
\end{eqnarray}

\subsection{Factorization formulae for the Bow-tie diagrams}

\begin{figure}[tb]
\begin{center}
\begin{tabular}{ccc}
\hspace{-2 cm}
\includegraphics[scale=0.70]{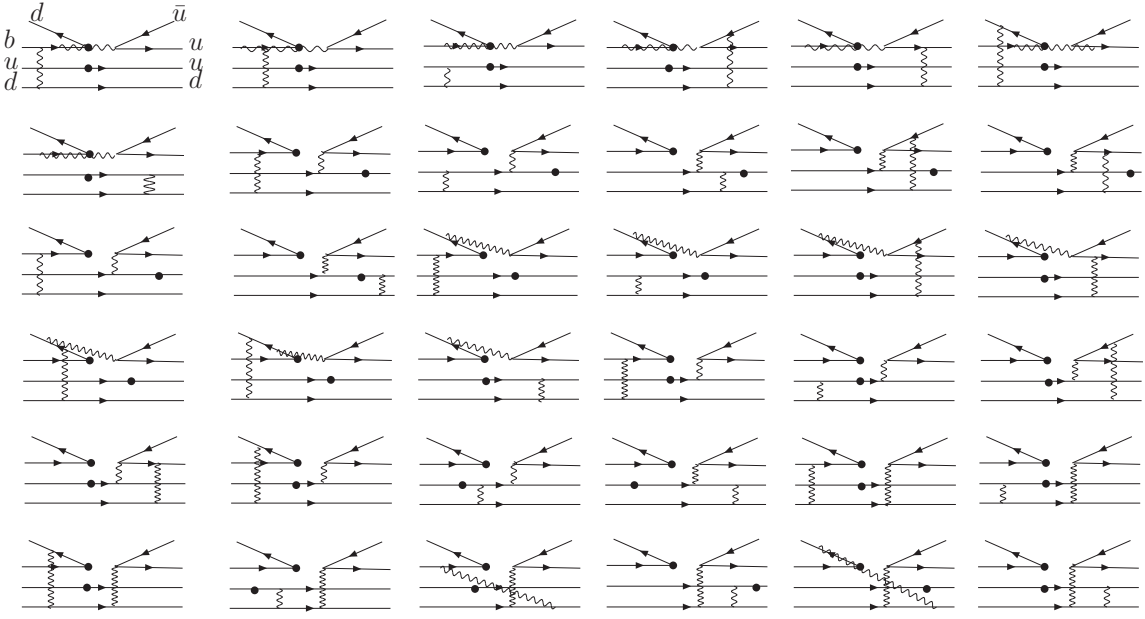}
\end{tabular}
\caption{Bow-tie ($B$) diagrams for the  $\Lambda_b \to p \pi$ decay to
lowest order in the pQCD approach where the  dots denote the weak
interactions vertices. As in the preceding figures, the two hard gluons are
 needed to transfer the large momentum to the light quarks in the
 initial state.}\label{Bow-tie-diagrams}
\end{center}
\end{figure}

For the 17th diagram in Fig. \ref{Bow-tie-diagrams} (labeled
as $B_{17}$), we have:
\begin{eqnarray}
f_1^{B_{17}}&=& G_F  { \pi^2 \over 216  \sqrt{3}} f_{\Lambda_b} f_p
\int [d x ] \int [d x^{\prime}] \int dy \,
[\alpha_s(t^{B_{17}})]^2  \psi_{\Lambda_b}(x) \nonumber\\
&\times &\bigg\{\bigg[-16 M_{\Lambda_b}^5 [(-(C_5+
C_7)+(C_6+C_{8})) V_{tb}V_{td}^* ] (x_3^{\prime}+x_1^{\prime}(y-1) -y+1 ) \phi_M^A(y)\bigg]\psi_p^{V}(x^{\prime})\nonumber\\
&&+\bigg[-32 m_0 M_{\Lambda_b}^4 [(-C_1+C_2)V_{ub}V_{ud}^*+(-(C_3+
C_9)+(C_4+C_{10})) V_{tb}V_{td}^* ] (y-1) (\phi_M^P (y)+\phi_M^T (y)) \nonumber\\
&&\;\;\;\;-16 M_{\Lambda_b}^5 [(-(C_5+
C_7)+(C_6+C_{8})) V_{tb}V_{td}^* ] (x_3^{\prime}+x_1^{\prime}(y-1)-y+1 ) \phi_M^A(y) \bigg]\psi_p^A(x^{\prime})\nonumber\\
&&+ \bigg[32 M_{\Lambda_b}^5 [(-C_1+C_2)V_{ub}V_{ud}^*+(-(C_3+
C_9)+(C_4+C_{10})) V_{tb}V_{td}^* ] x_3^{\prime} \psi_M^{A}(y)\nonumber\\
&&\;\;\;\;+32m_0 M_{\Lambda_b}^4 [(-(C_5+
C_7)+(C_6+C_{8})) V_{tb}V_{td}^* ] (y-1) (\phi_M^P (y)+\phi_M^T (y)) \bigg]\psi_p^{T}(x^{\prime}) \bigg \}\nonumber\\
&\times &\frac{1}{32 \pi^2}\int b_1^{\prime} db_1^{\prime} \int b_3
db_3 \int b_q db_q \int d\theta_1 \int d\theta_2 \, {\rm exp}
[-S^{B_{17}} (x, x^{\prime},y, b, b^{\prime}, b_q)] \,
\int_0^1\frac{dz_1dz_2
dz_3}{z_1(1-z_1)z_2(1-z_2)} \nonumber\\
&& \sqrt{ \frac{X_3^{B_{17}}} {|Z_3^{B_{17}}|}} \bigg \{ K_1
(\sqrt{X_3^{B_{17}}Z_3^{B_{17}}})\Theta(Z_3^{B_{17}})
+\frac{\pi}{2}[J_1(\sqrt{X_3^{B_{17}}|Z_3^{B_{17}}|})+iN_1(\sqrt{X_3^{B_{17}}|Z_3^{B_{17}}|})]
\Theta(-Z_3^{B_{17}})\bigg\} . \nonumber \\
\end{eqnarray}
where the auxiliary functions in the above expression are defined as
\begin{eqnarray}
A^{B_{17}}&=&x_3^{\prime} (y-1) M_{\Lambda_b}^2,
B^{B_{17}}=(x_1^{\prime}-1) M_{\Lambda_b}^2,
C^{B_{17}}=(y-1)(1-x_1^{\prime}) M_{\Lambda_b}^2,\,\,
D^{B_{17}}=x_3 x_3^{\prime}M_{\Lambda_b}^2\nonumber\\
X_3^{B_{17}}&=&((-b_1^{\prime}+b_3)-(-b_3+b_q)z_2-(-b_1^{\prime}+b_3-b_q)z_1(1-z_2))^2
\nonumber \\ &&+\frac{z_2(1-z_2)}{z_3}((-b_3+b_q)-z_1
(-b_1^{\prime}+b_3-b_q) )^2 +\frac{z_1(1-z_1)z_2(1-z_2)}{z_2z_3} (-b_1^{\prime}+b_3-b_q)^2,\nonumber\\
Z_3^{B_{17}}&=& A(1-z_3)+\frac{z_3}{z_2(1-z_2)}\bigg [B
(1-z_2)+\frac{z_2}{z_1(1-z_1)}[C(1-z_1)+Dz_1] \bigg ] ,
\nonumber\\
t^{B_{17}} &=&
max(\sqrt{|A^{B_{17}}|},\sqrt{|B^{B_{17}}|},\sqrt{|C^{B_{17}}|},\sqrt{|D^{B_{17}}|},
\omega,\omega^{\prime}, \omega_q).
\end{eqnarray}

Similarly, the factorization formula for the form factor $f_2$
contributed by fig. $B_{17}$ can be written as:
\begin{eqnarray}
f_2^{B_{17}}&=& G_F  { \pi^2 \over 216  \sqrt{3}} f_{\Lambda_b} f_p
\int [d x ] \int [d x^{\prime}] \int dy \,
[\alpha_s(t^{B_{17}})]^2  \psi_{\Lambda_b}(x) \nonumber\\
&\times &\bigg\{\bigg[-32 m_0 M_{\Lambda_b}^4
[(-C_1+C_2)V_{ub}V_{ud}^*+(-(C_3+
C_9)+(C_4+C_{10})) V_{tb}V_{td}^* ] (y-1) (\phi_M^P (y)+\phi_M^T (y)) \nonumber\\
&&\;\;\;\;+16 M_{\Lambda_b}^5 [(-(C_5+
C_7)+(C_6+C_{8})) V_{tb}V_{td}^* ] (x_3^{\prime}+x_1^{\prime}(y-1)-y+1 ) \phi_M^A(y) \bigg]\psi_p^V(x^{\prime})\nonumber\\
&&+\bigg[16 M_{\Lambda_b}^5 [(-(C_5+ C_7)+(C_6+C_{8}))
V_{tb}V_{td}^* ] (x_3^{\prime}+x_1^{\prime}(y-1) -y+1 )
\phi_M^A(y)\bigg]\psi_p^{A}(x^{\prime})\nonumber\\
&&+ \bigg[32 M_{\Lambda_b}^5 [(-C_1+C_2)V_{ub}V_{ud}^*+(-(C_3+
C_9)+(C_4+C_{10})) V_{tb}V_{td}^* ] x_3^{\prime} \psi_M^{A}(y)\nonumber\\
&&\;\;\;\;-32m_0 M_{\Lambda_b}^4 [(-(C_5+
C_7)+(C_6+C_{8})) V_{tb}V_{td}^* ] (y-1) (\phi_M^P (y)+\phi_M^T (y)) \bigg]\psi_p^{T}(x^{\prime}) \bigg \}\nonumber\\
&\times &\frac{1}{32 \pi^2}\int b_1^{\prime} db_1^{\prime} \int b_3
db_3 \int b_q db_q \int d\theta_1 \int d\theta_2 \, {\rm exp}
[-S^{B_{17}} (x, x^{\prime},y, b, b^{\prime}, b_q)] \,
\int_0^1\frac{dz_1dz_2
dz_3}{z_1(1-z_1)z_2(1-z_2)} \nonumber\\
&& \sqrt{ \frac{X_3^{B_{17}}} {|Z_3^{B_{17}}|}} \bigg \{ K_1
(\sqrt{X_3^{B_{17}}Z_3^{B_{17}}})\Theta(Z_3^{B_{17}})
+\frac{\pi}{2}[J_1(\sqrt{X_3^{B_{17}}|Z_3^{B_{17}}|})+iN_1(\sqrt{X_3^{B_{17}}|Z_3^{B_{17}}|})]
\Theta(-Z_3^{B_{17}})\bigg\} . \nonumber \\
\end{eqnarray}

For the 19th diagram in Fig. \ref{Bow-tie-diagrams} (labeled
as $B_{19}$ ), we have:
\begin{eqnarray}
f_1^{B_{19}}&=& G_F  { \pi^2 \over 27  \sqrt{3}} f_{\Lambda_b} f_p
\int [d x ] \int [d x^{\prime}] \int dy \,
[\alpha_s(t^{B_{19}})]^2  \psi_{\Lambda_b}(x) \nonumber\\
&\times &\bigg\{\bigg[-16 M_{\Lambda_b}^5 [(-(C_5+
C_7)+(C_6+C_{8})) V_{tb}V_{td}^* ] x_2^{\prime} \phi_M^A(y)\bigg]\psi_p^{V}(x^{\prime})\nonumber\\
&&+\bigg[-64 m_0 M_{\Lambda_b}^4 [(-C_1+C_2)V_{ub}V_{ud}^*+(-(C_3+
C_9)+(C_4+C_{10})) V_{tb}V_{td}^* ] \phi_M^P (y) \nonumber\\
&&\;\;\;\;-16 M_{\Lambda_b}^5 [(-(C_5+
C_7)+(C_6+C_{8})) V_{tb}V_{td}^* ] x_2^{\prime} \phi_M^A(y) \bigg]\psi_p^A(x^{\prime})\nonumber\\
&&+ \bigg[32 M_{\Lambda_b}^5 [(-C_1+C_2)V_{ub}V_{ud}^*+(-(C_3+
C_9)+(C_4+C_{10})) V_{tb}V_{td}^* ] x_2^{\prime} \psi_M^{A}(y)\nonumber\\
&&\;\;\;\;-64 m_0 M_{\Lambda_b}^4 [(-(C_5+
C_7)+(C_6+C_{8})) V_{tb}V_{td}^* ] (x_1^{\prime}-2) \phi_M^P (y) \bigg]\psi_p^{T}(x^{\prime}) \bigg \}\nonumber\\
&\times &\frac{1}{16 \pi^2}\int b_1^{\prime} db_1^{\prime} \int
b_2^{\prime} db_2^{\prime} \int b_q db_q  \int d\theta_1 \int
d\theta_2 \, {\rm exp} [-S^{B_{19}} (x, x^{\prime}, y,  b,
b^{\prime}, b_q)] \, \nonumber \\
&& \{ K_0(\sqrt{C^{B_{19}}}|b_q|) \theta(C^{B_{19}}) +{ i \pi \over
2} [ N_0(\sqrt{C^{B_{19}}}|b_q|)+ iK_0(\sqrt{C^{B_{19}}}|b_q|) ]
\theta(-C^{B_{19}}) \}
\int_0^1\frac{dz_1dz_2}{z_1(1-z_1)} \nonumber\\
&&\sqrt{\frac{X_2^{B_{19}}}{|Z_2^{B_{19}}|}} \bigg\{ K_1
(\sqrt{X_2^{B_{19}} Z_2^{E_{26}}})\Theta(Z_2^{B_{19}})
+\frac{\pi}{2}[J_1(\sqrt{X_2^{B_{19}}
|Z_2^{B_{19}}|})+iN_1(\sqrt{X_2^{B_{19}}
|Z_2^{B_{19}}|})]\Theta(-Z_2^{B_{19}})
\bigg \} , \nonumber \\
\end{eqnarray}
where the auxiliary functions in the above expression are defined as:
\begin{eqnarray}
&& A^{B_{19}}=-x_2^{\prime} M_{\Lambda_b}^2,
B^{B_{19}}=(x_1^{\prime}-1) M_{\Lambda_b}^2,
C^{B_{19}}=x_2^{\prime}(y-1) M_{\Lambda_b}^2,
D^{B_{19}}=x_3 x_3^{\prime}M_{\Lambda_b}^2\nonumber\\
&&Z_2^{B_{19}} = A^{B_{19}}(1-z_2)+\frac{z_2}{z_1(1-z_1)}[B^{B_{19}}(1-z_1)+D^{B_{19}} z_1] , \nonumber\\
&&X_2^{B_{19}} =[(b_2^{\prime}+b_q) - z_1b_1^{\prime}]^2  +
\frac{z_1(1-z_1)}{z_2} {b_1^{\prime}}^2
\nonumber\\
&& t^{B_{19}} =
max(\sqrt{|A^{B_{19}}|},\sqrt{|B^{B_{19}}|},\sqrt{|C^{B_{19}}|},\sqrt{|D^{B_{19}}|},
\omega,\omega^{\prime}, \omega_q).
\end{eqnarray}

Similarly, the factorization formula for the form factor $f_2$
contributed by $B_{19}$ can be written as:
\begin{eqnarray}
f_2^{B_{19}}&=& G_F  { \pi^2 \over 27  \sqrt{3}} f_{\Lambda_b} f_p
\int [d x ] \int [d x^{\prime}] \int dy \,
[\alpha_s(t^{B_{19}})]^2  \psi_{\Lambda_b}(x) \nonumber\\
&\times &\bigg\{\bigg[-64 m_0 M_{\Lambda_b}^4
[(-C_1+C_2)V_{ub}V_{ud}^*+(-(C_3+
C_9)+(C_4+C_{10})) V_{tb}V_{td}^* ] \phi_M^P (y) \nonumber\\
&&\;\;\;\;+16 M_{\Lambda_b}^5 [(-(C_5+
C_7)+(C_6+C_{8})) V_{tb}V_{td}^* ] x_2^{\prime} \phi_M^A(y) \bigg]\psi_p^V(x^{\prime})\nonumber\\
&&+\bigg[16 M_{\Lambda_b}^5 [(-(C_5+ C_7)+(C_6+C_{8}))
V_{tb}V_{td}^* ] x_2^{\prime}
\phi_M^A(y)\bigg]\psi_p^{A}(x^{\prime})\nonumber\\
&&+ \bigg[32 M_{\Lambda_b}^5 [(-C_1+C_2)V_{ub}V_{ud}^*+(-(C_3+
C_9)+(C_4+C_{10})) V_{tb}V_{td}^* ] x_2^{\prime} \psi_M^{A}(y)\nonumber\\
&&\;\;\;\;+64 m_0 M_{\Lambda_b}^4 [(-(C_5+
C_7)+(C_6+C_{8})) V_{tb}V_{td}^* ] (x_1^{\prime}-2) \phi_M^P (y) \bigg]\psi_p^{T}(x^{\prime}) \bigg \}\nonumber\\
&\times &\frac{1}{16 \pi^2}\int b_1^{\prime} db_1^{\prime} \int
b_2^{\prime} db_2^{\prime} \int b_q db_q  \int d\theta_1 \int
d\theta_2 \, {\rm exp} [-S^{B_{19}} (x, x^{\prime}, y,  b,
b^{\prime}, b_q)] \, \nonumber \\
&& \{ K_0(\sqrt{C^{B_{19}}}|b_q|) \theta(C^{B_{19}}) +{ i \pi \over
2} [ N_0(\sqrt{C^{B_{19}}}|b_q|)+ iK_0(\sqrt{C^{B_{19}}}|b_q|) ]
\theta(-C^{B_{19}}) \}
\int_0^1\frac{dz_1dz_2}{z_1(1-z_1)} \nonumber\\
&&\sqrt{\frac{X_2^{B_{19}}}{|Z_2^{B_{19}}|}} \bigg\{ K_1
(\sqrt{X_2^{B_{19}} Z_2^{E_{26}}})\Theta(Z_2^{B_{19}})
+\frac{\pi}{2}[J_1(\sqrt{X_2^{B_{19}}
|Z_2^{B_{19}}|})+iN_1(\sqrt{X_2^{B_{19}}
|Z_2^{B_{19}}|})]\Theta(-Z_2^{B_{19}})
\bigg \} . \nonumber \\
\end{eqnarray}

\subsection{Factorization formulae for the penguin annihilation diagrams}

\begin{figure}[tb]
\begin{center}
\begin{tabular}{ccc}
\hspace{-2 cm}
\includegraphics[scale=0.70]{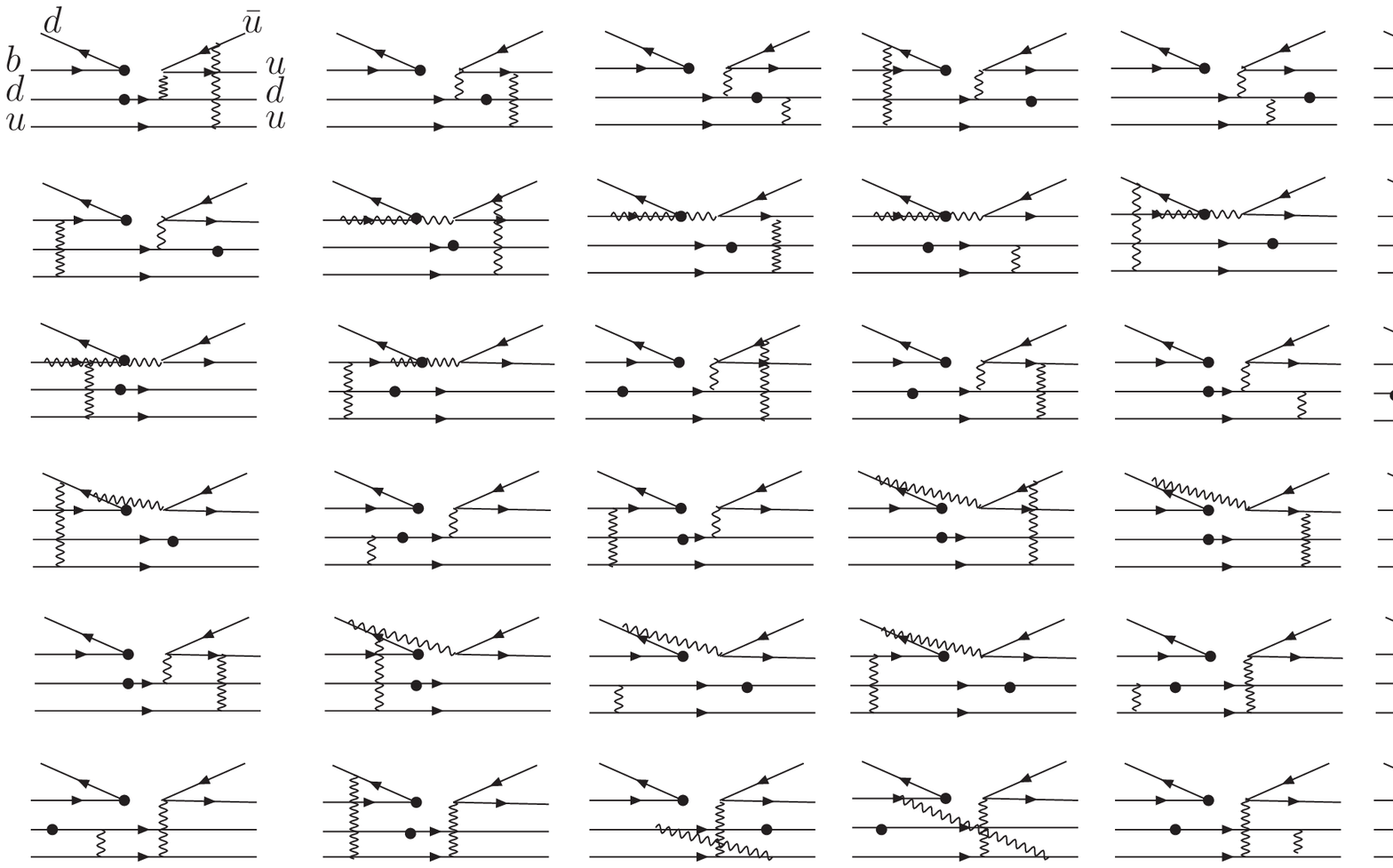}
\end{tabular}
\caption{Penguin annihilation ($P$) diagrams for the  $\Lambda_b \to p
\pi$ decay to lowest order in the pQCD approach where the  dots
denote the weak interactions vertices. As before, the two hard gluons are
essential to transfer the large momentum to the light quarks in the
initial state. These diagrams are called $P_1, P_2,...,P_{36}$.
}\label{penguin-annihilation-diagrams}
\end{center}
\end{figure}

For the 14th diagram in Fig. \ref{penguin-annihilation-diagrams}
 (labeled as $P_{14}$), we have:
\begin{eqnarray}
f_1^{P_{14}}&=& G_F  { \pi^2 \over 54  \sqrt{3}} f_{\Lambda_b} f_p
\int [d x ] \int [d x^{\prime}] \int dy \,
[\alpha_s(t^{P_{14}})]^2  \psi_{\Lambda_b}(x) \nonumber\\
&\times &\bigg\{\bigg[48 M_{\Lambda_b}^4 [(C_3+C_4)-{1 \over
2}(C_9+C_{10}) ]V_{tb}V_{td}^*
(M_{\Lambda_b} (y-1)\phi_M^A (y) +m_0 (\phi_M^P (y)-\phi_M^T (y)) ) \nonumber\\
&&\;\;\;\;-32  M_{\Lambda_b}^4  [(C_5-{1 \over 2} C_7)+{1 \over
2}(C_6-{1 \over 2}C_{8}) ] V_{tb}V_{td}^*  y ( M_{\Lambda_b}
x_2^{\prime}\phi_M^A(y) +m_0 (\phi_M^T(y)-\phi_M^P(y)))\nonumber\\
&&\;\;\;\;-32  M_{\Lambda_b}^4  [{1 \over 2}(C_5-{1 \over 2}
C_7)+(C_6-{1 \over 2}C_{8}) ] V_{tb}V_{td}^*   ( M_{\Lambda_b}
x_2^{\prime}\phi_M^A(y) +m_0 (y-2) \phi_M^P(y)+ m_0 y \phi_M^T(y)) \bigg]\psi_p^V(x^{\prime})\nonumber\\
&&+\bigg[48 M_{\Lambda_b}^4 [(C_3+C_4)-{1 \over 2}(C_9+C_{10})
]V_{tb}V_{td}^*(M_{\Lambda_b} (y-1)\phi_M^A (y) - m_0 (\phi_M^P (y)-\phi_M^T (y)) ) \nonumber\\
&&\;\;\;\; + 32  M_{\Lambda_b}^4  [(C_5-{1 \over 2} C_7)+{1 \over
2}(C_6-{1 \over 2}C_{8}) ] V_{tb}V_{td}^*  y ( M_{\Lambda_b}
x_2^{\prime}\phi_M^A(y) - m_0 (\phi_M^T(y)-\phi_M^P(y)))\nonumber\\
&&\;\;\;\;+32  M_{\Lambda_b}^4  [{1 \over 2}(C_5-{1 \over 2}
C_7)+(C_6-{1 \over 2}C_{8}) ] V_{tb}V_{td}^*   ( M_{\Lambda_b}
x_2^{\prime}\phi_M^A(y)- m_0 (y-2) \phi_M^P(y)- m_0 y \phi_M^T(y))
\bigg]\psi_p^A(x^{\prime})\nonumber \\
&&+\bigg[96 m_0 M_{\Lambda_b}^4 [(C_3+C_4)-{1 \over 2}(C_9+C_{10})
]V_{tb}V_{td}^* y  (\phi_M^P (y)-\phi_M^T (y)) \nonumber\\
&&\;\;\;\; + 64  M_{\Lambda_b}^5 y [(C_5-{1 \over 2} C_7)+{1 \over
2}(C_6-{1 \over 2}C_{8}) ] V_{tb}V_{td}^*  (y-1) \phi_M^A(y) \nonumber\\
&&\;\;\;\;- 64  M_{\Lambda_b}^5  [{1 \over 2}(C_5-{1 \over 2}
C_7)+(C_6-{1 \over 2}C_{8}) ] V_{tb}V_{td}^*   (x_3^{\prime}-1)
\phi_M^A(y) \bigg]\psi_p^T(x^{\prime})\nonumber \\
&\times &\frac{1}{16 \pi^2} \int b_2 db_2 \int b_q db_q \int
b_2^{\prime} db_2^{\prime} \int d\theta_1 \int d\theta_2 \, {\rm
exp} [-S^{P_{14}} (x, x^{\prime}, y,  b,
b^{\prime}, b_q)] \, \nonumber \\
&& \{ K_0(\sqrt{C^{P_{14}}}|b_2|) \theta(C^{P_{14}}) +{ i \pi \over
2} [ N_0(\sqrt{|C^{P_{14}}|}|b_2|)+ iK_0(\sqrt{|C^{P_{14}}|}|b_2|) ]
\theta(-C^{P_{14}}) \}
\int_0^1\frac{dz_1dz_2}{z_1(1-z_1)} \nonumber\\
&&\sqrt{\frac{X_2^{P_{14}}}{|Z_2^{P_{14}}|}} \bigg\{ K_1
(\sqrt{X_2^{P_{14}} Z_2^{E_{26}}})\Theta(Z_2^{P_{14}})
+\frac{\pi}{2}[J_1(\sqrt{X_2^{P_{14}}
|Z_2^{P_{14}}|})+iN_1(\sqrt{X_2^{P_{14}}
|Z_2^{P_{14}}|})]\Theta(-Z_2^{P_{14}})
\bigg \} , \nonumber \\
\end{eqnarray}
where the auxiliary functions in the above expression are defined as:
\begin{eqnarray}
&& A^{P_{14}}=x_2^{\prime} M_{\Lambda_b}^2,
B^{P_{14}}=(1-x_3^{\prime}y) M_{\Lambda_b}^2,
C^{P_{14}}=x_2x_2^{\prime}M_{\Lambda_b}^2,\,\,
D^{P_{14}}=x_1^{\prime} (y-1) M_{\Lambda_b}^2\nonumber\\
&&Z_2^{P_{14}} = A^{P_{14}}(1-z_2)+\frac{z_2}{z_1(1-z_1)}[B^{P_{14}}(1-z_1)+D^{P_{14}} z_1] , \nonumber\\
&&X_2^{P_{14}} =[(b_2+b_2^{\prime}) - z_1 b_q]^2  +
\frac{z_1(1-z_1)}{z_2} {b_q}^2 ,
\nonumber\\
&& t^{P_{14}} =
max(\sqrt{|A^{P_{14}}|},\sqrt{|B^{P_{14}}|},\sqrt{|C^{P_{14}}|},\sqrt{|D^{P_{14}}|},
\omega,\omega^{\prime}, \omega_q).
\end{eqnarray}

Similarly, the factorization formula for the form factor $f_2$
contributed by $P_{14}$ can be written as:
\begin{eqnarray}
f_2^{P_{14}}&=& G_F  { \pi^2 \over 54  \sqrt{3}} f_{\Lambda_b} f_p
\int [d x ] \int [d x^{\prime}] \int dy \,
[\alpha_s(t^{P_{14}})]^2  \psi_{\Lambda_b}(x) \nonumber\\
&\times &\bigg\{\bigg[48 M_{\Lambda_b}^4 [(C_3+C_4)-{1 \over
2}(C_9+C_{10}) ]V_{tb}V_{td}^*
(M_{\Lambda_b} (y-1)\phi_M^A (y) +m_0 (\phi_M^P (y)-\phi_M^T (y)) ) \nonumber\\
&&\;\;\;\;+32  M_{\Lambda_b}^4  [(C_5-{1 \over 2} C_7)+{1 \over
2}(C_6-{1 \over 2}C_{8}) ] V_{tb}V_{td}^*  y ( M_{\Lambda_b}
x_2^{\prime}\phi_M^A(y) +m_0 (\phi_M^T(y)-\phi_M^P(y)))\nonumber\\
&&\;\;\;\;-32  M_{\Lambda_b}^4  [{1 \over 2}(C_5-{1 \over 2}
C_7)+(C_6-{1 \over 2}C_{8}) ] V_{tb}V_{td}^*   ( M_{\Lambda_b}
x_2^{\prime}\phi_M^A(y) +m_0 (y-2) \phi_M^P(y)+ m_0 y \phi_M^T(y)) \bigg]\psi_p^V(x^{\prime})\nonumber\\
&&+\bigg[48 M_{\Lambda_b}^4 [(C_3+C_4)-{1 \over 2}(C_9+C_{10})
]V_{tb}V_{td}^*(M_{\Lambda_b} (y-1)\phi_M^A (y) - m_0 (\phi_M^P (y)-\phi_M^T (y)) ) \nonumber\\
&&\;\;\;\; - 32  M_{\Lambda_b}^4  [(C_5-{1 \over 2} C_7)+{1 \over
2}(C_6-{1 \over 2}C_{8}) ] V_{tb}V_{td}^*  y ( M_{\Lambda_b}
x_2^{\prime}\phi_M^A(y) - m_0 (\phi_M^T(y)-\phi_M^P(y)))\nonumber\\
&&\;\;\;\;+32  M_{\Lambda_b}^4  [{1 \over 2}(C_5-{1 \over 2}
C_7)+(C_6-{1 \over 2}C_{8}) ] V_{tb}V_{td}^*   ( M_{\Lambda_b}
x_2^{\prime}\phi_M^A(y)- m_0 (y-2) \phi_M^P(y)- m_0 y \phi_M^T(y))
\bigg]\psi_p^A(x^{\prime})\nonumber \\
&&+\bigg[96 m_0 M_{\Lambda_b}^4 [(C_3+C_4)-{1 \over 2}(C_9+C_{10})
]V_{tb}V_{td}^* y  (\phi_M^P (y)-\phi_M^T (y)) \nonumber\\
&&\;\;\;\; + 64  M_{\Lambda_b}^5 y [(C_5-{1 \over 2} C_7)+{1 \over
2}(C_6-{1 \over 2}C_{8}) ] V_{tb}V_{td}^*  (y-1) \phi_M^A(y) \nonumber\\
&&\;\;\;\; + 64  M_{\Lambda_b}^5  [{1 \over 2}(C_5-{1 \over 2}
C_7)+(C_6-{1 \over 2}C_{8}) ] V_{tb}V_{td}^*   (x_3^{\prime}-1)
\phi_M^A(y) \bigg]\psi_p^T(x^{\prime})\nonumber \\
&\times &\frac{1}{16 \pi^2} \int b_2 db_2 \int b_q db_q \int
b_2^{\prime} db_2^{\prime} \int d\theta_1 \int d\theta_2 \, {\rm
exp} [-S^{P_{14}} (x, x^{\prime}, y,  b,
b^{\prime}, b_q)] \, \nonumber \\
&& \{ K_0(\sqrt{C^{P_{14}}}|b_2|) \theta(C^{P_{14}}) +{ i \pi \over
2} [ N_0(\sqrt{|C^{P_{14}}|}|b_2|)+ iK_0(\sqrt{|C^{P_{14}}|}|b_2|) ]
\theta(-C^{P_{14}}) \}
\int_0^1\frac{dz_1dz_2}{z_1(1-z_1)} \nonumber\\
&&\sqrt{\frac{X_2^{P_{14}}}{|Z_2^{P_{14}}|}} \bigg\{ K_1
(\sqrt{X_2^{P_{14}} Z_2^{E_{26}}})\Theta(Z_2^{P_{14}})
+\frac{\pi}{2}[J_1(\sqrt{X_2^{P_{14}}
|Z_2^{P_{14}}|})+iN_1(\sqrt{X_2^{P_{14}}
|Z_2^{P_{14}}|})]\Theta(-Z_2^{P_{14}})
\bigg \} . \nonumber \\
\end{eqnarray}

For the 26th diagram in Fig. \ref{penguin-annihilation-diagrams}
 (labeled as $P_{26}$), we have:
\begin{eqnarray}
f_1^{P_{26}}&=& G_F  { \pi^2 \over 27  \sqrt{3}} f_{\Lambda_b} f_p
\int [d x ] \int [d x^{\prime}] \int dy \,
[\alpha_s(t^{P_{26}})]^2  \psi_{\Lambda_b}(x) \nonumber\\
&\times &\bigg\{\bigg[16  M_{\Lambda_b}^4  [(C_5-{1 \over 2}
C_7)-(C_6-{1 \over 2}C_{8}) ] V_{tb}V_{td}^*  ( M_{\Lambda_b}
x_1^{\prime}\phi_M^A(y) - 2 m_0 (x_3^{\prime} -2) \phi_M^P(y) )\nonumber\\
&&\;\;\;\; -16  M_{\Lambda_b}^4  [(C_5-{1 \over 2} C_7)-(C_6-{1
\over 2}C_{8}) ] V_{tb}V_{td}^*   ( M_{\Lambda_b}
x_1^{\prime}\phi_M^A(y) +2 m_0  \phi_M^P(y)) \bigg]\psi_p^V(x^{\prime})\nonumber\\
&&+\bigg[-16  M_{\Lambda_b}^4  [(C_5-{1 \over 2} C_7)-(C_6-{1 \over
2}C_{8}) ] V_{tb}V_{td}^*  ( M_{\Lambda_b}
x_1^{\prime}\phi_M^A(y) + 2 m_0 (x_3^{\prime} -2) \phi_M^P(y) )\nonumber\\
&&\;\;\;\; +16  M_{\Lambda_b}^4  [(C_5-{1 \over 2} C_7)-(C_6-{1
\over 2}C_{8}) ] V_{tb}V_{td}^*   ( M_{\Lambda_b}
x_1^{\prime}\phi_M^A(y) - 2 m_0  \phi_M^P(y)) \bigg]\psi_p^A(x^{\prime}) \bigg \}\nonumber\\
&\times &\frac{1}{16 \pi^2} \int b_1^{\prime} db_1^{\prime}   \int
b_2 db_2 \int b_q db_q \int d\theta_1 \int d\theta_2 \, {\rm exp}
[-S^{P_{26}} (x, x^{\prime}, y,  b,
b^{\prime}, b_q)] \, \nonumber \\
&& \{ K_0(\sqrt{D^{P_{26}}}|b_q|) \theta(D^{P_{26}}) +{ i \pi \over
2} [ N_0(\sqrt{|D^{P_{26}}|}|b_q|)+ iK_0(\sqrt{|D^{P_{26}}|}|b_q|) ]
\theta(-D^{P_{26}}) \}
\int_0^1\frac{dz_1dz_2}{z_1(1-z_1)} \nonumber\\
&&\sqrt{\frac{X_2^{P_{26}}}{|Z_2^{P_{26}}|}} \bigg\{ K_1
(\sqrt{X_2^{P_{26}} Z_2^{E_{26}}})\Theta(Z_2^{P_{26}})
+\frac{\pi}{2}[J_1(\sqrt{X_2^{P_{26}}
|Z_2^{P_{26}}|})+iN_1(\sqrt{X_2^{P_{26}}
|Z_2^{P_{26}}|})]\Theta(-Z_2^{P_{26}})
\bigg \} , \nonumber \\
\end{eqnarray}
where the auxiliary functions in the above expression are defined as:
\begin{eqnarray}
&& A^{P_{26}}=-x_1^{\prime} M_{\Lambda_b}^2,
B^{P_{26}}=(x_3^{\prime}-1) M_{\Lambda_b}^2,\,\,
C^{P_{26}}=x_2x_2^{\prime}M_{\Lambda_b}^2,
D^{P_{26}}=x_1^{\prime} (y-1) M_{\Lambda_b}^2\nonumber\\
&&Z_2^{P_{26}} = A^{P_{26}}(1-z_2)+\frac{z_2}{z_1(1-z_1)}[B^{P_{26}}(1-z_1)+C^{P_{26}} z_1] , \nonumber\\
&&X_2^{P_{26}} =[(b_1^{\prime}+b_q) - z_1 b_2]^2  +
\frac{z_1(1-z_1)}{z_2} {b_2}^2 ,
\nonumber\\
&& t^{P_{26}} =
max(\sqrt{|A^{P_{26}}|},\sqrt{|B^{P_{26}}|},\sqrt{|C^{P_{26}}|},\sqrt{|D^{P_{26}}|},
\omega,\omega^{\prime}, \omega_q).
\end{eqnarray}

Similarly, the factorization formula for the form factor $f_2$
contributed by $P_{26}$ can be written as:
\begin{eqnarray}
f_2^{P_{26}}&=& G_F  { \pi^2 \over 27  \sqrt{3}} f_{\Lambda_b} f_p
\int [d x ] \int [d x^{\prime}] \int dy \,
[\alpha_s(t^{P_{26}})]^2  \psi_{\Lambda_b}(x) \nonumber\\
&\times &\bigg\{\bigg[-16  M_{\Lambda_b}^4  [(C_5-{1 \over 2}
C_7)-(C_6-{1 \over 2}C_{8}) ] V_{tb}V_{td}^*  ( M_{\Lambda_b}
x_1^{\prime}\phi_M^A(y) - 2 m_0 (x_3^{\prime} -2) \phi_M^P(y) )\nonumber\\
&&\;\;\;\; -16  M_{\Lambda_b}^4  [(C_5-{1 \over 2} C_7)-(C_6-{1
\over 2}C_{8}) ] V_{tb}V_{td}^*   ( M_{\Lambda_b}
x_1^{\prime}\phi_M^A(y) +2 m_0  \phi_M^P(y)) \bigg]\psi_p^V(x^{\prime})\nonumber\\
&&+\bigg[16  M_{\Lambda_b}^4  [(C_5-{1 \over 2} C_7)-(C_6-{1 \over
2}C_{8}) ] V_{tb}V_{td}^*  ( M_{\Lambda_b}
x_1^{\prime}\phi_M^A(y) + 2 m_0 (x_3^{\prime} -2) \phi_M^P(y) )\nonumber\\
&&\;\;\;\; +16  M_{\Lambda_b}^4  [(C_5-{1 \over 2} C_7)-(C_6-{1
\over 2}C_{8}) ] V_{tb}V_{td}^*   ( M_{\Lambda_b}
x_1^{\prime}\phi_M^A(y) - 2 m_0  \phi_M^P(y)) \bigg]\psi_p^A(x^{\prime}) \bigg \}\nonumber\\
&\times &\frac{1}{16 \pi^2} \int b_1^{\prime} db_1^{\prime}   \int
b_2 db_2 \int b_q db_q \int d\theta_1 \int d\theta_2 \, {\rm exp}
[-S^{P_{26}} (x, x^{\prime}, y,  b,
b^{\prime}, b_q)] \, \nonumber \\
&& \{ K_0(\sqrt{D^{P_{26}}}|b_q|) \theta(D^{P_{26}}) +{ i \pi \over
2} [ N_0(\sqrt{|D^{P_{26}}|}|b_q|)+ iK_0(\sqrt{|D^{P_{26}}|}|b_q|) ]
\theta(-D^{P_{26}}) \}
\int_0^1\frac{dz_1dz_2}{z_1(1-z_1)} \nonumber\\
&&\sqrt{\frac{X_2^{P_{26}}}{|Z_2^{P_{26}}|}} \bigg\{ K_1
(\sqrt{X_2^{P_{26}} Z_2^{E_{26}}})\Theta(Z_2^{P_{26}})
+\frac{\pi}{2}[J_1(\sqrt{X_2^{P_{26}}
|Z_2^{P_{26}}|})+iN_1(\sqrt{X_2^{P_{26}}
|Z_2^{P_{26}}|})]\Theta(-Z_2^{P_{26}})
\bigg \} . \nonumber \\
\end{eqnarray}

\subsection{Factorization formulae for the three-gluon-vertex
diagrams}

\begin{figure}[tb]
\begin{center}
\begin{tabular}{ccc}
\hspace{-2 cm}
\includegraphics[scale=0.70]{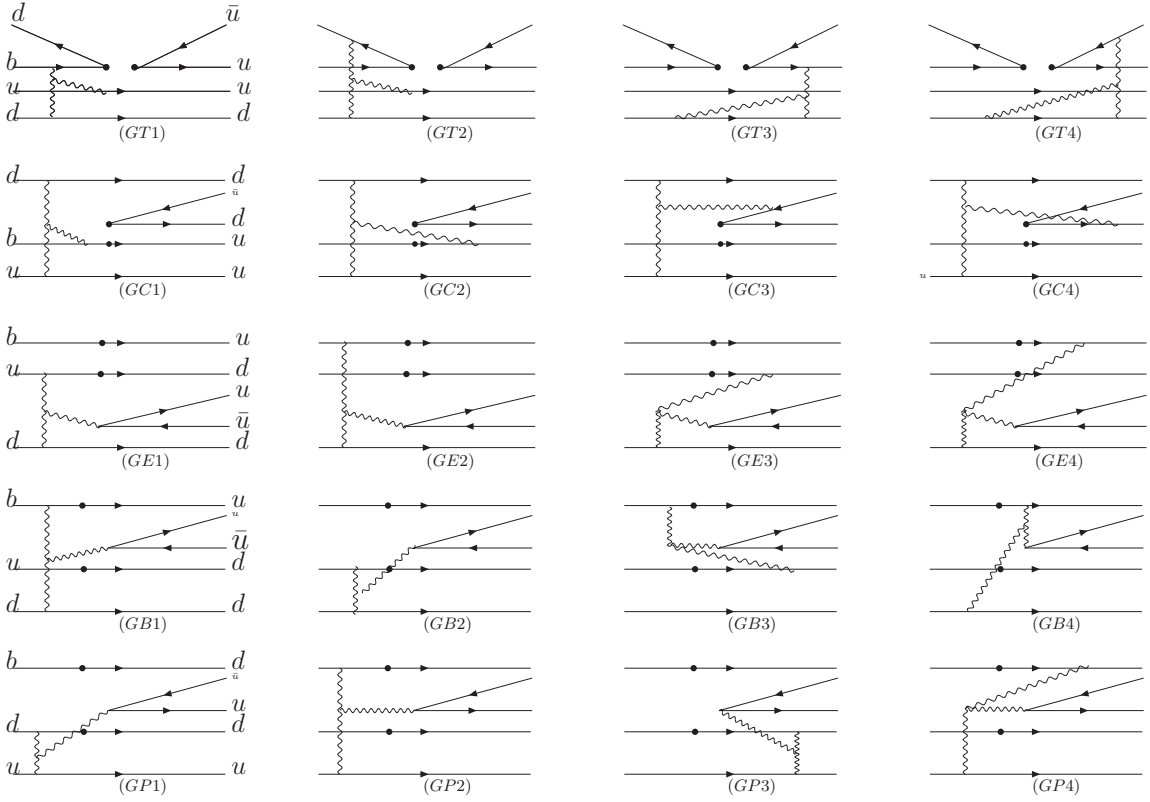}
\end{tabular}
\caption{Feynman diagrams responsible for the $\Lambda_b \to p \pi$
decay with three-gluon-vertex to the lowest order in the pQCD
approach where the dots denote the weak interactions
vertices.}\label{three-gluon-diagrams}
\end{center}
\end{figure}

Now, we can focus on the hard amplitudes contributed by the
topological diagrams shown in Fig. \ref{all-Feynman-diagrams} with
the insertion of the  three-gluon-vertex, which have been grouped in
Fig. \ref{three-gluon-diagrams}. It needs to be pointed out
that the insertion of the three-gluon-vertex to the external and
internal $W$ emission diagrams, namely the diagrams  $GTi$ and $GCi$
$(i=1-4)$ in Fig. \ref{three-gluon-diagrams} have null
effect on the decay amplitude, since the color factors in these
diagrams are proportional to $\epsilon_{i j k} \epsilon_{i^{\prime}
j^{\prime} k^{\prime}} f^{a b c} (T^a)_{i^{\prime} i} (T^b
)_{j^{\prime} j} (T^c)_{k^{\prime} k}$, which equals zero taking
into account the symmetry property of the structure constant $f^{a b
c}$. This is also the reason why the Feynman
diagrams with the three-gluon-vertex are neglected in computing the hard
amplitudes for the semi-leptonic decays of the $\Lambda_b$ baryon
\cite{h.n.li_proton, Shih:1999yh, He:2006ud}.

For the 1st diagram in Fig. \ref{three-gluon-diagrams}
(labeled as $GE1$), we have:
\begin{eqnarray}
f_1^{GE1}&=& G_F  { \pi^2 \over 24 \sqrt{3}} f_{\Lambda_b} f_p \int
[d x ] \int [d x^{\prime}] \int dy \,
[\alpha_s(t^{GE1})]^2  \psi_{\Lambda_b}(x) \nonumber\\
&\times &\bigg\{ \bigg[16  M_{\Lambda_b}^4((C_5+
C_7)-(C_6+C_8))V_{tb}V_{td}^* (M_{\Lambda_b} (3 x_2^{\prime}+x_3)
\phi_M^A(y) \nonumber \\
&& + m_0 ((3-2 x_3)\phi_M^P(y)-( 1- 2 y -2  x_2^{\prime}) ) \phi_M^T(y)  \bigg]\psi_p^T(x^{\prime})\nonumber\\
&\times &\frac{1}{16 \pi^2}\int b_1db_1 \int b_2^{\prime}
db_2^{\prime} \int b_q db_q \int d\theta_1 \int d\theta_2 \, {\rm
exp} [-S^{GE1} (x,
x^{\prime}, y,  b, b^{\prime}, b_q)] \,  \nonumber \\
&&\{ K_0(\sqrt{A^{GE1}}|b_1|) \theta(A^{GE1})+ {\pi i \over 2}
[J_0(\sqrt{|A^{GE1}|}|b_1|) + i N_0(\sqrt{|A^{GE1}|}|b_1|)]
\theta(-A^{GE1}) \}
\int_0^1\frac{dz_1dz_2}{z_1(1-z_1)} \nonumber\\
&&\sqrt{\frac{X_2^{GE1}}{|Z_2^{GE1}|}} \bigg\{ K_1 (\sqrt{X_2^{GE1}
Z_2^{GE1}})\Theta(Z_2^{GE1}) +\frac{\pi}{2}[J_1(\sqrt{X_2^{GE1}
|Z_2^{GE1}|})+iN_1(\sqrt{X_2^{GE1} |Z_2^{GE1}|})]\Theta(-Z_2^{GE1})
\bigg \} , \nonumber \\
\end{eqnarray}
where the auxiliary functions in the expression above are defined as
\begin{eqnarray}
&&A^{GE1}=-x_2^{\prime} M_{\Lambda_b}^2,
B^{GE1}=-x_2^{\prime}M_{\Lambda_b}^2,
C^{GE1}=x_2^{\prime}(y-1)M_{\Lambda_b}^2,\,\,
D^{GE1}=x_3 y  M_{\Lambda_b}^2, \nonumber\\
&&Z_2^{GE1} = B^{GE1}(1-z_2)+\frac{z_2}{z_1(1-z_1)}[C^{GE1}(1-z_1)+D^{GE1} z_1] , \nonumber\\
&&X_2^{GE1} = [(-b_1 +b_2^{\prime}+b_q)-z_1 b_q]^2 +
\frac{z_1(1-z_1)}{z_2} b_q^2 ,
\nonumber\\
&& t^{GE1} =
max(\sqrt{|A^{GE1}|},\sqrt{|B^{GE1}|},\sqrt{|C^{GE1}|},\sqrt{|D^{GE1}|},
\omega,\omega^{\prime}, \omega_q).
\end{eqnarray}
Similarly, the factorization formula for the form factor $f_2$
contributed by $GE1$ are written as:
\begin{eqnarray}
f_1^{GE1}=-f_2^{GE1}.
\end{eqnarray}

For the 3rd diagram in Fig. \ref{three-gluon-diagrams}
(labeled as $GE3$), we have:
\begin{eqnarray}
f_1^{GE3}&=& - G_F  { \pi^2 \over 24 \sqrt{3}} f_{\Lambda_b} f_p
\int [d x ] \int [d x^{\prime}] \int dy \,
[\alpha_s(t^{GE1})]^2 \psi_{\Lambda_b}(x) \nonumber\\
&\times &\bigg\{ \bigg[-8 M_{\Lambda_b}^4
[(C_1-C_2)V_{ub}V_{ud}^*+((C_3+ C_9)-(C_4+C_{10})) V_{tb}V_{td}^*]
\nonumber \\ && \hspace{0.5 cm} \times (2 M_{\Lambda_b} x_2^{\prime}
\phi_M^A(y) +m_0 (3
\phi_M^P(y)+(1- 2 y) \phi_M^T(y))  \nonumber\\
&&\;\;\;\;-8 M_{\Lambda_b}^4((C_5+ C_7)-(C_6+C_8))V_{tb}V_{td}^* (2
M_{\Lambda_b} x_2^{\prime} \phi_M^A(y) +m_0 (3 \phi_M^P(y) +(1- 2y)
\phi_M^T(y))  \bigg]\psi_p^V(x^{\prime})\nonumber\\
&&+\bigg[-8 M_{\Lambda_b}^4 [(C_1-C_2)V_{ub}V_{ud}^*+((C_3+
C_9)-(C_4+C_{10})) V_{tb}V_{td}^*] \nonumber \\
&& \hspace{0.5 cm} \times (2 M_{\Lambda_b} x_2^{\prime}
\phi_M^A(y) +m_0 (3 \phi_M^P(y) +(1- 2y) \phi_M^T(y))  \nonumber\\
&&\;\;\;\;+8 M_{\Lambda_b}^4((C_5+ C_7)-(C_6+C_8))V_{tb}V_{td}^* (2
M_{\Lambda_b} x_2^{\prime} \phi_M^A(y) +m_0 (3 \phi_M^P(y) +(1- 2y)
\phi_M^T(y))  \bigg] \psi_p^{A}(x^{\prime}) \bigg \}\nonumber\\
&&+\bigg[16 M_{\Lambda_b}^4((C_5+ C_7)-(C_6+C_8))V_{tb}V_{td}^*
(x_3^{\prime}-1)( M_{\Lambda_b} (2 y-1)\phi_M^A(y) \nonumber \\
&& \hspace{0.5 cm} - m_0 (3 (y-1) \phi_M^P(y) +(1+y) \phi_M^T(y))
\bigg]
\psi_p^{T}(x^{\prime}) \bigg \}\nonumber\\
&\times &\frac{1}{16 \pi^2}\int b_2^{\prime} db_2^{\prime} \int b_3
db_3 \int b_q db_q \int d\theta_1 \int d\theta_2 \, {\rm exp}
[-S^{GE3} (x,
x^{\prime}, y,  b, b^{\prime}, b_q)] \,  \nonumber \\
&&\{ K_0(\sqrt{A^{GE3}}|b_2^{\prime}|) \theta(A^{GE3})+ {\pi i \over
2} [J_0(\sqrt{|A^{GE3}|}|b_2^{\prime}|) + i
N_0(\sqrt{|A^{GE3}|}|b_2^{\prime}|)] \theta(-A^{GE3}) \}
\int_0^1\frac{dz_1dz_2}{z_1(1-z_1)} \nonumber\\
&&\sqrt{\frac{X_2^{GE3}}{|Z_2^{GE3}|}} \bigg\{ K_1 (\sqrt{X_2^{GE3}
Z_2^{GE3}})\Theta(Z_2^{GE3}) +\frac{\pi}{2}[J_1(\sqrt{X_2^{GE3}
|Z_2^{GE3}|})+iN_1(\sqrt{X_2^{GE3} |Z_2^{GE3}|})]\Theta(-Z_2^{GE3})
\bigg \} , \nonumber \\
\end{eqnarray}
where the auxiliary functions in the above expression are defined as:
\begin{eqnarray}
&&A^{GE3}=(x_3^{\prime}-1)M_{\Lambda_b}^2,
B^{GE3}=-x_2^{\prime}M_{\Lambda_b}^2,
C^{GE3}=x_2^{\prime}(y-1)M_{\Lambda_b}^2,
D^{GE3}=x_3 yM_{\Lambda_b}^2, \nonumber\\
&&Z_2^{GE3} = B^{GE3}(1-z_2)+\frac{z_2}{z_1(1-z_1)}[C^{GE3}(1-z_1)+D^{GE3} z_1] , \nonumber\\
&&X_2^{GE3} = [({b_2}^{\prime} +b_q)-z_1b_q]^2 +
\frac{z_1(1-z_1)}{z_2} b_q^2 ,
\nonumber\\
&& t^{GE3} =
max(\sqrt{|A^{GE3}|},\sqrt{|B^{GE3}|},\sqrt{|C^{GE3}|},\sqrt{|D^{GE3}|},
\omega,\omega^{\prime}, \omega_q).
\end{eqnarray}

Similarly, the factorization formula for the form factor $f_2$
contributed by  $GE3$ can be written as
\begin{eqnarray}
f_2^{GE3}&=& - G_F  { \pi^2 \over 24 \sqrt{3}} f_{\Lambda_b} f_p
\int [d x ] \int [d x^{\prime}] \int dy \,
[\alpha_s(t^{GE3})]^2 \psi_{\Lambda_b}(x) \nonumber\\
&\times &\bigg\{ \bigg[-8 M_{\Lambda_b}^4
[(C_1-C_2)V_{ub}V_{ud}^*+((C_3+ C_9)-(C_4+C_{10}))
V_{tb}V_{td}^*]\nonumber \\
&&\hspace{0.5 cm} \times (2 M_{\Lambda_b} x_2^{\prime} \phi_M^A(y)
+m_0 (3 \phi_M^P(y)
+(1- 2 y) \phi_M^T(y))  \nonumber\\
&&\;\;\;\;-8 M_{\Lambda_b}^4((C_5+ C_7)-(C_6+C_8))V_{tb}V_{td}^* (2
M_{\Lambda_b} x_2^{\prime} \phi_M^A(y) +m_0 (3 \phi_M^P(y) +(1- 2y)
\phi_M^T(y))  \bigg]\psi_p^V(x^{\prime})\nonumber\\
&&+\bigg[-8 M_{\Lambda_b}^4 [(C_1-C_2)V_{ub}V_{ud}^*+((C_3+
C_9)-(C_4+C_{10})) V_{tb}V_{td}^*] \nonumber \\
&&\hspace{0.5 cm} \times (2 M_{\Lambda_b} x_2^{\prime} \phi_M^A(y) +m_0 (3 \phi_M^P(y) +(1- 2y) \phi_M^T(y))  \nonumber\\
&&\;\;\;\;+8 M_{\Lambda_b}^4((C_5+ C_7)-(C_6+C_8))V_{tb}V_{td}^* (2
M_{\Lambda_b} x_2^{\prime} \phi_M^A(y) +m_0 (3 \phi_M^P(y) +(1- 2y)
\phi_M^T(y))  \bigg] \psi_p^{A}(x^{\prime}) \bigg \}\nonumber\\
&&+\bigg[-16 M_{\Lambda_b}^4((C_5+ C_7)-(C_6+C_8))V_{tb}V_{td}^*
(x_3^{\prime}-1)( M_{\Lambda_b} (2 y-1)\phi_M^A(y) \nonumber \\
&& \hspace{0.5 cm} - m_0 (3 (y-1) \phi_M^P(y) +(1+y) \phi_M^T(y))
\bigg]
\psi_p^{T}(x^{\prime}) \bigg \}\nonumber\\
&\times &\frac{1}{16 \pi^2}\int b_2^{\prime} db_2^{\prime} \int b_3
db_3 \int b_q db_q \int d\theta_1 \int d\theta_2 \, {\rm exp}
[-S^{GE3} (x,
x^{\prime}, y,  b, b^{\prime}, b_q)] \,  \nonumber \\
&&\{ K_0(\sqrt{A^{GE3}}|b_2^{\prime}|) \theta(A^{GE3})+ {\pi i \over
2} [J_0(\sqrt{|A^{GE3}|}|b_2^{\prime}|) + i
N_0(\sqrt{|A^{GE3}|}|b_2^{\prime}|)] \theta(-A^{GE3}) \}
\int_0^1\frac{dz_1dz_2}{z_1(1-z_1)} \nonumber\\
&&\sqrt{\frac{X_2^{GE3}}{|Z_2^{GE3}|}} \bigg\{ K_1 (\sqrt{X_2^{GE3}
Z_2^{GE3}})\Theta(Z_2^{GE3}) +\frac{\pi}{2}[J_1(\sqrt{X_2^{GE3}
|Z_2^{GE3}|})+iN_1(\sqrt{X_2^{GE3} |Z_2^{GE3}|})]\Theta(-Z_2^{GE3})
\bigg \} . \nonumber \\
\end{eqnarray}

%%%%%%%%%%%%%%%%%%%%%%%%%%%%%%%%%%%%%%%%%%%%%%%%%%%%%%

\end{document}